\documentclass[fleqn,usenatbib]{mnras}

\usepackage{newtxtext,newtxmath}

\usepackage[T1]{fontenc}

\DeclareRobustCommand{\VAN}[3]{#2}
\let\VANthebibliography\thebibliography
\def\thebibliography{\DeclareRobustCommand{\VAN}[3]{##3}\VANthebibliography}

\usepackage{graphicx}	
\usepackage{amsmath}	

\usepackage{float}
\usepackage{orcidlink}
\usepackage{longtable} 
\usepackage{pdflscape} 
\usepackage{rotating}
\usepackage{booktabs}
\usepackage{siunitx}

\DeclareRobustCommand{\ion}[2]{%
  \relax\ifmmode
    {\mathrm{#1\,\textsc{#2}}}%
  \else
    \textup{#1\,{\mdseries\textsc{#2}}}%
  \fi}

\newcommand{\lam}{$\lambda$}
\newcommand{\ha}{H$\alpha$}
\newcommand{\hb}{H$\beta$}
\newcommand{\hg}{H$\gamma$}

\newcommand{\hi}{\ion{H}{i}}
\newcommand{\hii}{\ion{H}{ii}}
\newcommand{\hh}{\ion{H}{ii}~}
\newcommand{\hei}{\ion{He}{i}}
\newcommand{\heii}{\ion{He}{ii}}

\newcommand{\oii}{[\ion{O}{ii}]}
\newcommand{\oiii}{[\ion{O}{iii}]}
\newcommand{\nii}{[\ion{N}{ii}]}
\newcommand{\sii}{[\ion{S}{ii}]}
\newcommand{\siii}{[\ion{S}{iii}]}
\newcommand{\cliii}{[\ion{Cl}{iii}]}
\newcommand{\ariii}{[\ion{Ar}{iii}]}
\newcommand{\ariv}{[\ion{Ar}{iv}]}
\newcommand{\neiii}{[\ion{Ne}{iii}]}
\newcommand{\feii}{[\ion{Fe}{ii}]}
\newcommand{\feiii}{[\ion{Fe}{iii}]}
\newcommand{\hgi}{\ion{Hg}{i}}
\newcommand{\x}{\ensuremath{\mathbin{\vcenter{\hbox{\scalebox{0.8}{$\times$}}}}}~}

\newcommand{\Rhat}{$\hat{R}$}
\newcommand{\RNe}{$\widehat{RNe}$}
\newcommand{\NA}{\multicolumn{1}{c}{--}}

\title[The DESIRED strong-line calibrations I]{The DESIRED strong-line calibrations: I. New empirical metallicity relations for the local and high-redshift universe}

\author[F. F. Rosales-Ortega et al.]{
F. Fabián Rosales-Ortega$^{1}$\thanks{E-mail: frosales@inaoep.mx}\orcidlink{0000-0002-3642-9146},
J. Eduardo Méndez-Delgado$^{2}$\thanks{E-mail: jmendez@astro.unam.mx}\orcidlink{0000-0002-6972-6411}, 
Jonhatan U. Guerrero-González$^{1}$\orcidlink{0009-0002-2044-3274},
\and
César Esteban$^{3,4}$\orcidlink{0000-0002-5247-5943},
Jorge García-Rojas$^{3,4}$\orcidlink{0000-0002-6138-1869},
Karla Z. Arellano-Córdova$^{5}$\orcidlink{0000-0002-2644-3518},
Alejandra Z. Lugo-Aranda$^{6}$\orcidlink{0000-0001-9226-9178},
\and
Omar Espíndola-Camacho$^{1}$\orcidlink{0009-0007-6272-7108}, 
J. C. López-Gutiérrez$^{2}$\orcidlink{0009-0004-3936-7800},
L. E. Martínez-Rivero$^{2}$\orcidlink{0009-0002-7592-2950},
Christophe Morisset$^{7}$\orcidlink{0000-0001-5801-6724},
\and
Maialen Orte-García$^{3,4}$\orcidlink{0000-0002-0539-1720},
Elena Reyes-Rodríguez$^{3,4}$\orcidlink{0000-0003-1192-6987},
L. Toribio San Cipriano$^{8}$\orcidlink{0000-0002-8313-7875},
Kathryn Kreckel$^{9}$\orcidlink{0000-0001-6551-3091}, 
\and
Oleg Egorov$^{9}$\orcidlink{0000-0002-4755-118X},
Igor A. Zinchenko$^{9,10}$\orcidlink{0000-0002-2944-2449},
Sebastián F. Sánchez$^{2,3}$\orcidlink{0000-0001-6444-9307}
and
José M. Vílchez$^{11}$\orcidlink{0000-0001-7299-8373}
\\\\
$^{1}$Instituto Nacional de Astrofísica, Óptica y Electrónica (INAOE SECIHTI), Luis E. Erro 1, 72840, Tonantzintla, Puebla, México\\
$^{2}$Instituto de Astronomía, Universidad Nacional Autónoma de México, A.P. 70-264, 04510 CDMX, México\\
$^{3}$Instituto de Astrofísica de Canarias, E-38205, La Laguna, Tenerife, Spain\\
$^{4}$Departamento de Astrofísica, Universidad de La Laguna, E-38206 La Laguna, Tenerife, Spain\\
$^{5}$Institute for Astronomy, University of Edinburgh, Royal Observatory, Edinburgh, EH9 3HJ, United Kingdom\\
$^{6}$Instituto de Astronomía, Universidad Nacional Autónoma de México, A.P. 106, Ensenada 22800, BC, México\\
$^{7}$Instituto de Ciencias Físicas, Universidad Nacional Autónoma de México, Av. Universidad s/n, 62210 Cuernavaca, Morelos, Mexico\\
$^{8}$Centro de Investigaciones Energéticas, Medioambientales y Tecnológicas (CIEMAT), Madrid, Spain\\
$^{9}$Astronomisches Rechen-Institut, Zentrum für Astronomie der Universität Heidelberg, Mönchhofstraße 12-14, D-69120 Heidelberg, Germany\\
$^{10}$Main Astronomical Observatory, National Academy of Sciences of Ukraine, 27 Akademika Zabolotnoho Street, UA-03680 Kyiv, Ukraine\\
$^{11}$Instituto de Astrofísica de Andalucía (IAA-CSIC), Glorieta de la Astronomía s/n, 18008 Granada, Spain
}

\date{Accepted XXX. Received YYY; in original form ZZZ}

\pubyear{\the\year{}}

\begin{document}
\label{firstpage}
\pagerange{\pageref{firstpage}--\pageref{lastpage}}
\maketitle

\begin{abstract}
We present the most comprehensive set of empirical optical strong-line metallicity calibrations to date, based on the DEep Spectra of Ionised REgions Database (DESIRED), the largest compilation of \hh regions and galaxies with direct electron-temperature determinations. We construct a high-quality calibration sample of 2392 spectra: 1029 extragalactic \hh regions, 1296 local star-forming galaxies, and 67 high-redshift ($z > 2$) galaxies, drawn from 201 independent literature references, spanning $12+\log({\rm O/H}) \in [6.79, 9.07]$. Physical conditions and chemical abundances are derived homogeneously using up-to-date atomic data. We derive 27 strong-line calibrations covering oxygen-, nitrogen-, sulphur-, argon-, and neon-based line ratios, including 4 previously uncalibrated diagnostics, with reported validity ranges and intrinsic dispersions (typically $\sim0.15-0.35$~dex). For the first time in a systematic calibration framework, all relations are presented for both the homogeneous temperature case ($t^2 = 0$) and a scenario including temperature inhomogeneities ($t^2 > 0$), thereby reconciling abundances from recombination lines (RLs) and collisionally excited lines (CELs) and directly tackling the abundance discrepancy problem. A comparison with previous calibrations shows that the DESIRED relations span the broadest validity intervals while remaining anchored to the empirical data. 
Comparisons with $z>2$ galaxies show that several DESIRED calibrations remain consistent with independent high-redshift relations within their intrinsic scatter, whereas others display systematic offsets associated with elevated N/O or harder ionising conditions. Thus, DESIRED calibrations can provide reliable metallicity estimates at high redshift for a subset of diagnostics, while highlighting the need to account for diagnostic-specific redshift evolution.
\end{abstract}




\section{Introduction}
\label{sec:1}

The chemical evolution of galaxies is fundamental to our understanding of their formation and growth. Unlike dark matter, heavy elements (or metals) trace the baryonic processes that shape the visible Universe. Metals are produced by stars, redistributed by feedback, and diluted by inflows; their spatial and temporal patterns therefore trace the baryon cycle (accretion, star formation, enrichment, and outflows) and provide stringent constraints on cosmological simulations \citep{FinlatorDave:08, Dave:12, Lilly:13}. Metallicity affects the cooling efficiency of the interstellar medium (ISM), dust formation, star-formation regulation, and the emergent ionising spectrum of massive stars, thereby coupling stellar evolution to the observable properties of \hh regions and star-forming galaxies (SFGs) across cosmic time \citep{MaiolinoMannucci:19, Kewley:19, PerouxHowk:20, Sanders:26}. The metal content of the gas phase is therefore one of the key physical quantities in galaxy evolution studies.

The interplay of the baryon cycle that governs the flow of material into and out of galaxies is reflected, empirically, in scaling relations that encode information about galaxy evolution. For example, the mass--metallicity relation (MZR) demonstrates that more massive systems are systematically more chemically enriched \citep{Lequeux:79, Tremonti:04}, showing a clear evolution with redshift, with higher-redshift galaxies exhibiting systematically lower metallicities at fixed stellar mass \citep{Erb:06, Maiolino:08, Sanders:21, Curti:24}. The scatter and secondary trends (with star-formation rate, stellar mass and gas content) are commonly interpreted as the interplay between metal-poor inflows from the intergalactic medium, metal-enriched outflows driven by stellar feedback and active galactic nuclei, and changes in star-formation efficiency \citep{Brooks:07, Mannucci:10, Lilly:13, Dave:11b, Dave:12, Chisholm:15, Kewley:19, Curti:17}. On smaller scales, metallicity gradients within galaxies serve as a dynamical clock for interaction and star-formation histories. Isolated spirals typically exhibit steep, negative gradients consistent with inside-out disc formation, whereas galaxy mergers flatten these gradients by driving pristine gas into central regions \citep{Kewley:10, Sanchez:14, Ho:15}.

A robust determination of the gas-phase metallicity is therefore essential for interpreting these scaling relations and, more broadly, for constraining the chemical evolution of galaxies. Metallicity can be inferred from a wide range of nebular emission lines, with the classical approach relying on the electron temperature (T$_{\rm e}$) derived from temperature-sensitive auroral lines in the optical spectrum. In ionised nebulae, the flux of an emission line depends on the abundance of the ionic species and its emissivity, which is sensitive to the $T_{\rm e}$ and the electron density ($n_{\rm e}$). Collisionally excited lines (CELs) are typically bright but depend exponentially on $T_{\rm e}$, whereas recombination lines (RLs) depend much more weakly on $T_{\rm e}$ but are, in the case of heavy elements, orders of magnitude fainter, and thus only measurable in a few nearby regions \citep[e.g.,][]{GarciaRojasEsteban:07, Peimbert:17}. Electron temperatures can be inferred from the ratio of collisionally excited transitions of the same ion originating from different upper energy levels. The so-called ``auroral'' lines--transitions from high-lying quantum states whose excitation is extremely sensitive to temperature--enable this measurement, classically \oiii $\lambda$4363 relative to \oiii $\lambda\lambda$4959,5007, but also
the ratios of \nii~$\lambda$5755, \siii~$\lambda$6312, and \oii~$\lambda\lambda$7320,7330 with respect to their corresponding strong-line counterparts. This temperature-based approach is often referred to as the ``direct'' method \citep{Dinerstein:90, OsterbrockFerland:06, Kewley:19}. 

In practice, however, it is observationally demanding: auroral lines are intrinsically faint (often hundreds of times fainter than the strongest Balmer lines), and become increasingly difficult to detect in metal-rich nebulae, where enhanced cooling lowers $T_{\rm e}$ and further suppresses auroral emission. This challenge is compounded, particularly for \oiii~$\lambda$4363, by contamination from \hgi~$\lambda$4358 sky emission or \feii~$\lambda$4360 blends in low-resolution spectra of high-metallicity objects \citep[e.g.,][]{Curti:17, ArellanoCordovaRodriguez:20}, whereas other auroral lines such as \nii~$\lambda$5755 are less affected by these specific issues, though they present their own detection challenges at low metallicity.

As a result, robust $T_{\rm e}$-based abundances have historically been limited to relatively small samples of nearby \hh regions and galaxies, or have required statistical approaches such as stacking large numbers of spectra \citep[e.g.][]{AndrewsMartini:13, Curti:17, MaiolinoMannucci:19}. 

The practical difficulty of measuring auroral lines has motivated a vast literature of alternative metallicity estimators based on strong nebular lines. Optical nebular line ratios are most widely exploited because some of the strongest lines fall in this wavelength range, are easily accessible from ground-based facilities, and have been observed for enormous samples, from local surveys to deep high-redshift spectroscopy. Strong-line calibrations have been developed empirically through the direct method \citep[e.g.,][]{Pagel:79, PilyuginGrebel:16, Curti:17, Curti:20}, through photoionisation models \citep[e.g.,][]{McGaugh:91, KewleyDopita:02, Dopita:16, PerezMontero:21}, or using hybrid approaches \citep[e.g.,][]{PerezMonteroDiaz:05, Maiolino:08}. Each of these methods has its own strengths and weaknesses, they rely on different samples of \hh regions, models, parameters, simplified assumptions, and corrections for unseen stages of ionisation \citep{Stasinska:05, LopezSanchez:12, Peimbert:17}.

For empirical calibrations, the methodology involves measuring precise metallicities through the direct method, comparing these abundances with strong-line flux ratios, and calibrating the relationships through suitable functional forms. Yet, systematic discrepancies among calibrations remain substantial. Different methods can disagree by several tenths of a dex, affecting inferred scaling relations and their redshift evolution \citep{KewleyEllison:08, Moustakas:10, LopezSanchez:12, Curti:17, Kewley:19}. Some strong-line diagnostics are also degenerate with other nebular parameters (e.g., the ionisation parameter, the ratio of ionising photon density to particle density;  pressure, $P \approx nkT_{\rm e}$; or density), or encode indirect correlations--most notably those involving nitrogen, whose abundance depends on both primary and secondary nucleosynthetic channels and can vary at fixed O/H \citep{Pilyugin:12a, PerezMonteroContini:09, PerezMontero:21,Scholte:25}. These issues motivate the use of multiple diagnostics and, where possible, self-consistent approaches that mitigate sensitivity to ionisation conditions \citep[e.g.][]{Dopita:16,PilyuginGrebel:16,Brazzini:24}.

A further fundamental limitation lies in the thermal structure of nebulae. Treating \hh regions as isothermal and homogeneous is a convenient approximation, but it may not be physically realistic. The common two-zone (or sometimes three-zone) approach assigns separate temperatures to low- and high-ionisation zones, typically anchored by $T_2 \equiv T_{\rm e}$(\nii) or $T_{\rm e}$(\oii) for the low-ionisation region and $T_3 \equiv T_{\rm e}$(\oiii) for the high-ionisation region, with $T_3$--$T_2$ relations used when only one auroral diagnostic is available \citep{Campbell:86, Garnett:92, Kewley:19, ArellanoCordovaRodriguez:20, OrteGarcia:26}. This step is not merely technical: the chosen $T_3$--$T_2$ relation propagates directly into ionic abundances and thus into the calibration of most strong-line indices \citep{PerezMontero:21}.

Beyond zone-averaged temperatures, a more fundamental issue arises from the presence of temperature inhomogeneities within the ionised gas. For over 80 years, a systematic discrepancy of order a factor of two has been reported between heavy-element abundances derived from CELs and those inferred from RLs \citep{Wyse:42, Peimbert:17}. Studies of O/H abundances from ionising stars in Galactic \hh regions show systematically higher abundances compared to nebular determinations using CELs \citep{SimonDiazStasinska:11, GarciaRojas:14, MartinezHernandez:2026}. When extremely faint heavy-element RLs are employed instead, these discrepancies are largely mitigated, suggesting an inhomogeneous nebular temperature structure within \hh regions \citep{Peimbert:67, GarciaRojasEsteban:07, MendezDelgado:22a}. 
However, the picture is less clear in the extragalactic domain, where some studies find good agreement between chemical abundances in massive stars and CEL-based nebular metallicities \citep[e.g.][]{Bresolin:25}, while others do not. The origin of these discrepancies remains debated, and may reflect systematic uncertainties in stellar atmosphere modelling, difficulties in deriving abundances from blue supergiants, or genuine variations in the nebular temperature structure across different galaxy environments. 
Such temperature inhomogeneities can bias CEL-based abundances relative to RL-based abundances because CEL emissivities depend exponentially on the local $T_{\rm e}$, causing the hottest regions of the nebula to dominate the observed emission. This is central to the long-standing ``abundance discrepancy'' problem, in which abundances inferred from RLs exceed those from CELs, casting doubt on the absolute calibration of nebular abundance scales.

Recently, \citet{MendezDelgado:23a} presented observational evidence that internal temperature inhomogeneities, quantified by the root mean square temperature fluctuations parameter \citep[$t^2$,][]{Peimbert:67}, may contribute significantly to the abundance discrepancy problem in star-forming regions. Their results suggest that metallicities derived from CELs may be systematically underestimated relative to those obtained when temperature fluctuations are taken into account, with the magnitude of the effect depending on the physical conditions of the ionised gas \citep{Peimbert:67, PeimbertCostero:69, LopezSanchez:12, Toribio:17, MendezDelgado:23a}. 

In parallel, recent studies have shown that density inhomogeneities can also bias abundance determinations by affecting \oii{} and \sii{} emission-line ratios to Balmer lines, while having a comparatively minor impact on \nii\ diagnostics \citep{MendezDelgado:23b, Cataldi:25, Martinez:25}. If strong-line calibrations are anchored to CEL-based ``direct'' abundances under the implicit assumption of a homogeneous temperature structure ($t^2=0$), any systematic bias can propagate into secondary metallicity diagnostics, potentially introducing offsets in the inferred abundance scale. These physical effects--temperature and density inhomogeneities--have so far been only partially or implicitly accounted for in most widely used calibrations, and their omission introduces systematic errors in gas-phase abundance determinations that are increasingly relevant in the context of precision studies of galaxy evolution \citep{Peimbert:07}. Among the few attempts to explicitly correct strong-line methods for temperature inhomogeneities, \citet{PenaGuerrero:17} applied a correction function to the R23 calibration of \citet{PilyuginThuan:05} to account for the effects of $t^2$, demonstrating that such corrections can systematically shift derived abundances by several tenths of a dex.

Moreover, because strong-line relations are empirical mappings between observed emission-line ratios and metallicity, the properties of the calibration sample play a central role in setting their accuracy, robustness, and domain of applicability. Systematic biases can arise if the objects with detected auroral lines are not representative of the full range of ionisation conditions, or if the sample lacks sufficient diversity and coverage across the multi-dimensional parameter space defined by excitation, electron density, and chemical abundance patterns \citep{Curti:17, PilyuginGrebel:16}. In such cases, the resulting calibrations may perform well within the limited regime sampled by the calibrators, yet yield biased or unreliable metallicities when applied to broader or physically distinct galaxy populations \citep{Stasinska:10}.

Historically, many widely used strong-line calibrations have been derived from relatively modest samples, typically comprising a few hundred objects. For instance, the classical N2 and O3N2 calibrations evolved from early compilations based on limited datasets \citep[e.g.][]{Denicolo:02, PettiniPagel:04} to later efforts incorporating larger $T_{\rm e}$-based samples and integral-field spectroscopy of nearby galaxies \citep[e.g.][]{PerezMontero:09, Marino:13, PilyuginGrebel:16}, while alternative strategies have combined low-metallicity detections with stacked spectra to extend $T_{\rm e}$-based calibration coverage towards high metallicity \citep{Curti:17, Curti:20}. Nevertheless, several influential and relatively recent studies still rely on calibration samples containing a few hundred ($<$400) objects \citep[e.g.][]{Pilyugin:12b, Steidel:14, Yates:20, Nakajima:22, Brazzini:24}. 
Among calibrations built from heterogeneous literature sources, the compilations of \citet{Marino:13} (603 spectra) and \citet{Ho-Ting:19} ($\sim$950 spectra) remain the largest, the latter also exploring artificial neural networks as a flexible, non-linear fitting framework\footnote{Although \citet{Pilyugin:12b} and \citet{PilyuginGrebel:16} report parent compilations comprising 715 and 965 spectra, respectively, the final calibration relations in those works are based on smaller subsamples, containing 414 and 313 spectra.}. By contrast, \citet{PerezMontero:21} presented the largest calibration sample drawn from a single, homogeneous source, using 1268 spatially unresolved, extreme emission-line galaxies (EELGs) from the SDSS to calibrate multiple strong-line diagnostics, albeit within a restricted region of parameter space. More recently, \citet{Scholte:25} extended this effort by combining 782 local \oiii~$\lambda$4363-detected galaxies from the DESI Early Data Release with a sample of high-redshift sources from JWST/NIRSpec.


Indeed, the sensitivity and wavelength coverage of JWST/NIRSpec have enabled the detection of auroral lines in growing samples of galaxies at $z>1$, extending direct-method metallicity measurements to $z\sim10$ \citep{Schaerer:22, ArellanoCordova:22, Curti:23, Sanders:24, Laseter:24, Scholte:25, ArellanoCordova:26}. This observational breakthrough has motivated a new generation of strong-line calibrations specifically intended for high-redshift galaxies, based on the premise that the physical conditions of the ionised ISM evolve with redshift. In particular, high-redshift galaxies are thought to exhibit higher electron densities and harder ionising spectra than their local counterparts \citep{Steidel:14, Shapley:15, Sanders:16a, Sanders:20, Sanders:26}.
This has prompted empirical calibrations from surveys including JADES \citep{Laseter:24, Isobe:2026}, PRIMAL \citep{Chakraborty:25}, MARTA \citep{Cataldi:25}, EXCELS \citep{Scholte:25} and AURORA \citep{Sanders:26}. However, current high-redshift samples with auroral detections remain limited (typically $20$--$140$ galaxies) and predominantly cover moderately low metallicities regimes, limiting their leverage across the full parameter space relevant for galaxy evolution studies.

In this context, there is a clear need for a new generation of calibration datasets that are simultaneously large, built explicitly on high-quality spectroscopy, analysed in a fully homogeneous manner, and designed to span the relevant parameter space rather than optimising for auroral-line detectability alone. Over the past decade, a vast amount of deep optical data has been obtained with 8--10\,m class telescopes, yet these observations remain fragmented across the literature and have not been consolidated into a modern, consistently analysed calibration sample. As a result, the level of systematic disagreement among published strong-line calibrations has not been substantially reduced, despite the substantial improvement in observational capabilities. Two fundamental limitations underpin this stagnation: first, the widespread assumption of negligible temperature inhomogeneities in nebular abundance determinations, and second, the absence of a sufficiently large, up-to-date calibration sample explicitly selected for data quality rather than being driven primarily by auroral-line detectability.

The goal of this work is to provide a new generation of optical strong-line metallicity calibrations based on the most extensive and high-quality compilation of $T_{\rm e}$-based data assembled to date, while explicitly addressing for the first time the role of temperature variations ($t^2>0$) in the abundance determinations. For this purpose, we employ spectra of over 2,300 \hh regions and SFGs with optical auroral-line measurements from the DESIRED project (DEep Spectra of Ionised REgions Database; \citealt{MendezDelgado:23a,MendezDelgado:24b, Esteban:25}), a collaborative project focused on the homogeneous statistical analysis of spectra of ionised nebulae.



This first paper of the DESIRED strong-line calibration series is organised as follows. Sec.~\ref{sec:2} describes the DESIRED catalogue of ionised nebulae, the selection criteria applied to define the calibration sample, and the methodology adopted to derive physical conditions and chemical abundances in a fully homogeneous fashion using up-to-date atomic data.
In Sec.~\ref{sec:3} we presents 27 empirical optical strong-line metallicity calibrations derived from the high-quality DESIRED calibration sample---23 previously reported in the literature and 4 introduced in this work---covering oxygen-, nitrogen-, sulphur-, argon-, and neon-based line ratios, with their corresponding validity ranges and intrinsic dispersions. All calibrations are derived for both the temperature-homogeneous case ($t^2=0$) and a scenario accounting for temperature inhomogeneities ($t^2>0$), directly addressing the abundance discrepancy problem by reconciling metallicities derived from RLs and CELs, and are applicable to both the local and high-redshift Universe.
In Sec.~\ref{sec:4} we evaluate the performance of the derived calibrations by comparing the inferred abundances against the $T_{\rm e}$-based metallicities of the DESIRED sample, quantifying their empirical scatter. We also compare the new calibrations with previous determinations from the literature and discuss their implications. We further assess the applicability of the DESIRED calibrations at high redshift, arguing that the ionisation conditions of early-Universe galaxies are already represented within a sufficiently diverse calibration sample, such that redshift-specific recalibration may be unnecessary when the calibrating data span the full range of relevant physical conditions. The section closes with practical guidelines for the correct application of the DESIRED calibrations to spectroscopic observations of star-forming regions. Finally, we summarise our results and present future perspectives in Sec.~\ref{sec:5}.

\section{Calibration sample}
\label{sec:2}

\subsection{The DESIRED sample}
\label{sec:desired}

For this work, we use the latest compilation of the DEep Spectra of Ionised REgions Database (DESIRED) project \citep{MendezDelgado:23a,MendezDelgado:24b}. The DESIRED collaboration aims to investigate the properties of photoionised regions in unprecedented detail by compiling and analysing a large number of high-quality optical spectra of ionised nebulae using a homogeneous methodology.

Originally, DESIRED consisted of a collection of nearly 200 deep, high signal-to-noise (S/N) optical spectra of Galactic and extragalactic \hh regions, SFGs, Galactic planetary nebulae, and other ionised gaseous objects observed by the DESIRED collaboration over the past 20 years.

Subsequently, the original compilation was extended to include high-quality UV--optical--NIR spectra of \hh regions, local and high-redshift SFGs, and planetary nebulae from the literature, provided that at least one of the following auroral-to-nebular intensity ratios of collisionally excited lines (CELs) was detected: \oiii{} $\lambda$4363/$\lambda$5007, \nii{} $\lambda$5755/$\lambda$6584, or \siii{} $\lambda$6312/$\lambda$9069. These line ratios enable a direct determination of the electron temperature, $T_{\rm e}$. This expanded dataset is referred to as DESIRED-E (DESIRED Extended; \citealt{MendezDelgado:24b, Esteban:25}). From this point onward, and for simplicity, any reference to DESIRED in the remainder of the text should be understood as referring specifically to the DESIRED-E sample.

For this purpose, we performed a comprehensive search of the literature for emission-line regions presenting at least one of the auroral-to-nebular intensity ratios listed above. For each spectrum, we compiled the extinction-corrected intensity ratios of all emission lines reported in the reference studies. For a subset of high-redshift galaxies observed with JWST for which the original studies do not report extinction-corrected line intensities \citep{Stiavelli:25, Tang:25, Scholtz:25, Pollock:25, Harikane:25, Bhattacharya:25}, we recomputed the reddening correction homogeneously, adopting the \citet{Fitzpatrick:99} extinction law with $R_V=3.1$ and assuming an intrinsic \ha/\hb\ ratio of 2.86 for non-obscured ionised regions; objects with observed \ha/\hb\ <2.86 were assigned zero reddening.

Unlike previous compilations, we also collected the reported uncertainties in the line intensities, which allowed us to select only the highest-quality spectra. In all cases, we considered exclusively those line ratios with reported observational uncertainties below 40 per cent (S/N $> 2.5$).
We refer to this compilation as the DESIRED sample. As of 31 March 2026, it comprises 3236 high-quality spectra of ionised nebulae. By object type, the sample consists of 40 per cent \hh regions (Galactic and extragalactic), 47 per cent local and high-redshift SFGs, 12 per cent planetary nebulae (PNe), and 1 per cent other ionised gaseous objects\footnote{Ring nebulae (RNe) around evolved massive stars, Herbig--Haro objects, and protoplanetary discs in the Orion Nebula.}. 

For the purposes of this paper, which aims to define a calibration sample for metallicity indicators, we restrict our analysis to a subsample comprising extragalactic \hh regions (individual star-forming nebulae in local spiral, starburst or irregular galaxies) and local and high-redshift ($z = 1-10$) SFGs, whose general spectroscopic properties are similar from giant extragalactic \hh regions \citep[][]{Sargent:70, Melnick:85, Telles:97}, for a total of 2762 spectra. 

\subsection{Nebular physical conditions and oxygen abundances}
\label{sec:nebular}

The physical conditions of the gas---namely the electron density, $n_{\rm e}$, and the electron temperature, $T_{\rm e}$,---as well as the ionic and total oxygen abundances of the DESIRED subsample described above were derived using PyNeb version 1.1.18 \citep{Luridiana:15, Morisset:20, Mendoza:23}, adopting the atomic data listed in Appendix~\ref{app:atomic} and the \hi\ effective recombination coefficients from \citet{StoreyHummer:95}, following the prescriptions described in \citet{MendezDelgado:23b}, restricted for CELs. Below, we summarise the procedure.

\begin{table}
\centering
\caption{Number of available auroral lines on the DESIRED calibration sample.}
\label{tab:desired}
\begin{tabular}{lcc}
\toprule
Auroral lines / diagnostics & Regions & Percentage\\
\midrule
\oiii~ $\lambda$4363/$\lambda$5007 & 1973 & 82\% \\
\nii~ $\lambda$5755/$\lambda$6584  &  593 & 25\% \\
\siii~ $\lambda$6312/$\lambda$9069 & 1271 & 53\% \\
\midrule
Multiple\\
\midrule
\oiii~ \& \nii  & 232 & 10\% \\
\oiii~ \& \siii & 993 & 42\% \\
\nii~  \& \siii & 418 & 17\% \\
\oiii~ \& \nii~ \& \siii & 198 & 8\% \\
\bottomrule
\end{tabular}
\end{table}

To calculate the emissivity of collisional excitation lines, PyNeb calculates the relative population of the atomic levels of the ions of interest by solving the statistical equilibrium equations. Depending on the available emission lines of each spectrum of the DESIRED subsample, density-sensitive line intensity ratios (e.g., \sii{} $\lambda$6731/$\lambda$6717, \oii{} $\lambda$3726/$\lambda$3729) are cross-correlated with temperature-sensitive diagnostics (\nii{} $\lambda$5755/$\lambda$6584, \oiii{} $\lambda$4363/$\lambda$5007, \siii{} $\lambda$6312/$\lambda$9069), propagating uncertainties in line intensities using a Monte Carlo simulation. For each convergence, a value of $n_{\rm e}$ and $T_{\rm e}$ and their associated uncertainties are obtained. The average $n_{\rm e}$, weighted by the inverse square of the error of the different convergences, is adopted as the representative value of the diagnostic (e.g., $n_{\rm e}$(\sii{} $\lambda$6731/$\lambda$6717), this approach takes into account small temperature dependence of density diagnostics under typical nebular conditions). Once the density for each available diagnostic is obtained, the criteria outlined by \citet{MendezDelgado:23b} is applied to estimate a global average density representative of each nebulae: 
a) if $n_{\rm e}$(\sii{} $\lambda$6731/$\lambda$6716) $<$ 100 cm$^{-3}$, $n_{\rm e} = 100\pm100$ cm$^{-3}$; 
b) if $n_{\rm e}$(\sii{} $\lambda$6731/$\lambda$6716) $\in$ [100,1000] cm$^{-3}$, $n_{\rm e}$ is the average between $n_{\rm e}$(\sii{} $\lambda$6731/$\lambda$6716) and $n_{\rm e}$(\oii{} $\lambda$3726/$\lambda$3729).
c) if $n_{\rm e}$(\sii{} $\lambda$6731/$\lambda$6716) $>$ 1000 cm$^{-3}$, $n_{\rm e}$ is the average between the available diagnostics among $n_{\rm e}$(\sii{} $\lambda$6731/$\lambda$6716), $n_{\rm e}$(\oii{} $\lambda$3726/$\lambda$3729), $n_{\rm e}$(\cliii{} $\lambda$5518/$\lambda$5538), $n_{\rm e}$(\ariv{} $\lambda$4711/$\lambda$4740), and $n_{\rm e}$(\feiii{} $\lambda$4658/$\lambda$4702).In those cases where a value of $n_{\rm e}$ could not be calculated from the source references, a value $n_{\rm e} = 100\pm100$ cm$^{-3}$ was adopted.

Then, the electron temperatures $T_{2}$(\nii{} $\lambda$5755/$\lambda$6584), $T_{3}$(\oiii{} $\lambda$4363/$\lambda$5007) and $T_{\rm e}$(\siii{} $\lambda$6312/$\lambda$9069) are calculated depending on the available auroral lines for each spectrum, using the average $n_{\rm e}$. Uncertainties in the line ratios are propagated through Monte Carlo simulations to estimate the errors associated with each $T_{\rm e}$ diagnostic.

We also account for potential biases in the determination of $T_3$, as the \oiii\ $\lambda$4363 auroral line can be blended with \feii\ $\lambda$4360 in spectra of intermediate or low spectral resolution \citep{Curti:17}. In cases where this blend is identified, we discard the use of \oiii\ $\lambda$4363 to avoid introducing spurious overestimates of $T_3$. The impact of this blending is expected to be more significant in regions of high metallicity and low ionisation degree \citep{ArellanoCordovaRodriguez:20}. Most of the star-forming regions with these properties included in this work are drawn from the CHAOS survey \citep{Berg:15}, for which particular care has been taken to reliably separate the \oiii\ $\lambda$4363 and \feii\ $\lambda$4360 features \citep{Berg:20, Rogers:21, Rogers:22}. Consequently, this observational effect is not expected to introduce systematic biases in the electron temperature determinations presented here.


Ionic abundances are estimated under two scenarios: (a) a homogeneous nebular temperature structure ($t^2=0$; the ``direct method'') and b) accounting for the effect of an inhomogeneous temperature structure ($t^2>0$, \citealt{Peimbert:67}). The \oii{} $\lambda\lambda$3726,3729 doublet is used to determine the ionic abundances of O$^+$. When this doublet is unavailable owing to spectral coverage limitations or observational defects, the auroral lines doublets \oii{} $\lambda\lambda$7319,7320 and \oii{} $\lambda\lambda$7330,7331 are used instead. When both lines of a given pair are resolved (due to high spectral resolution), the sum of their line intensities is adopted.

For $t^2=0$, $T_{2}$(\nii{} $\lambda$5755/$\lambda$6584) and the adopted value of $n_{\rm e}$ are used to compute the abundance of the low-ionisation O$^+$ ion propagating the uncertainties in $n_{\rm e}$, $T_{\rm e}$ and the line ratios through via Monte Carlo simulations. If $T_{2}$ is not available for a given object, the updated DESIRED temperature relations by \citet{OrteGarcia:26} are used to estimate $T_2$ from $T_{3}$(\oiii{} $\lambda$4363/$\lambda$5007) or, when this diagnostic is absent, from $T_{\rm e}$(\siii{} $\lambda$6312/$\lambda$9069).

The abundance of the O$^{2+}$ ion is derived from the sum of the bright \oiii{} $\lambda\lambda$4959, 5007 nebular lines, adopting $T_{3}$ as the temperature representative of the high-ionisation zone and the adopted value of $n_{\rm e}$. When $T_{3}$ is not available, the temperature relations of \citet{OrteGarcia:26} are applied along with the values of $T_{2}$ and/or $T_{\rm e}$(\siii{}), according to the diagnostics available. 

For $t^2>0$, the treatment of the low-ionisation ion O$^+$ is identical to the one described above. For the high-ionisation ion O$^{2+}$, a correction for temperature inhomogeneities is applied following the empirical results of \citet{MendezDelgado:23a}. In this approach, the average electron temperature of the highly ionised gas, $T_0$(O$^{2+}$), is estimated from $T_{2}$(\nii) using their Eq.~(4), reproduced below as Eq.~\eqref{eq:T0vsTN2}:

\begin{equation}
    \label{eq:T0vsTN2}
    T_0(\text{O}^{2+})= (1.17 \pm 0.05) \times T_{2}(\nii) - (3340 \pm 470) \text{ [K]}.
\end{equation}

To roughly estimate the effect of $t^2$ when $T_{2}$ is unavailable, $T_0$(O$^{2+}$) is determined by combining Eq.~\eqref{eq:T0vsTN2} with the temperature relations between $T_{\rm e}$(\nii{}) and $T_{\rm e}$(\oiii{}) proposed by \citet{OrteGarcia:26}, which are in very good agreement with those of \citet{Garnett:92}. The determination of abundances in the presence of temperature fluctuations requires only the knowledge of $T_0$(O$^{2+}$), whose derivation has been described above. From this value, $t^2$(O$^{2+}$) can be explicitly estimated, as both quantities are related through Eq.~18 of \citet{Peimbert:67}.
Note that the $t^2>0$ metallicities derived through this methodology represent a first-order correction based on the adopted empirical parametrisation.

The total oxygen abundance is computed as the sum of the ionic O$^+$/H$^+$ and O$^{2+}$/H$^+$ ratios. As shown in Table~\ref{tab:desired}, 25\% of the sample, in the case of $t^2>0$, is based exclusively on $T_{\rm e}$(\nii{}), which has been shown to be a reliable procedure \citep{Kreckel:2025}. Observational studies of metal-poor regions have shown that the contribution of O$^{3+}$/H$^+$ to the total O/H ratio is small, typically of order 5 per cent ($\approx$0.02 dex at most), and therefore negligible compared to the typical uncertainties in O/H determinations \citep[e.g.][]{Izotov:99a, Amayo:21, Berg:21, DominguezGuzman:22}. In extremely metal-poor \hh regions (12+log(O/H) $<$ 7.5), ionisation conditions may be harder than predicted by standard photoionisation models, potentially leading to detectable \heii\ emission and a non-negligible O$^{3+}$ component. In such cases, an ionisation correction factor (ICF) can in principle be applied, as commonly done for planetary nebulae (PNe, \citealt{TorresPeimbertPeimbert:77}). However, direct measurements in metal-poor systems, including Extreme Emission Line Galaxies with O$^{3+}$/H$^+$ abundances derived from UV O~IV] lines, confirm that the O$^{3+}$ contribution remains at the $\lesssim$5 per cent level \citep{Berg:21}. Given that this contribution is negligible for the purpose of defining a robust calibration sample, the O$^{3+}$/H$^+$ term is not included in the total oxygen abundance.

\subsection{The DESIRED calibration sample}
\label{sec:calib_sample}

\begin{figure}
	\includegraphics[width=\columnwidth]{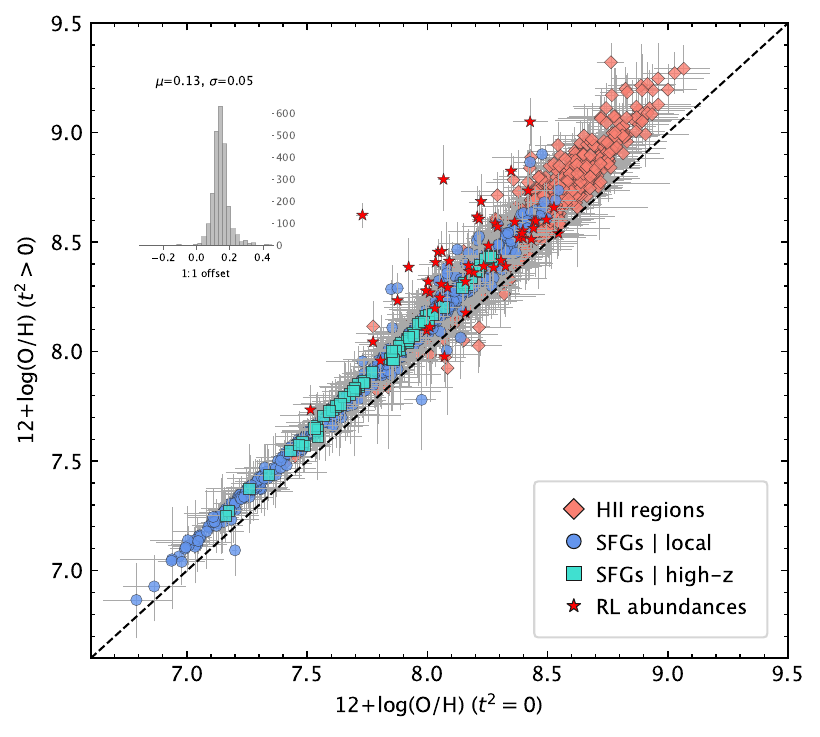}
    \caption{Total oxygen abundance for the DESIRED calibration sample derived under the homogeneous temperature structure assumption ($t^2=0$, x-axis) and corrected for temperature inhomogeneities ($t^2>0$, y-axis). The $t^2>0$ values are systematically higher, as expected from the effect of temperature fluctuations on CEL-based abundance determinations. The dashed line denotes the one-to-one relation and error bars correspond to 1$\sigma$ uncertainties. Symbols denote object type: extragalactic \hh regions (diamonds), local (circles), and high-$z$ (squares) SFGs. A subsample of O/H abundance determinations based on RLs is shown as red stars; these correspond to measurements from the O\,II V1 multiplet and are largely independent of the gas temperature structure. The inset histogram shows the distribution of $\Delta$(O/H) = (O/H)$_{t^2>0}$ $-$ (O/H)$_{t^2=0}$, with a mean offset of 0.13\,dex and a dispersion, $\sigma_{\rm O/H}$, of 0.054\,dex.}
    \label{fig:OH}
\end{figure}

\begin{figure}
	\includegraphics[width=\columnwidth]{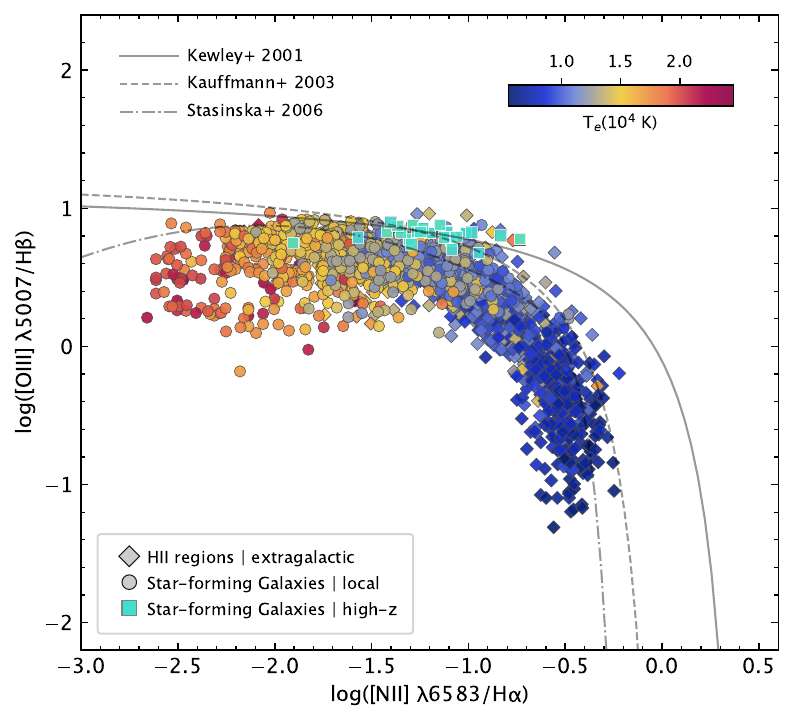}
     \caption{BPT diagnostic diagram of the DESIRED calibration catalogue, colour-coded by their electron temperature, $T_{\rm e} \equiv T_{3}$(\oiii), in units of 10$^4$ K. Solid, dashed, and dash-dotted curves indicate standard demarcation lines separating star-forming regions from other ionisation sources: the theoretical upper limit for photoionised nebulae from \citet{Kewley:01} (solid), the empirical relation of \citet{Kauffmann:03} (dashed), and the dividing line between purely SFGs and AGN hosts from \citet{Stasinska:06b} (dash-dotted). 
     }
    \label{fig:bpt}
\end{figure}

As discussed in Sec.~\ref{sec:1}, a paramount requirement in assembling a sample for metallicity calibrations is the quality of the underlying data. An effective way to constrain sample quality is to make use of the emission-line intensity uncertainties reported in the literature, together with the propagated errors associated with the determination of physical conditions and the total chemical abundances derived in Sec.~\ref{sec:nebular}. With the aim of constructing the highest-quality compilation of $T_{\rm e}$-based literature data for calibration purposes, we base our analysis from the DESIRED subsample of 2762 extragalactic \hh regions and local and high-redshift ($z = 1-10$) SFGs described in Sec.~\ref{sec:desired}, and apply the following high-quality selection criteria: we exclude regions for which: (a) the auroral lines have relative uncertainties larger than 40 per cent (S/N $< 2.5$); (b) the observed nebular line intensity ratios of transitions arising from the same upper atomic level differ by more than 20 per cent from their theoretical values (e.g. \oiii $\lambda$5007/$\lambda$4959, \nii $\lambda$6584/$\lambda$6548, and \siii $\lambda$9531/$\lambda$9069\footnote{In the specific case of the \siii $\lambda\lambda$9531,9069 lines, when the observed ratio $\lambda$9531/$\lambda$9069 $<$ 2.47 \citep{Noll:12}, we assume that the \siii $\lambda$9531 line is affected by telluric absorption bands, whereas \siii $\lambda$9069 remains unaffected.}; \citealt{StoreyZeippen:00}); (c) the uncertainty in the derived oxygen abundance exceeds 0.3 dex; (d) the oxygen abundance derived allowing for temperature fluctuations ($t^2>0$) is lower than that obtained under the assumption of a homogeneous temperature structure ($t^2=0$); (e) the derived electron temperature is higher than 25,000 K, this temperature regime is typically dominated by spurious detections of \oiii{} $\lambda$4363 or by observational defects that artificially increase the derived electron temperature\footnote{While we acknowledge that a fraction of the regions in the DESIRED sample in this regime may correspond to genuine detections, the statistics appear to be largely governed by measurement uncertainties. A larger number of observations in this temperature range is therefore required to support a study of this nature on statistical grounds.}; and (f) the electron density $n_e$ derived from the \sii $\lambda$6731/$\lambda$6716 ratio exceeds 1000 cm$^{-3}$, as the $T_{\rm e}$ diagnostics become increasingly sensitive to electron densities above $n_{\rm e} \approx 10^4$ cm$^{-3}$ \citep{FroeseFischerTachiev:04, Tayal:11}.


\begin{figure*}
    \includegraphics[width=0.9\textwidth]{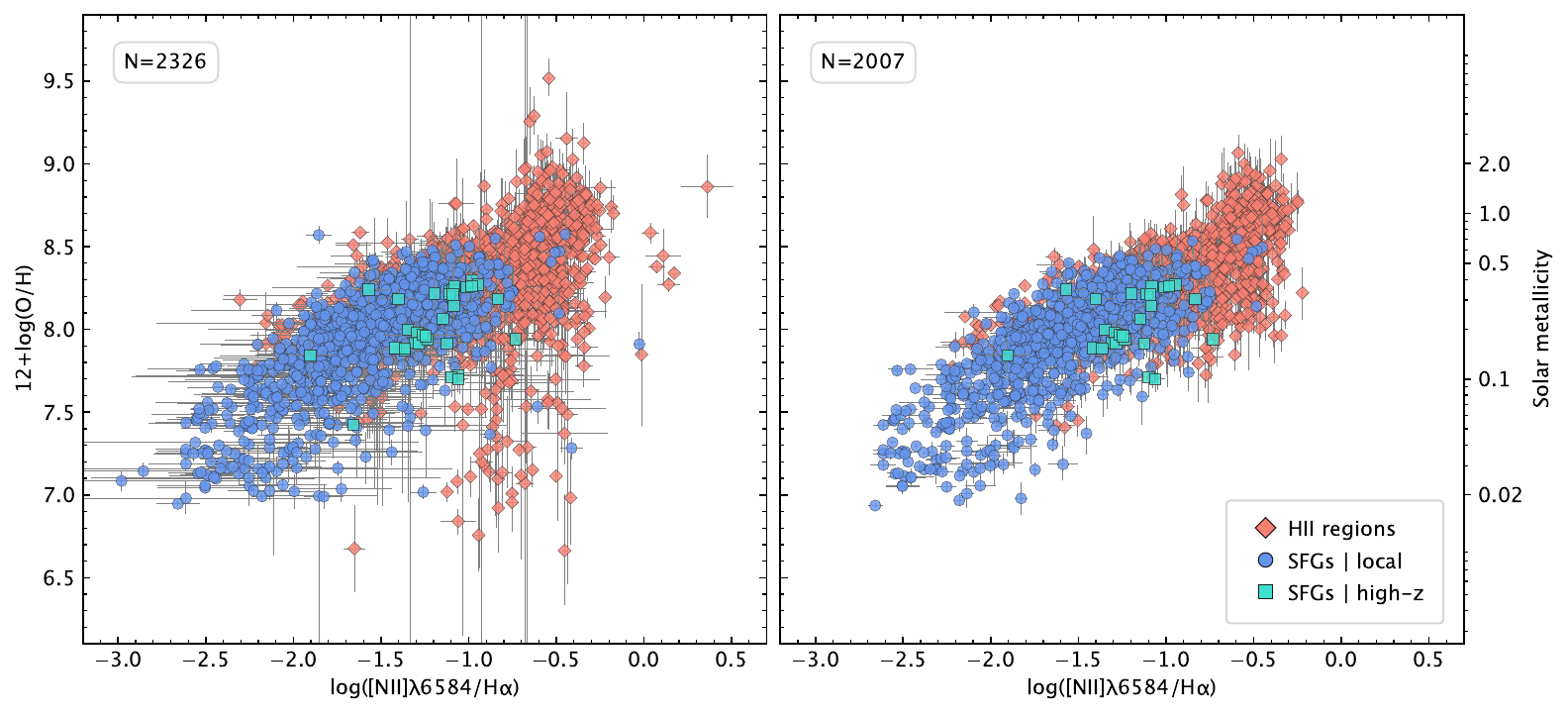}
     \caption{Impact of data quality on metallicity calibration in the 12+log(O/H) versus N2 $\equiv$ log([N II] $\lambda6584$/H$\alpha$) plane. The left panel includes all regions in our literature compilation with measurable N2 and detected auroral lines, allowing direct metallicity determinations without imposing quality or relative-error cuts; oxygen abundances correspond to the $t^2=0$ assumption and error bars denote 1$\sigma$ uncertainties. The right panel shows the same plane restricted to the DESIRED calibration sample after applying the high-quality selection criteria described in Secs.~\ref{sec:desired} and \ref{sec:calib_sample}. The number of spectra in each panel is indicated in the upper-left corner, and object types are identified in the right panel. The right-hand axis indicates the corresponding gas-phase metallicity expressed in solar units, $Z/Z_\odot$, obtained by assuming that the stellar metallicity traces the current ISM abundance in star-forming regions, adopting a solar oxygen abundance of 12+log(O/H)$_\odot = 8.70~(\pm0.04)$ \citep{AmarsiGrevesse:26}.}
    \label{fig:bad_good}
\end{figure*}

The final DESIRED calibration sample, fulfilling the high-quality selection criteria described above, comprises 2392 spectra drawn from 201 independent literature references, including 1029 \hh region, 1296 local SFG, and 67 high-redshift SFG spectra, corresponding to 2237 unique objects.
This constitutes the largest compilation to date of ionised star-forming regions with direct electron temperature determinations and a homogeneous, careful treatment of observational uncertainties, enabling highly precise chemical abundance measurements. The number of available auroral lines in the sample is summarised in Table~\ref{tab:desired}. 
The mean (median) auroral-line S/N is 15.8 (9.1) for \oiii\ $\lambda$4363, 6.7 (5.0) for \nii\ $\lambda$5755, and 11.4 (7.7) for \siii\ $\lambda$6312.
The references corresponding to the published spectra comprising the full DESIRED calibration sample are listed in Appendix~\ref{app:references}. For each spectrum in the calibration sample, Table~\ref{app:sample} provides the object name, classification, derived physical conditions ($n_{\rm e}$ and $T_{\rm e}$), ionic and total oxygen abundances computed for both $t^2=0$ and $t^2>0$, together with the corresponding bibliographic reference. 

Fig.~\ref{fig:OH} shows the total oxygen abundance derived under the assumption of a homogeneous temperature structure ($t^2=0$) compared to the values corrected for temperature inhomogeneities ($t^2>0$) for the full DESIRED calibration sample. As expected, the $t^2>0$ abundances are systematically higher than their $t^2=0$ counterparts, with a mean offset of 0.13\,dex and a dispersion of 0.05\,dex, reflecting the range of physical conditions across the sample. Interestingly, in the DESIRED sample, the difference between the $t^2=0$ and $t^2>0$ abundance scales is, on average, larger at higher metallicities and decreases towards the low-metallicity regime. The metallicity coverage spans 12+log(O/H) $\in$ [6.79, 9.07] for $t^2=0$ and 12+log(O/H) $\in$ [6.87, 9.32] for $t^2>0$. Throughout this paper, \hii\ regions are represented by diamonds, local SFGs by circles, and high-redshift SFGs by squares. Unless otherwise stated, references to the electron temperature correspond to the high-ionisation-zone temperature, $T_{\rm e}$ $\equiv$ $T_3$. In this figure, a subsample of O/H abundance determinations based on RLs is shown in red. These abundances are derived from the fluxes of the O\,II V1 multiplet, $\lambda\lambda$4638.86, 4641.81, 4649.13, 4650.84, 4661.63, 4673.73, 4676.23, 4696.35, which are measurable in the deepest spectra of this sample, together with the effective recombination coefficients from \citet{Storey:2017}. Abundance determinations based on these recombination lines are practically independent of the gas temperature structure and are robust against temperature fluctuations. For the subsample of 51 objects with RL-based abundances, the mean offset with respect to the CEL-based abundances is 0.25\,dex, with a dispersion of 0.18\,dex.

Fig.~\ref{fig:bpt} presents the Baldwin–Phillips–Terlevich (BPT) diagnostic diagram \citep{BPT:81} for the DESIRED calibration catalogue, colour-coded by $T_{\rm e}$, in units of 10$^4$ K. 
A small number of objects lie marginally above the upper boundary for photoionised nebulae defined by \citet{Kewley:01}; however, all of them remain within 0.1 dex of the upper envelope set by this demarcation. Consistently, only a very small fraction of sources with hard ionising radiation fields---typically located to the right of the \citet{Kauffmann:03} demarcation---are included in the sample. 

The impact of high-quality data on metallicity calibration efforts is illustrated in Fig.~\ref{fig:bad_good} in the 12+log(O/H) versus N2 $\equiv$ log(\nii $\lambda$6584/H$\alpha$) plane. The left panel displays all spectra from our literature compilation for which the N2 index \citep{Raimann:00,Denicolo:02} can be measured and auroral lines are available, enabling a direct determination of the metallicity, without imposing any quality or relative-error cuts\footnote{This sample, which originally comprised 3439 regions, includes spectra that are later deprecated in the full DESIRED compilation, prior to restricting the analysis exclusively to line ratios with reported observational uncertainties below 40 percent, as described in Sec.~\ref{sec:desired}.}. The 12+log(O/H) values shown correspond to the homogeneous temperature structure assumption ($t^2=0$). Error bars represent 1$\sigma$ uncertainties, highlighting the large errors that affect both the metallicity estimates and the N2 index for a significant fraction of the regions. The right panel shows the same plane restricted to the DESIRED calibration sample, after applying the high-quality selection criteria described above. Notably, a substantial number of outliers disappear after the quality cuts are applied, revealing more clearly the well-known monotonic increase of metallicity with increasing values of the N2 index.

\section{Analysis and results}
\label{sec:3}

\begin{table*}
\centering
\caption{Adopted diagnostic notations and corresponding emission-line ratio definitions for the 27 DESIRED empirical metallicity relations presented in this work; 23 adopted from the literature and 4 newly introduced in this study (separated by a horizontal line). Definition ($x$): Explicit definition of the emission-line ratio adopted for each calibration, including the constituent lines and their rest-frame wavelengths in \AA. Note that all line ratios are expressed in logarithmic units. $N_{\rm regions}$: Number of objects in the full DESIRED calibration sample with detections of the emission lines required to compute the corresponding ratio and used in the fitting procedure. The last column lists the seminal references for each diagnostic, corresponding to the work that introduced the emission-line ratio and/or first employed it as an abundance indicator.}
\label{tab:calibrations}
\begin{tabular}{l l c c l }
\toprule
Diagnostic & Definition ($x$) & N$_{\rm regions}$ & Notes & References \\
\midrule
R2      & log( \oii~\lam3727/\hb~)  &  2111   & \NA & {\scriptsize \citet{Maiolino:08}} \\
R3      & log( \oiii~\lam5007/\hb~) &  2392   & \NA & {\scriptsize \citet{Jensen:76}} \\
R23     & log( (\oii~\lam3727 + \oiii~\lam\lam4959,5007)/\hb~) &  2111   & \NA & {\scriptsize \citet{Pagel:79}} \\
O32     & log( \oiii~\lam5007 / \oii~\lam3727 ) &  2111   &  \NA & {\scriptsize \citet{Diaz:00,Nagao:06}} \\
\Rhat   & 0.47 \x R2 + 0.88 \x R3 &  2111   &  (a,b) & {\scriptsize \citet{Laseter:24}} \\
\RNe    & 0.47 \x log(\oii~\lam3727/\hg) + 0.88 \x log(\neiii~\lam3869/\hg) &  1338   & (b) & {\scriptsize \citet{Scholte:25}} \\
Ne3O2   & log( \neiii~\lam3869 / \oii~\lam3727 ) &  1500   & \NA & {\scriptsize \citet{Nagao:06,PerezMontero:07}} \\
Ne3     & log( \neiii~\lam3869/\hb~) &  1680   & (c) & {\scriptsize \citet{Sanders:26}} \\
R2Ne3   & log[ (\oii~\lam3727 + \neiii~\lam3869)/\hb~] &  1500   & (c) & {\scriptsize \citet{Sanders:26}}  \\
O3S2    & log( (\oiii~\lam5007/\hb) / (\sii~\lam\lam6717,6731/\ha~) ) &  2033   & \NA & {\scriptsize \citet{Curti:20}} \\
R3S2    & log( \oiii~\lam5007/\hb~ + \sii~\lam\lam6717,6731/\ha~) &  2033   &  \NA & {\scriptsize \citet{Curti:20}}\\
N2      & log( \nii~\lam6584/\ha~) &  2007   & \NA  & {\scriptsize \citet{StorchiBergmann:94, Denicolo:02}} \\
O3N2    & log( \oiii~\lam5007/\hb~ / \nii~\lam6584/\ha~) &  2007   & \NA & {\scriptsize \citet{Alloin:79,PettiniPagel:04}} \\
R3N2    & log( \oiii~\lam5007 / \nii~\lam6584 ) &  2007   & \NA & {\scriptsize \citet{Alloin:79, Maiolino:08}} \\
N2O2    & log( \nii~\lam6584 / \oii~\lam3727 ) &  1779   & \NA & {\scriptsize \citet{Jensen:76, KewleyDopita:02}}  \\
S2      & log( \sii~\lam\lam6717,6731/\ha~) &  2033   & \NA & {\scriptsize \citet{Curti:20}} \\
N2S2    & log( \nii~\lam6584 / \sii~\lam\lam6717,6731 ) &  1940   & \NA & {\scriptsize \citet{Jensen:76, KewleyDopita:02}}  \\
N2S2\ha & log( \nii~\lam6584 / \sii~\lam\lam6717,31) + 0.264 \x log( \nii~\lam6584/\ha~) &  1940   & \NA & {\scriptsize \citet{Dopita:16}} \\
S3      & log( \siii~9069,9531/\ha~) &  831   & \NA & {\scriptsize \citet{Sanders:26}} \\
S3O3    & log( \siii~\lam9069 / \oiii~\lam5007 ) &  831   & \NA & {\scriptsize \citet{Stasinska:06a}} \\
S23     & log[ (\sii~\lam\lam6717,6731 + \siii~\lam\lam9069,9531)/\ha~] &  821   & \NA  & {\scriptsize \citet{VilchezEsteban:96, DiazPM:00}}\\
Ar3     & log( \ariii~\lam7136/\ha~) &  1315   & \NA  & {\scriptsize \citet{Sanders:26}}\\
Ar3O3   & log( \ariii~\lam7136 / \oiii~\lam5007 ) &  1315   & \NA & {\scriptsize \citet{Stasinska:06a}} \\
\hline
O3HeI   & log( \oiii~\lam5007 / \hei~\lam5876 ) &  1590   & \NA & {\scriptsize This work}   \\
Ar3N2   & log( \ariii~\lam7136 / \nii~\lam6584 ) &  1263   & \NA & {\scriptsize This work}  \\
Ne3S2   & log( \neiii~\lam3869 / \sii~\lam\lam6717,6731 ) &  1406   & \NA & {\scriptsize This work}  \\
Ne3S3   & log( \neiii~\lam3869 / \siii~\lam9069 ) &  530   & \NA & {\scriptsize This work} \\
\bottomrule
\end{tabular}
\begin{description}
    \item Notes -- The \oii~$\lambda3727$ intensity corresponds to the sum of the \oii~$\lambda\lambda3726,3729$ doublet.
    \item (a) \Rhat\ is constructed from the R2 and R3 definitions and is therefore already expressed in logarithmic form.
    \item (b) Rotation angle verified against the DESIRED calibration sample; original coefficients retained for comparability with the literature (see Sect.~\ref{sec:calibrations} and Appendix~\ref{app:new_Rhat_RNe}).
    \item (c) Note that the definitions adopted here differ from the closely related diagnostics Ne3 $\equiv$ \neiii $\lambda3869$/H$\delta$ and RO2Ne3 $\equiv$ (\neiii $\lambda3869$ + \oii $\lambda3727$)/H$\delta$ proposed by \citet{Sanders:26}; for consistency and uniformity across other calibrations, in this work these indices are normalised to H$\beta$.

\end{description}

\end{table*}

\subsection{The DESIRED metallicity calibrations}
\label{sec:calibrations}

We employ the high-quality DESIRED calibration sample described in Sec.~\ref{sec:calib_sample} to derive new strong-line metallicity relations linking emission-line ratios to the gas-phase oxygen abundance. Our analysis revisits 23 calibrations previously reported in the literature and introduces 4 additional line ratios, resulting in a total of 27 metallicity calibrations. These relations are derived for both the homogeneous temperature structure assumption ($t^2=0$) and, for the first time in a systematic way, including the effect of temperature fluctuations ($t^2>0$).

To this end, we carried out a comprehensive literature survey to identify strong-line metallicity indicators. Although we do not claim completeness, the selected set encompasses the most widely used diagnostics, as well as several less frequently adopted indices—whether due to limited spectral coverage or to the small number of calibrating regions available when they were originally derived.

It is important to note that these calibrations are purely empirical mappings between the observed emission-line ratios and the $T_{\rm e}$-based oxygen abundance, without any explicit correction for potential biases introduced by diffuse ionised gas (DIG) contamination or by variations in the nitrogen-to-oxygen abundance ratio (N/O) at fixed O/H. The DIG can enhance low-ionisation line ratios---particularly those involving \sii\ and, to a lesser extent, \nii---in spatially integrated spectra, while deviations from the mean N/O--O/H relation of the calibrating sample can bias the inferred O/H for any diagnostic that includes nitrogen lines. These effects are inherently folded into the empirical scatter of the calibrations rather than being corrected for individually. The practical implications of these limitations, as well as specific recommendations for the application of the DESIRED calibrations, are discussed in Sec.~\ref{sec:4}.

Table~\ref{tab:calibrations} summarises the line-ratio definitions and abbreviations adopted in this work, together with the number of regions used in each calibration and relevant notes. The number of calibrating objects varies from one relation to another, depending on the availability of the emission lines required to compute the corresponding index.

In this work, the gas-phase oxygen abundance is parametrised as a polynomial function of the empirically observed emission-line ratio:

\begin{equation}
    12 + \log(\mathrm{O/H}) = \sum_{i} a_{i} x^{i},
	\label{eq:polynomial}
\end{equation}

\noindent
where $x$ denotes the strong-line ratio as defined in Table~\ref{tab:calibrations}, and $a_i$ are the best-fitting calibration coefficients obtained for both the $t^2=0$ and $t^2>0$ cases, listed in Tables~\ref{tab:coeff_t2eq0} and \ref{tab:coeff_t2ge0}, respectively.

For each diagnostic, we systematically explored multiple fitting strategies: ordinary least-squares polynomial fits, orthogonal distance regression \citep[ODR;][]{BoggsRogers:90}, and density-equalised weighted fits, applied both directly [$12+\log(\mathrm{O/H}) = f(x)$] and inversely [$x = f(12+\log(\mathrm{O/H}))$, numerically inverted], using both individual data points and binned medians. The fitting strategy that minimises the total dispersion while avoiding overfitting was adopted for each calibrator, as assessed by comparing the Bayesian Information Criterion \citep[BIC;][]{Schwarz:78} and the degree of the polynomial. 
For diagnostics exhibiting the classical two-branch behaviour (bivalued), each branch is obtained as the numerical inverse of a polynomial fitted in the form $x = f(12+\log(\mathrm{O/H}))$, evaluated up to the metallicity at which the derivative is zero, which defines the turnover. In some cases (e.g. R3 and R3S2), a single fit reproduces both branches and yields a unique turnover metallicity. In the remaining cases, separate fits are adopted for the upper and lower branches, resulting in two closely spaced turnover values, with the narrow intervening metallicity interval excluded from both calibrations. 
The adopted fitting strategy for each diagnostic is indicated in Table~\ref{tab:coeff_t2eq0}; the fitting methodology for the $t^2>0$ case is identical to the $t^2=0$ fit. Full details of the fitting methodology are provided in Appendix~\ref{app:fits}.

Tables~\ref{tab:coeff_t2eq0} and \ref{tab:coeff_t2ge0} also report the line-ratio range over which each calibration is valid, the total dispersion (root-mean-square scatter of the residuals) of the best-fitting solution ($\sigma_{\rm fit}$), the intrinsic dispersion ($\sigma_{\rm int}$), which quantifies the residual scatter in metallicity at fixed strong-line index after subtracting in quadrature the contribution of measurement uncertainties from the observed dispersion (see Appendix~\ref{app:fits}), and for bivalued diagnostics (R3, R23, \Rhat, \RNe, Ne3, R3S2, Ar3 and O3HeI), the turnover values of 12+log(O/H). The corresponding values for each branch are labelled as ``upper'' and ``lower''. The validity range is defined as the interval in the line-ratio axis for which at least five calibrating data points are present within bins of 0.1~dex, ensuring adequate statistical sampling towards the edges of the relation. Data falling outside this range are excluded from the fit. The minimum and maximum oxygen abundances attained by each calibration within its validity range, in 12+log(O/H) units, are also listed in these tables.

Traditional strong-line calibrations report a single global dispersion that applies uniformly across the full range of the diagnostic. However, the calibration uncertainty is not constant: it may vary significantly depending on the metallicity regime, due to uneven sampling, changing sensitivity of the line ratio, or the presence of secondary physical parameters. To address this, we additionally provide a quadratic parametrisation of the dispersion as a function of the line ratio:

\begin{equation}
    \sigma(x) = e_0 + e_1\, x + e_2\, x^2,
    \label{eq:sigma_profile}
\end{equation}

\noindent
where the coefficients $e_i$ are listed alongside the calibration coefficients in Tables~\ref{tab:coeff_t2eq0} and \ref{tab:coeff_t2ge0}, and describe the total, rather than intrinsic, dispersion (see Appendix~\ref{app:fits}). This index-dependent dispersion profile enables the user to estimate the expected calibration uncertainty at any value of the emission-line ratio, rather than relying on a single $\sigma_{\rm fit}$ value that may underestimate the uncertainty in sparsely sampled regimes and overestimate it where the calibration is best constrained.

\begin{figure*}
    \includegraphics[width=\textwidth]{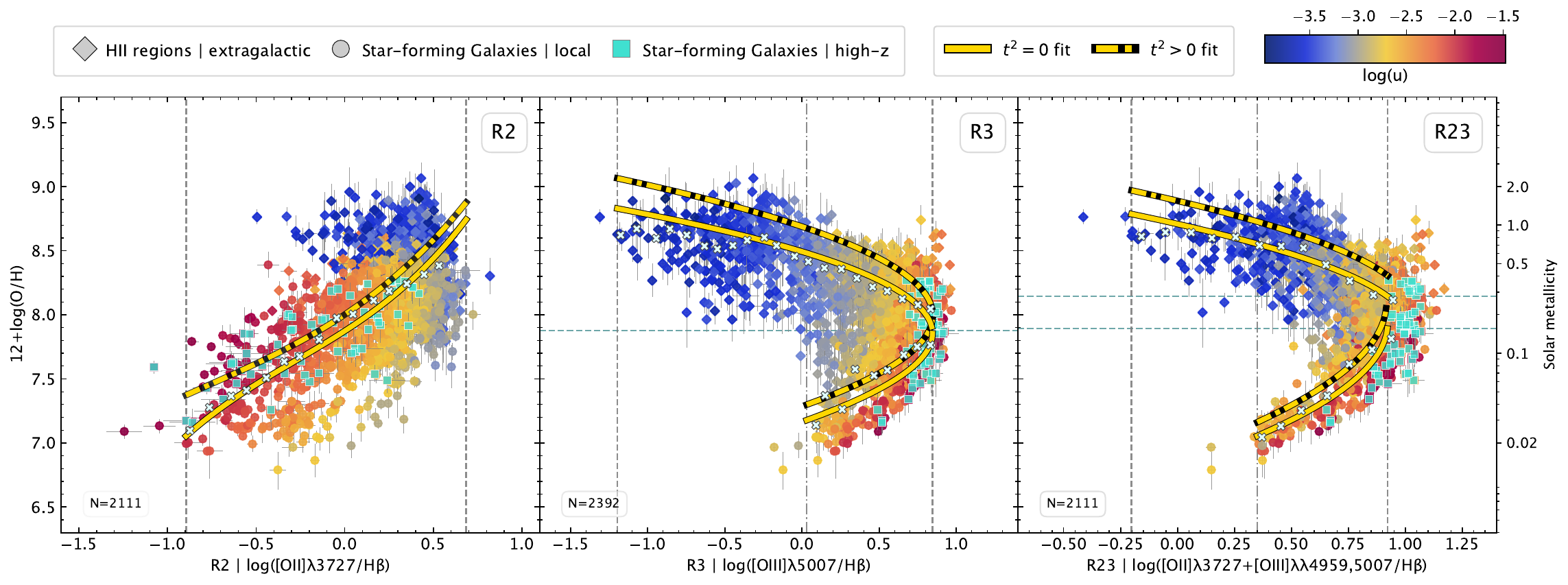}
    \includegraphics[width=\textwidth]{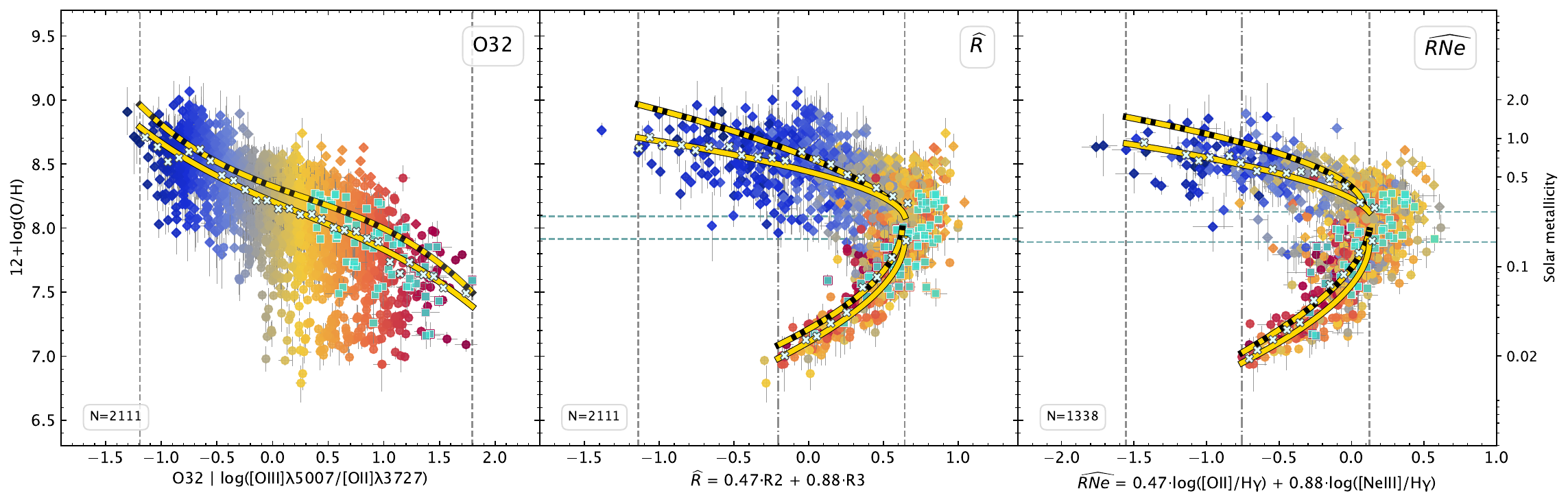}
    \includegraphics[width=\textwidth]{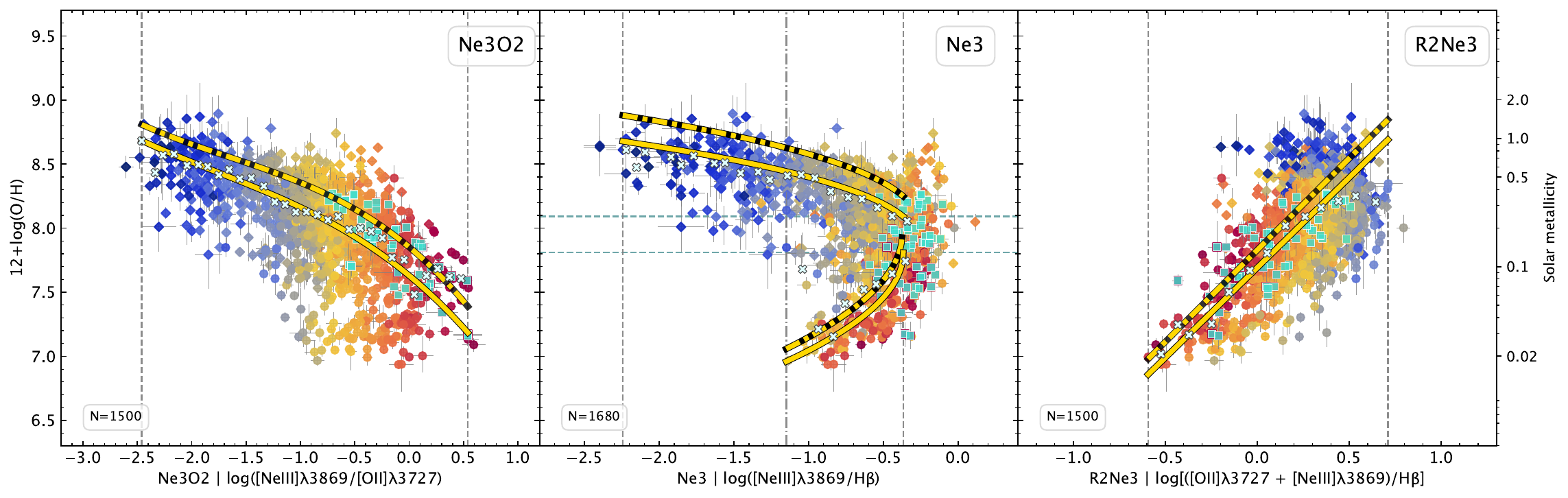}
    \caption{Gas-phase metallicity calibrations derived from the DESIRED calibration sample. The top row (from left to right) shows R2, R3, and R23; the middle row presents O32, \Rhat, and \RNe; and the bottom row displays Ne3O2, Ne3, and R2Ne3. The notation and index definitions for each diagnostic are given in Table~\ref{tab:calibrations} and indicated on the x-axis of the corresponding panels. The yellow solid curve represents the best-fitting relation for the classical \textit{direct method} calibration ($t^2=0$), while the yellow–black dashed curve shows the calibration that accounts for an inhomogeneous temperature structure ($t^2>0$; \citealt{Peimbert:67}). For monotonic diagnostics (e.g. R2), the vertical grey dashed lines indicate the validity range of each calibration, as defined in the text. For double-valued diagnostics (e.g. R23), the grey dashed lines mark the validity range of the upper branch, whereas the grey dash-dotted line indicates the minimum index value defining the lower branch, as summarised in Tables~\ref{tab:coeff_t2eq0} and \ref{tab:coeff_t2ge0}. The horizontal dashed lines mark the adopted separation (turnover) between lower and upper branches. 
    For these diagnostics, the plotted curves correspond to the exact numerical inversion of the fitted $x=f(12+\log(\mathrm{O/H}))$ relation for each branch, whereas the polynomial coefficients listed in Tables~\ref{tab:coeff_t2eq0} and \ref{tab:coeff_t2ge0} provide analytical approximations for practical use.
    Data points are colour-coded according to their ionisation parameter, $\log(u)$ (following the parametrisation by \citealt{Diaz:00}), and symbols denote the object type as indicated at the top of the figure. High-redshift objects are highlighted in cyan to distinguish them from local Universe sources. Error bars correspond to 1$\sigma$ uncertainties. 
    The light-blue crosses show the mean metallicity in 0.1 dex bins of the line ratio. 
    The number of regions used in each calibration is given in the lower-left corner of each panel, as reported in Table~\ref{tab:calibrations}. The right-hand axis shows the equivalent gas-phase metallicity expressed in solar units, $Z/Z_\odot$, following the definition described in Fig.~\ref{fig:bpt}.}
    \label{fig:calib_1}
\end{figure*}

The DESIRED metallicity calibrations are presented in Figs.~\ref{fig:calib_1} to \ref{fig:calib_1d}. For clarity, only the classical ``direct method'' oxygen abundance ($t^2=0$) is shown for each diagnostic, colour-coded according to the ionisation parameter, $\log u$, as parametrised by \citet{Diaz:00}:

\begin{equation}
    \log u = -0.80 \log(\oii~\lambda3727/\oiii~\lambda\lambda4959,5007) - 3.02.
	\label{eq:log_u}
\end{equation}


The yellow solid curve represents the calibration derived under the assumption of homogeneous temperature structure ($t^2=0$), while the yellow-black dashed curve corresponds to the relation that accounts for temperature fluctuations ($t^2>0$). As shown in Fig.~\ref{fig:OH}, the $t^2>0$ solutions yield systematically higher metallicities, with a mean offset of $\sim 0.13$ dex in 12+log(O/H) at fixed line ratio, consistent with the average difference found across the DESIRED calibration sample, although the exact offset depends on the specific diagnostic and metallicity regime. For brevity, throughout the remainder of this paper, any quoted values of the metallicity, 12+log(O/H), or the dispersion, $\sigma_{\rm fit}$, refer to the $t^2=0$ case, unless explicitly stated otherwise


\subsection{Oxygen-based metallicity diagnostics}
\label{sec:oxygen}

Fig.~\ref{fig:calib_1} presents both classical and more recent diagnostics based on \hi\ recombination lines and oxygen collisionally excited lines.
The R2 calibration \citep{Maiolino:08}, constructed from 2111 spectra, exhibits a monotonic behaviour in which metallicity increases with the line ratio, albeit with one of the largest dispersions among all calibrations presented here (together with S3), with a RMS ($\sigma_{\rm fit}$) of 0.33~dex, reflecting the intrinsically complex morphology of this line ratio. This is consistent with previous works indicating that this diagnostic is strongly sensitive to the ionisation parameter and to the ionisation state of the gas \citep{MaiolinoMannucci:19}, which significantly limits its precision as a standalone metallicity indicator. Notably, the 67 high-redshift spectra are distributed throughout the locus of low-redshift objects, with no clear systematic offset. We note that the sensitivity of R2 to metallicity decreases towards the upper end of its validity range, above 12+log(O/H) $\sim$ 8.3, consistent with the saturation of this diagnostic expected from photoionisation models at solar and super-solar metallicities (e.g. \citealt{Kewley:19}).

The R3 calibration \citep{Jensen:76}, based on 2392 regions, displays the classical bivalued structure with a turnover at 12+log(O/H) $\approx$ 7.91, primarily driven by the interplay between metallicity and ionisation parameter, as illustrated by the colour-coding of the data points. The upper branch spans a broad dynamic range and is particularly useful for R3 < 0.0, corresponding to 12+log(O/H) $\approx 8.4-8.9$, with a dispersion of 0.18 dex. In contrast, the lower-metallicity branch is strongly degenerate, resulting in a larger dispersion of 0.21 dex.

The classical R23 diagnostic, first proposed by \citet{Pagel:79}, is constructed from 2111 spectra and displays the characteristic double-valued behaviour, with closely spaced turnover values at 12+log(O/H) $\approx$ 8.14 (upper branch) and $\approx$ 7.89 (lower branch); in this regime the line ratio becomes essentially insensitive to metallicity, so that the intervening $\sim$0.25 dex interval falls outside the validity range of both branches. The upper and lower branches show dispersions of 0.17 and 0.18 dex, respectively. The combination of \oii and \oiii lines partially cancels the opposing dependences of each ion on the ionisation parameter, making R23 less sensitive to ionisation conditions than either R2 or R3 individually \citep{MaiolinoMannucci:19}. As a consequence, the ionisation-parameter stratification visible in the colour-coding of the data points is less pronounced than for R3, yet still present across the full extent of the R23 index.  Note that, for R3 and R23, the high-redshift spectra are concentrated at the highest values of the ionisation parameter ($\log u \gtrsim -2.0$), consistent with the harder ionising radiation fields and elevated ionisation states characteristic of early-Universe SFGs \citep{Chakraborty:25, Sanders:26}, yet they remain intermixed with low-redshift spectra exhibiting similar properties.

The O32 diagnostic, primarily regarded as a tracer of the ionisation parameter \citep{Diaz:00, KewleyDopita:02}, as evident from the colour-coding of the data points, exhibits only a secondary dependence on metallicity, arising from the well-established anticorrelation between ionisation parameter and oxygen abundance \citep{Nagao:06}. The DESIRED calibration of O32, constructed from 2111 spectra, follows a monotonic trend in which metallicity increases as the line ratio decreases. Similarly to R2, the O32 calibration shows a large scatter ($\sigma_{\rm fit} = 0.29$ dex), which becomes more pronounced in the low-metallicity regime (12+log(O/H) < 8.0, corresponding to O32 $> -0.2$), likely reflecting the increased diversity of ionisation conditions at sub-solar abundances.
Consequently, O32 may be more reliable at higher metallicities (12+log(O/H) $\approx 8.0-8.9$), or alternatively as a diagnostic to distinguish between low- and high-metallicity regimes when used in conjunction with bivalued calibrators. The larger dispersion and reduced sampling at low metallicity have been interpreted as evidence for a flattening and/or intrinsic bivalued behaviour of the relation, leading to more complex functional forms \citep[e.g.][]{Nakajima:22, Garg:24}. The increased scatter observed in oxygen-based diagnostics below 12+log(O/H) $\approx$ 7.8, particularly for R2 and O32, may reflect a broader distribution of ionisation parameters at fixed metallicity in metal-poor systems.

The \Rhat\ diagnostic is defined as a linear combination of R2 and R3 corresponding to a rotation around the O/H axis in the R2--R3--O/H space, designed to minimise the scatter at fixed metallicity \citep{Laseter:24}. This transformation is motivated by reducing the secondary dependence on ionisation parameter at fixed O/H. \citet{Laseter:24} found an optimal rotation of 61.82 degrees ($\widehat{R}=0.47\cdot R2+0.88\cdot R3$) using a compiled calibration sample of $z\sim0$ galaxies and \hh\ regions. 

We verified this rotation against the DESIRED calibration sample, which is more than four times larger than that used by \citet{Laseter:24}, and find a best-fitting rotation of 59.2 degrees ($\widehat{R}=0.51\cdot R2+0.86\cdot R3$), within a broad and shallow minimum spanning $\phi\approx57-63$ degrees that comfortably encompasses the \citet{Laseter:24} value (see Appendix~\ref{app:new_Rhat_RNe}). The total dispersion, $\sigma_{\rm tot}$, achieved with this DESIRED-specific rotation (0.162 and 0.177 dex for the upper and lower branches, respectively) differs from that obtained with the original \citet{Laseter:24} coefficients (0.175 and 0.178 dex) by only $\sim0.01$~dex or less in either branch, a difference with no practical impact on the resulting metallicities. We therefore retain the original \citet{Laseter:24} projection here, in order to preserve direct comparability with the broader literature.

The DESIRED calibration for \Rhat, based on 2111 regions, displays a bivalued structure qualitatively similar to that of R23, with turnover values at 12+log(O/H) $\approx$ 8.09 (upper) and $\approx$ 7.92 (lower); metallicities within the intervening $\sim$0.17 dex interval are excluded from both branches. The upper branch has a dispersion of 0.18~dex, comparable to that of R23 (0.17~dex). On the lower branch, the scatter is also 0.18~dex, comparable to that of R23 (0.18~dex), making \Rhat\ particularly well constrained at 12+log(O/H) $\lesssim7.8$. 
The high-redshift spectra are well intermixed with the low-redshift objects along the relation, consistent with the reduced sensitivity of \Rhat\ to the evolving ionisation conditions at high redshift. As for R3 and R23, however, most $z>2$ objects lie in the lower-sensitivity regime of the relation (\Rhat\ $\gtrsim$ 0.5), with essentially no coverage of the upper branch and many objects falling within the interval excluded from both branches. Nevertheless, \Rhat\ provides a well-constrained metallicity diagnostic on the lower branch, making it particularly effective for probing low-metallicity star-forming regions in both the local and high-redshift Universe. As with the other bivalued oxygen diagnostics, an additional line ratio is therefore required to distinguish between the two branches.

\subsection{Neon-based metallicity diagnostics}
\label{sec:neon}

The so-called $\alpha$-elements are those elements lighter than Fe whose most abundant isotopes have mass numbers that are multiples of four, corresponding to an integer number of $\alpha$ particles (He nuclei). Neon, sulphur, and argon are $\alpha$-elements and, in principle, share the same dominant nucleosynthetic origin as oxygen, namely core-collapse supernovae. They are therefore expected to exhibit similar behaviour in their cosmic evolution. As a consequence, the abundance ratios of these elements relative to oxygen are not expected to vary significantly as a function of redshift, metallicity, or across different star-formation histories, within the limits imposed by observational uncertainties and intrinsic dispersion. For this reason, $\alpha$-element emission lines are frequently used as proxies for gas-phase metallicity in the ionised interstellar medium \citep[e.g.][]{Izotov:06, PerezMontero:07, Croxall:16, Esteban:20, Esteban:25, Rogers:22, ArellanoCordova:24}.

In the specific case of neon, its nucleosynthetic origin closely parallels that of oxygen, as Ne is primarily produced during carbon burning and subsequent stages in massive stars. As a result, a nearly constant Ne/O ratio is generally expected \citep[e.g.][]{Iwamoto:99, Johnson:19, Kobayashi:20}. Nevertheless, some studies report a mild increase of Ne/O with metallicity, of the order of $\sim$0.1 dex, which has been attributed to oxygen depletion onto dust grains in metal-rich environments and to uncertainties in the adopted ICFs \citep[e.g.][]{Izotov:06,MesaDelgado:09,MendezDelgado:22b, Esteban:25}.

The neon emission lines accessible in the optical spectrum are \neiii\ $\lambda$3869 and \neiii\ $\lambda$3967, the two arising from the same upper $^1D_2$ level of the Ne$^{2+}$ ion, with $\lambda$3869 being the strongest. Ne$^{2+}$ is produced by the ionisation of Ne$^{+}$ by photons with energies between 40.96 and 63.45 eV. Given the relatively high ionisation threshold required to form Ne$^{2+}$, a substantial fraction of neon in typical star-forming \hh regions is expected to remain in the Ne$^+$ state, which is not observable in the optical, making ionisation correction factors necessary to recover the total neon abundance.

Fig.~\ref{fig:calib_1} presents the calibrations based on the \neiii\ $\lambda$3869 emission line, including combinations with \oii\ $\lambda$3727, H$\gamma$, and H$\delta$. These diagnostics benefit from the relatively small wavelength separation between the involved lines, allowing them to be observed within a narrow spectral window---an advantage for high-redshift studies. In addition, they are less sensitive to dust reddening than many commonly used ratios. The main limitation of these indicators is that \neiii\ $\lambda$3869 is typically an order of magnitude fainter than \oiii\ $\lambda$5007, which restricts its detectability to spectra of sufficient signal-to-noise.

Neon-based diagnostics can be regarded as analogues of the oxygen-based ratios and employed in a similar manner, with the high-ionisation \oiii\ lines (ionisation energy 35.1 eV) replaced by the comparably high-ionisation \neiii\ line. Our new version of the Ne3O2 calibration, first introduced by \citet{Nagao:06} and \citet{PerezMontero:07}, is constructed from 1500 calibrating spectra and exhibits a monotonic decrease with increasing O/H. As Ne3O2 is formally equivalent to O32, it primarily traces the ionisation parameter. Consistent with O32, it shows a relatively large metallicity dispersion ($\sigma_{\rm fit}$ = 0.28 dex), which becomes more pronounced in the low-metallicity regime (12+log(O/H) $\lesssim$ 7.5, corresponding to Ne3O2 $\gtrsim$ $-$1.3).

The Ne3 calibration, based on 1680 spectra, exhibits a double-valued structure with turnover values at 12+log(O/H) $\approx$ 8.09 (upper) and $\approx$ 7.81 (lower); so that the intervening $\sim$0.28~dex interval falls outside the validity range of both branches. The upper branch is well constrained, with a dispersion of 0.16~dex, while the lower branch shows a larger scatter of 0.24~dex. This diagnostic, recently proposed by \citet{Sanders:26} for high-redshift objects, appears promising for metallicity determinations in the high-abundance regime (12+log(O/H) $\gtrsim$ 8.1, corresponding to Ne3 $\lesssim$ $-$1.2).

The R2Ne3 diagnostic, analogous in construction to R23 or \Rhat, is derived from 1500 spectra in the DESIRED calibration sample and was also proposed by \citet{Sanders:26}. It exhibits a monotonic increase of O/H with the line ratio, with a relatively large dispersion (0.30~dex) that tends to increase towards higher metallicities. Owing to its monotonic behaviour, R2Ne3 offers improved discriminatory power across the full metallicity range and may be particularly useful for separating low- and high-metallicity regimes without invoking a branch degeneracy. Note that we adopt H$\beta$ as the normalization line for both the Ne3 and R2Ne3 indicators, in contrast to \citet{Sanders:26}, who use the relatively weak H$\delta$ line (common in the study of high-redshift objects), which is not always detected even when \neiii\ and \oii\ are measurable. In the absence of H$\beta$, but when H$\gamma$ or H$\delta$ is available, the H$\beta$ flux can be inferred from the theoretical Case B recombination ratios relative to H$\gamma$ or H$\delta$, assuming negligible stellar Balmer absorption and provided that the emission-line fluxes have been properly corrected for dust reddening.

The \RNe\ diagnostic was recently introduced by \citet{Scholte:25} as a neon-based analogue of the \Rhat\ index, adopting the same rotation found by \citet{Laseter:24} for R2--R3 (61.82 degrees), on the grounds that R3 and Ne3 are tightly correlated, and designed to reduce the secondary dependence on the ionisation parameter. This diagnostic is particularly suited for very high-redshift observations with JWST/NIRSpec, remaining accessible up to $z\sim11.2$ (and for comparable objects with very high-ionisation properties). It is especially valuable in the redshift interval $9.5 \lesssim z \lesssim 11.2$, where \oiii\ $\lambda$5007 is shifted beyond the wavelength coverage of JWST/NIRSpec, rendering the emission lines required for the \Rhat\ diagnostic unobservable. 

We verified this rotation against the DESIRED calibration sample and find a best-fitting rotation of 60.0 degrees, with corresponding coefficients $[0.50,0.86]$, within a broad and shallow minimum spanning $\phi\approx57-63$ degrees that comfortably encompasses the \citet{Laseter:24}/\citet{Scholte:25} value (see Appendix \ref{app:new_Rhat_RNe}). The total dispersion achieved with the DESIRED optimal rotation (0.130 and 0.190~dex for the upper and lower branches, respectively) differs from that obtained with the original coefficients (0.137 and 0.198~dex) by less than 0.01~dex in either branch, again with no practical impact on the resulting metallicities. As with \citet{Scholte:25}, who reached the same conclusion for this diagnostic, we retain the original projection here to preserve direct comparability with the literature.

The \RNe\ calibration, constructed from 1338 spectra, displays a bivalued structure with turnover values at 12+log(O/H) $\approx$ 8.13 (upper) and $\approx$ 7.89 (lower). The upper-metallicity branch is particularly well-constrained, with a dispersion of 0.14~dex, among the lowest of all the diagnostics presented here; the lower-metallicity branch shows a dispersion of 0.20~dex. As with \Rhat, the high-redshift spectra are concentrated near the turnover and along the lower branch, with essentially no coverage of the upper branch, limiting the applicability of this diagnostic to high-redshift objects in the well-constrained lower-branch regime (see Sec.~\ref{sec:highz}).

\begin{table*}
\centering
\caption{
Best-fitting polynomial coefficients for the homogeneous temperature structure case ($t^2=0$), to be used in conjunction with Eq.~\eqref{eq:polynomial}. For diagnostics displaying a two-branch behaviour, the corresponding solutions are denoted as ``upper'' and ``lower''.
The Validity columns list the interval in line-ratio space over which each calibration is formally applicable.
The Coefficients~$\sigma(x)$ columns list the quadratic polynomial coefficients $e_i$ of the index-dependent dispersion profile $\sigma(x)$, to be used in conjunction with Eq.~\eqref{eq:sigma_profile}, which parametrises the calibration uncertainty as a function of the line ratio.
The reported $\sigma_{\rm fit}$ corresponds to the root-mean-square (RMS) scatter of the residuals around the adopted best-fitting relation, whereas $\sigma_{\rm int}$, the intrinsic dispersion, quantifies the residual scatter in metallicity at fixed strong-line index after accounting for measurement uncertainties, $\sigma_{\rm cal}$ denotes the standard deviation of the metallicity residuals relative to the $T_{\rm e}$-based abundances, as defined in Sec.~\ref{sec:sigma}.
The 12+log(O/H) columns specify the metallicity range spanned by the calibration sample for each relation, and the turnover value separating the upper and lower branches for bivalued diagnostics; where the two branches are calibrated with independent fits, the corresponding upper and lower turnover values are both listed.
The final column specifies the fitting method adopted for each calibration (see Notes below and Appendix~\ref{app:fits}).}
\label{tab:coeff_t2eq0}

{\setlength{\tabcolsep}{4pt}  

\begin{tabular}{l SSSS SS SSS ccc ccc l}
\toprule
& \multicolumn{4}{c}{Coefficients | Fit}
& \multicolumn{2}{c}{Validity}
  &  \multicolumn{3}{c}{Coefficients | $\sigma(x)$}
  & & & & \multicolumn{3}{c}{12+log(O/H)} &  \\
\cmidrule(lr){2-5} \cmidrule(lr){6-7} \cmidrule(lr){8-10} \cmidrule(lr){14-16}
Calibration
  & {$a_0$} & {$a_1$} & {$a_2$} & {$a_3$}
  & {Min} & {Max}
  & {$e_0$} & {$e_1$} & {$e_2$}
  & {$\sigma_{\rm fit}$} & {$\sigma_{\rm int}$} & {$\sigma_{\rm cal}$}
  & {Min} & {Max}
  & {Turnover} & {Fit} \\
\midrule

R2 & 7.87 & 0.89 & 0.30 & 0.37 & -0.90 & 0.69 & -0.13 & -0.07 & 0.32 & 0.33 & 0.31 & 0.33 & 7.05 & 8.74 & \NA & (b$^*_3$) \\
R3 ~{\scriptsize upper} & 8.50 & -0.38 & -0.19 & -0.09 & -1.20 & 0.85 & -0.07 & -0.02 & 0.21 & 0.18 & 0.16 & 0.18 & 7.99 & 8.83 & 7.91 & (b$^*_3$) \\
\phantom{R3} ~{\scriptsize lower} & 7.16 & 0.39 & 0.14 & 0.24 & 0.03 & 0.85 & -0.22 & -0.10 & 0.34 & 0.21 & 0.21 & 0.21 & 7.17 & 7.74 & \NA & (b$^*_3$) \\
R23 ~{\scriptsize upper} & 8.70 & -0.37 & -0.01 & -0.28 & -0.20 & 0.92 & -0.30 & 0.31 & 0.12 & 0.17 & 0.14 & 0.15 & 8.13 & 8.78 & 8.14 & (b$^*_3$) \\
\phantom{R23} ~{\scriptsize lower} & 7.17 & -0.80 & 1.58 & \NA & 0.35 & 0.92 & -1.79 & 2.58 & -0.72 & 0.18 & 0.18 & 0.17 & 7.09 & 7.78 & 7.89 & (b$^*_2)^{\ddag}$ \\
O32 & 8.21 & -0.37 & 0.04 & -0.05 & -1.19 & 1.80 & -0.02 & 0.03 & 0.25 & 0.29 & 0.29 & 0.29 & 8.79 & 7.39 & \NA & (a$^*_3)^{\dag}$ \\
\Rhat ~{\scriptsize upper} & 8.45 & -0.32 & -0.18 & -0.10 & -1.14 & 0.64 & -0.01 & -0.01 & 0.19 & 0.18 & 0.15 & 0.16 & 8.14 & 8.72 & 8.09 & (a$^*_3)^{\dag}$ \\
\phantom{\Rhat} ~{\scriptsize lower} & 7.12 & 0.62 & 0.07 & 0.96 & -0.20 & 0.64 & -0.36 & 0.34 & 0.06 & 0.18 & 0.18 & 0.14 & 6.98 & 7.80 & 7.92 & (a$^*_3$) \\
\RNe ~{\scriptsize upper} & 8.22 & -0.59 & -0.40 & -0.14 & -1.55 & 0.12 & 0.07 & 0.09 & 0.15 & 0.14 & 0.11 & 0.13 & 8.14 & 8.69 & 8.13 & (a$^*_3)^{\dag}$ \\
\phantom{\RNe} ~{\scriptsize lower} & 7.57 & 1.49 & 1.80 & 1.25 & -0.76 & 0.12 & -0.79 & -0.27 & 0.18 & 0.20 & 0.20 & 0.18 & 6.92 & 7.78 & 7.89 & (a$^*_3)^{\ddag}$ \\
Ne3O2 & 7.63 & -0.70 & -0.21 & -0.04 & -2.46 & 0.54 & -0.03 & -0.02 & 0.24 & 0.28 & 0.28 & 0.29 & 8.68 & 7.18 & \NA & (c$^*_3)^{\dag}$ \\
Ne3 ~{\scriptsize upper} & 7.77 & -1.03 & -0.48 & -0.09 & -2.24 & -0.37 & -0.07 & -0.14 & 0.11 & 0.16 & 0.14 & 0.14 & 8.08 & 8.68 & 8.09 & (b$^*_3$) \\
\phantom{Ne3} ~{\scriptsize lower} & 8.26 & 1.93 & 0.71 & \NA & -1.15 & -0.37 & -1.40 & -1.90 & -0.36 & 0.24 & 0.24 & 0.22 & 6.98 & 7.65 & 7.81 & (b$^*_2)^{\ddag}$ \\
R2Ne3 & 7.69 & 1.41 & \NA & \NA & -0.59 & 0.71 & -0.49 & 0.11 & 0.28 & 0.30 & 0.29 & 0.30 & 6.86 & 8.69 & \NA & (b$_1$) \\
O3S2 & 8.63 & -0.42 & 0.13 & -0.05 & -0.48 & 2.84 & 0.00 & 0.00 & 0.22 & 0.27 & 0.26 & 0.27 & 8.87 & 7.44 & \NA & (a$^*_3)^{\dag}$ \\
R3S2 ~{\scriptsize upper} & 8.59 & -0.41 & -0.20 & -0.25 & -0.72 & 0.83 & -0.15 & 0.06 & 0.20 & 0.18 & 0.16 & 0.18 & 7.97 & 8.87 & 7.90 & (b$^*_3$) \\
\phantom{R3S2} ~{\scriptsize lower} & 7.10 & 0.45 & 0.33 & \NA & 0.10 & 0.83 & -1.07 & 0.81 & 0.09 & 0.19 & 0.19 & 0.18 & 7.14 & 7.69 & \NA & (b$^*_2$) \\
N2 & 9.12 & 1.49 & 0.85 & 0.22 & -2.66 & -0.22 & 0.06 & 0.16 & 0.27 & 0.20 & 0.18 & 0.20 & 6.98 & 8.83 & \NA & (b$_3)^{\dag}$ \\
O3N2 & 8.56 & -0.21 & 0.02 & -0.02 & -0.80 & 3.26 & 0.01 & -0.01 & 0.19 & 0.23 & 0.22 & 0.23 & 8.76 & 7.51 & \NA & (a$^*_3)^{\dag}$ \\
R3N2 & 8.48 & -0.21 & -0.00 & -0.02 & -1.26 & 2.79 & 0.01 & -0.01 & 0.18 & 0.23 & 0.22 & 0.23 & 8.77 & 7.53 & \NA & (a$^*_3)^{\dag}$ \\
N2O2 & 8.61 & 0.38 & 0.18 & 0.20 & -1.87 & 0.35 & 0.06 & 0.09 & 0.24 & 0.25 & 0.23 & 0.25 & 7.24 & 8.78 & \NA & (b$_3)^{\dag}$ \\
S2 & 9.03 & 0.86 & \NA & \NA & -2.07 & -0.24 & 0.11 & 0.28 & 0.41 & 0.27 & 0.25 & 0.27 & 7.25 & 8.82 & \NA & (b$_1$) \\
N2S2 & 8.37 & 0.54 & 0.04 & 0.89 & -0.89 & 0.50 & -0.12 & -0.03 & 0.20 & 0.24 & 0.21 & 0.24 & 7.29 & 8.76 & \NA & (b$_3)^{\dag}$ \\
N2S2\ha & 8.47 & 0.54 & 0.45 & 0.44 & -1.48 & 0.43 & 0.01 & 0.03 & 0.19 & 0.21 & 0.19 & 0.22 & 7.22 & 8.81 & \NA & (b$_3)^{\dag}$ \\
S3 & 9.90 & 3.14 & 0.83 & \NA & -1.55 & -0.28 & -0.79 & -1.56 & -0.38 & 0.35 & 0.33 & 0.35 & 7.04 & 9.09 & \NA & (b$_1$) \\
S3O3 & 8.72 & 0.33 & -0.08 & \NA & -2.34 & 0.43 & 0.03 & 0.03 & 0.17 & 0.22 & 0.19 & 0.22 & 7.50 & 8.85 & \NA & (b$^*_2$) \\
S23 & 8.99 & 1.70 & 0.25 & \NA & -1.36 & -0.17 & 0.25 & 0.39 & 0.32 & 0.23 & 0.19 & 0.23 & 7.14 & 8.70 & \NA & (b$_2)^{\dag}$ \\
Ar3 ~{\scriptsize upper} & 6.66 & -1.78 & -0.40 & \NA & -2.33 & -1.58 & -0.58 & -2.12 & -1.79 & 0.11 & 0.06 & 0.11 & 8.49 & 8.66 & 8.48 & (b$_2$) \\
\phantom{Ar3} ~{\scriptsize lower} & 12.74 & 4.20 & 0.77 & \NA & -2.55 & -1.58 & -0.50 & -1.95 & -1.70 & 0.19 & 0.19 & 0.17 & 7.00 & 8.01 & 8.22 & (a$^*_2$) \\
Ar3O3 & 8.72 & -0.04 & -0.23 & \NA & -2.45 & -0.56 & 0.06 & 0.10 & 0.18 & 0.23 & 0.22 & 0.24 & 7.45 & 8.67 & \NA & (a$^*_2)^{\dag}$ \\
O3HeI ~{\scriptsize upper} & 8.74 & -0.32 & 0.20 & -0.14 & 0.04 & 1.71 & -0.10 & 0.23 & 0.04 & 0.16 & 0.14 & 0.14 & 8.07 & 8.73 & 8.07 & (b$^*_3$) \\
\phantom{O3HeI} ~{\scriptsize lower} & 7.87 & -1.94 & 1.05 & \NA & 1.04 & 1.71 & -1.27 & 3.79 & -2.61 & 0.20 & 0.20 & 0.18 & 6.99 & 7.64 & 7.78 & (b$^*_2)^{\ddag}$ \\
Ar3N2 & 7.99 & -0.57 & -0.24 & -0.09 & -1.85 & 0.56 & 0.06 & 0.12 & 0.23 & 0.25 & 0.24 & 0.25 & 8.78 & 7.58 & \NA & (b$^*_3)^{\dag}$ \\
Ne3S2 & 8.07 & -0.39 & -0.10 & -0.02 & -2.06 & 1.09 & 0.00 & 0.06 & 0.22 & 0.26 & 0.25 & 0.26 & 8.66 & 7.50 & \NA & (a$_3$) \\
Ne3S3 & 8.27 & -0.45 & -0.15 & -0.04 & -1.60 & 1.10 & 0.06 & 0.08 & 0.13 & 0.21 & 0.20 & 0.21 & 8.76 & 7.54 & \NA & (b$^*_3$) \\

\bottomrule
\end{tabular}}
\begin{description}
    \item Notes -- Fitting method codes: the letter denotes the regression method: (a)~ordinary least-squares polynomial, (b)~orthogonal distance regression (ODR), (c)~density-equalised weighted fit; the subscript indicates the polynomial degree (1--3); an asterisk ($^*$) indicates that the fit was performed in the direction [$x = f(12+\log(\mathrm{O/H}))$] and the reported coefficients were obtained via numerical inversion;  a dagger (${\dag}$) indicates that the fit was performed on data binned in the line-ratio index, while a double dagger (${\ddag}$) indicates that the fit was performed on data binned in $12+\log(\mathrm{O/H})$.
\end{description}
\end{table*}

\begin{table*}
\centering
\caption{Best-fitting polynomial coefficients for the inhomogeneous temperature structure case ($t^2 > 0$), to be used in conjunction with Eq.~\eqref{eq:polynomial} and \eqref{eq:sigma_profile}. The columns and definitions are identical to those described in Table~\ref{tab:coeff_t2eq0}.}
\label{tab:coeff_t2ge0}

{\setlength{\tabcolsep}{4pt}  

\begin{tabular}{l SSSS SS SSS ccc ccc }
\toprule
& \multicolumn{4}{c}{Coefficients | Fit}
& \multicolumn{2}{c}{Validity}
  &  \multicolumn{3}{c}{Coefficients | $\sigma(x)$}
  & & & & \multicolumn{3}{c}{12+log(O/H)}   \\
\cmidrule(lr){2-5} \cmidrule(lr){6-7} \cmidrule(lr){8-10} \cmidrule(lr){14-16}
Calibration
  & {$a_0$} & {$a_1$} & {$a_2$} & {$a_3$}
  & {Min} & {Max}
  & {$e_0$} & {$e_1$} & {$e_2$}
  & {$\sigma_{\rm fit}$} & {$\sigma_{\rm int}$} & {$\sigma_{\rm cal}$}
  & {Min} & {Max}
  & {Turnover}  \\
\midrule

R2 & 8.00 & 0.92 & 0.40 & 0.17 & -0.90 & 0.69 & -0.19 & -0.06 & 0.36 & 0.36 & 0.35 & 0.36 & 7.37 & 8.88 & \NA \\
R3 ~{\scriptsize upper} & 8.70 & -0.41 & -0.21 & -0.12 & -1.20 & 0.85 & -0.14 & -0.03 & 0.29 & 0.23 & 0.23 & 0.21 & 8.13 & 9.08 & 8.06\\
\phantom{R3} ~{\scriptsize lower} & 7.28 & 0.43 & 0.21 & 0.12 & 0.03 & 0.85 & -0.51 & 0.28 & 0.19 & 0.19 & 0.17 & 0.21 & 7.29 & 7.86 & \NA\\
R23 ~{\scriptsize upper} & 8.88 & -0.43 & 0.17 & -0.43 & -0.20 & 0.92 & -0.15 & 0.18 & 0.18 & 0.23 & 0.21 & 0.18 & 8.29 & 8.98 & 8.31\\
\phantom{R23} ~{\scriptsize lower} & 6.90 & 0.52 & 0.58 & \NA & 0.35 & 0.92 & -1.81 & 2.43 & -0.63 & 0.16 & 0.14 & 0.18 & 7.15 & 7.88 & 8.05\\
O32 & 8.33 & -0.32 & 0.08 & -0.09 & -1.19 & 1.80 & -0.02 & 0.02 & 0.30 & 0.33 & 0.32 & 0.33 & 8.96 & 7.50 & \NA \\
\Rhat ~{\scriptsize upper} & 8.56 & -0.47 & -0.26 & -0.15 & -1.14 & 0.64 & -0.03 & -0.02 & 0.23 & 0.22 & 0.22 & 0.22 & 8.11 & 8.98 & 8.18\\
\phantom{\Rhat} ~{\scriptsize lower} & 7.20 & 0.70 & 0.43 & 0.23 & -0.20 & 0.64 & -0.02 & 0.08 & 0.12 & 0.22 & 0.22 & 0.15 & 7.07 & 7.89 & 8.00\\
\RNe ~{\scriptsize upper} & 8.31 & -0.77 & -0.53 & -0.18 & -1.55 & 0.12 & 0.07 & 0.14 & 0.23 & 0.20 & 0.17 & 0.18 & 8.21 & 8.90 & 8.19\\
\phantom{\RNe} ~{\scriptsize lower} & 7.71 & 1.34 & 0.71 & 0.21 & -0.76 & 0.12 & -0.96 & -0.47 & 0.13 & 0.18 & 0.18 & 0.18 & 7.02 & 7.89 & 7.98\\
Ne3O2 & 7.85 & -0.71 & -0.25 & -0.05 & -2.46 & 0.54 & -0.02 & -0.04 & 0.26 & 0.32 & 0.31 & 0.32 & 8.81 & 7.39 & \NA \\
Ne3 ~{\scriptsize upper} & 7.90 & -1.11 & -0.52 & -0.10 & -2.24 & -0.37 & -0.04 & -0.07 & 0.19 & 0.21 & 0.21 & 0.16 & 8.24 & 8.88 & 8.25\\
\phantom{Ne3} ~{\scriptsize lower} & 8.11 & 1.15 & 0.21 & \NA & -1.15 & -0.37 & -2.05 & -2.70 & -0.64 & 0.21 & 0.20 & 0.23 & 7.07 & 7.72 & 7.93\\
R2Ne3 & 7.83 & 1.43 & \NA & \NA & -0.59 & 0.71 & -0.56 & 0.14 & 0.31 & 0.34 & 0.32 & 0.33 & 6.98 & 8.85 & \NA \\
O3S2 & 8.79 & -0.49 & 0.17 & -0.05 & -0.48 & 2.84 & -0.01 & 0.01 & 0.27 & 0.31 & 0.30 & 0.31 & 9.06 & 7.64 & \NA \\
R3S2 ~{\scriptsize upper} & 8.72 & -0.39 & -0.18 & -0.27 & -0.72 & 0.83 & -0.22 & 0.06 & 0.26 & 0.23 & 0.22 & 0.20 & 8.12 & 9.01 & 8.07\\
\phantom{R3S2} ~{\scriptsize lower} & 7.25 & 0.48 & 0.25 & \NA & 0.10 & 0.83 & -1.25 & 0.99 & 0.02 & 0.17 & 0.16 & 0.18 & 7.30 & 7.81 & \NA\\
N2 & 9.27 & 1.48 & 0.86 & 0.23 & -2.66 & -0.22 & 0.06 & 0.17 & 0.33 & 0.23 & 0.21 & 0.24 & 7.09 & 8.98 & \NA \\
O3N2 & 8.73 & -0.27 & 0.04 & -0.02 & -0.80 & 3.26 & 0.00 & -0.01 & 0.24 & 0.27 & 0.25 & 0.27 & 8.99 & 7.69 & \NA \\
R3N2 & 8.61 & -0.26 & 0.04 & -0.02 & -1.26 & 2.79 & 0.01 & -0.02 & 0.23 & 0.27 & 0.25 & 0.27 & 9.05 & 7.73 & \NA \\
N2O2 & 8.84 & 0.61 & 0.42 & 0.29 & -1.87 & 0.35 & 0.06 & 0.10 & 0.28 & 0.27 & 0.25 & 0.27 & 7.29 & 9.12 & \NA \\
S2 & 9.17 & 0.87 & \NA & \NA & -2.07 & -0.24 & 0.11 & 0.29 & 0.45 & 0.30 & 0.29 & 0.30 & 7.38 & 8.96 & \NA \\
N2S2 & 8.52 & 0.59 & 0.19 & 1.07 & -0.89 & 0.50 & -0.12 & 0.01 & 0.23 & 0.26 & 0.24 & 0.26 & 7.39 & 8.99 & \NA \\
N2S2\ha & 8.64 & 0.60 & 0.50 & 0.47 & -1.48 & 0.43 & 0.01 & 0.03 & 0.22 & 0.24 & 0.21 & 0.24 & 7.33 & 9.02 & \NA \\
S3 & 10.30 & 3.57 & 0.98 & \NA & -1.55 & -0.28 & -0.76 & -1.48 & -0.32 & 0.38 & 0.35 & 0.38 & 7.13 & 9.38 & \NA \\
S3O3 & 8.91 & 0.37 & -0.09 & \NA & -2.34 & 0.43 & -0.04 & -0.06 & 0.21 & 0.26 & 0.23 & 0.26 & 7.55 & 9.05 & \NA \\
S23 & 9.13 & 1.64 & 0.19 & \NA & -1.36 & -0.17 & 0.09 & 0.21 & 0.32 & 0.27 & 0.23 & 0.27 & 7.25 & 8.85 & \NA \\
Ar3 ~{\scriptsize upper} & 6.63 & -1.86 & -0.40 & \NA & -2.33 & -1.58 & -0.59 & -2.22 & -1.87 & 0.19 & 0.17 & 0.14 & 8.58 & 8.81 & 8.60\\
\phantom{Ar3} ~{\scriptsize lower} & 13.40 & 4.64 & 0.85 & \NA & -2.55 & -1.58 & -0.62 & -2.48 & -2.27 & 0.19 & 0.19 & 0.18 & 7.06 & 8.18 & 8.31\\
Ar3O3 & 8.84 & -0.09 & -0.25 & \NA & -2.45 & -0.56 & 0.03 & 0.07 & 0.24 & 0.27 & 0.25 & 0.27 & 7.57 & 8.82 & \NA \\
O3HeI ~{\scriptsize upper} & 9.02 & -0.39 & 0.23 & -0.16 & 0.04 & 1.71 & -0.16 & 0.32 & 0.09 & 0.21 & 0.21 & 0.18 & 8.21 & 9.00 & 8.21\\
\phantom{O3HeI} ~{\scriptsize lower} & 6.55 & 0.21 & 0.27 & \NA & 1.04 & 1.71 & -1.69 & 4.93 & -3.39 & 0.18 & 0.17 & 0.19 & 7.07 & 7.71 & 7.89\\
Ar3N2 & 8.14 & -0.57 & -0.25 & -0.09 & -1.85 & 0.56 & 0.03 & 0.07 & 0.26 & 0.28 & 0.27 & 0.28 & 8.95 & 7.73 & \NA \\
Ne3S2 & 8.22 & -0.39 & -0.10 & -0.03 & -2.06 & 1.09 & 0.00 & 0.03 & 0.26 & 0.29 & 0.28 & 0.29 & 8.82 & 7.64 & \NA \\
Ne3S3 & 8.42 & -0.47 & -0.16 & -0.05 & -1.60 & 1.10 & 0.02 & 0.04 & 0.18 & 0.24 & 0.23 & 0.24 & 8.95 & 7.65 & \NA \\

\bottomrule
\end{tabular}}
\end{table*}

\begin{figure*}
	\includegraphics[width=\textwidth]{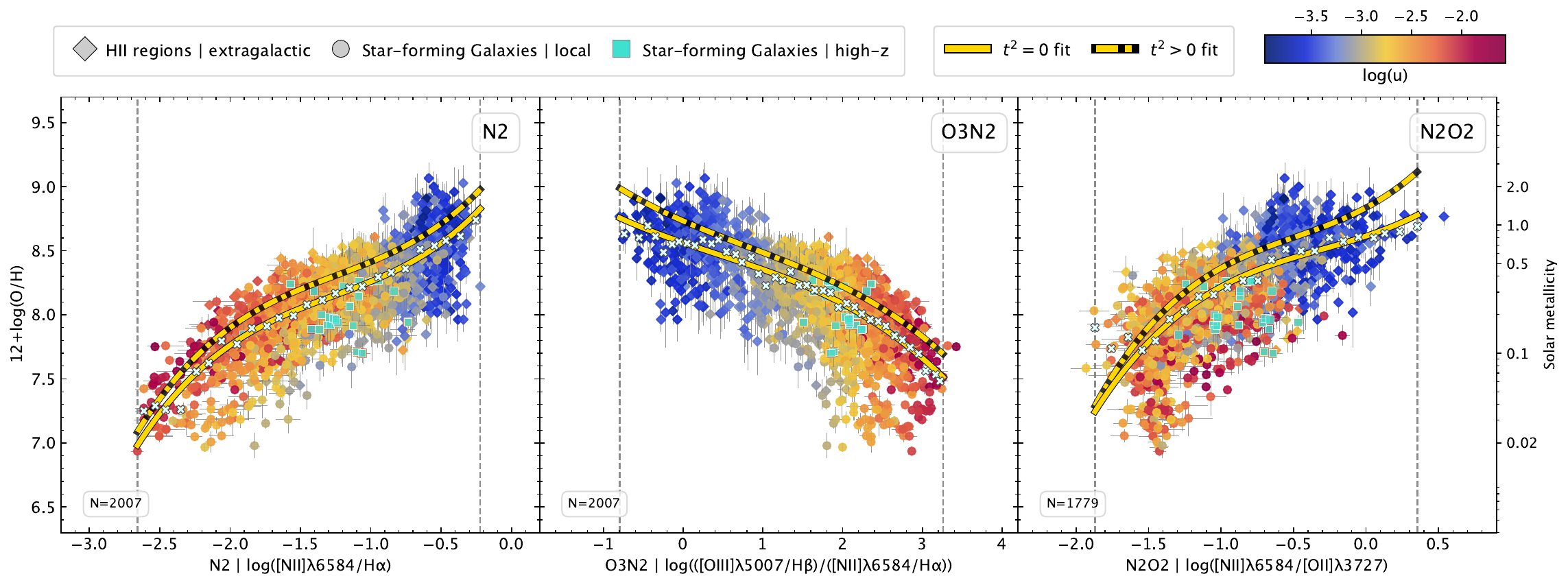}
	\includegraphics[width=\textwidth]{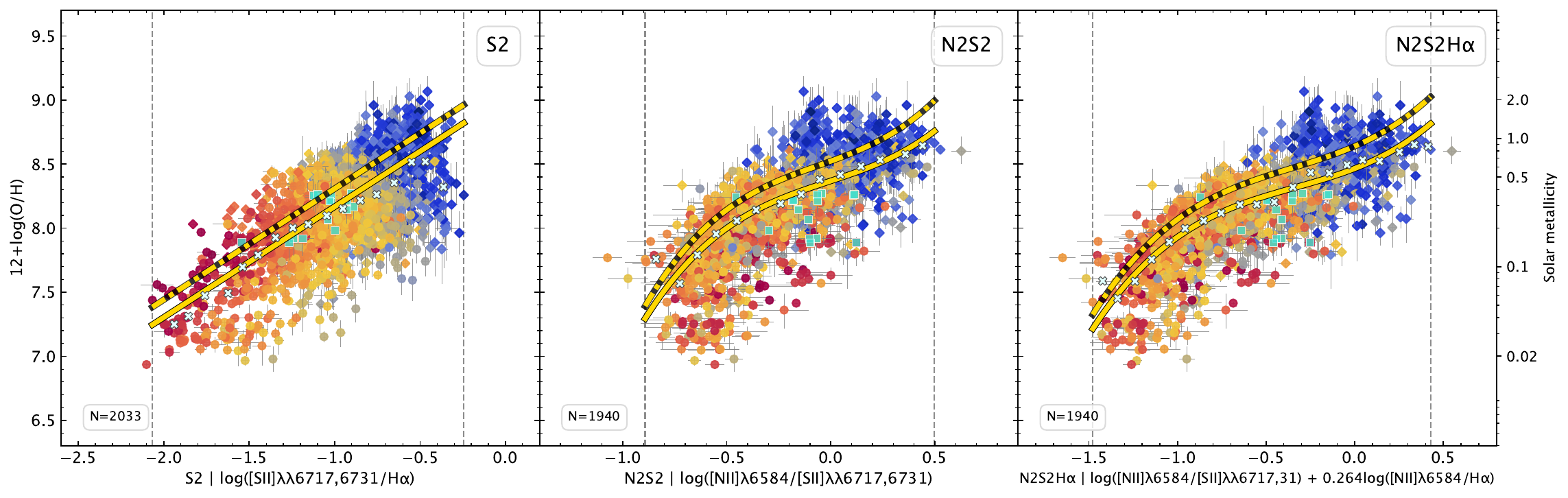}
	\includegraphics[width=\textwidth]{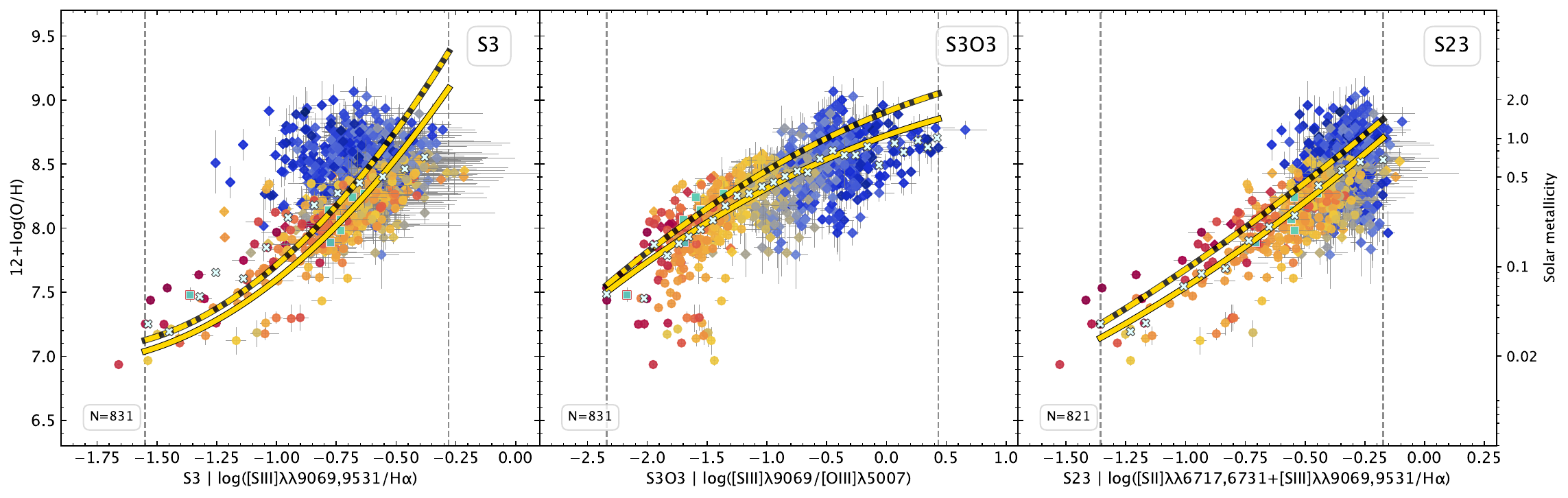}
    \caption{Gas-phase metallicity calibrations derived from the DESIRED calibration sample. Top row (from left to right): N2, O3N2, and N2O2; middle row: S2, N2S2, and N2S2H$\alpha$; bottom row: S3, S3O3 and S23. All symbols, definitions, and plotting conventions as in Fig.~\ref{fig:calib_1}.}
    \label{fig:calib_2}
\end{figure*}

\subsection{Nitrogen-based metallicity diagnostics}
\label{sec:nitrogen}

Nitrogen-based calibrators require particular care because the nucleosynthetic origin of nitrogen differs fundamentally from that of the $\alpha$-elements. While $\alpha$-element-to-oxygen ratios are expected to remain approximately constant, nitrogen exhibits both primary and secondary production channels. At low metallicities, nitrogen is predominantly produced as a primary element, mainly by massive stars, leading to an approximately constant N/O ratio with O/H \citep{EdmundsPagel:78, Alloin:79}. At higher metallicities (12+log(O/H) $\gtrsim$ 8.2), secondary production becomes dominant \citep{Kewley:19}, as nitrogen is synthesized from pre-existing C and O via the CNO cycle, mainly in low- and intermediate-mass stars. Because this secondary component depends on the previous metal content of the stellar population, N/O increases steeply with O/H in the metal-rich regime. Additionally, the delayed release of nitrogen due to stellar lifetimes, variations in star formation efficiency, and possible contributions from interacting massive binaries introduce significant dispersion in N/O at fixed metallicity \citep{Molla:06, Kobayashi:20, Romano:22, Farmer:23}. As a consequence, nitrogen-based diagnostics do not trace oxygen abundance uniquely, but are modulated by the chemical evolution history of the system.

Fig.~\ref{fig:calib_2} presents the nitrogen-based diagnostics N2, O3N2, N2O2, N2S2 and N2S2H$\alpha$, all based on the \nii\ $\lambda$6584 emission line and its combinations with \oii\ $\lambda$3727, \oiii\ $\lambda$5007, \sii\ $\lambda\lambda$6717,31 and H$\alpha$. 

The N2 index, originally proposed by \citet{StorchiBergmann:94} as an abundance estimator, is particularly convenient in practice, since the involved lines lie close in wavelength and therefore require minimal spectral coverage and are largely insensitive to dust reddening. However, this diagnostic essentially traces the nitrogen abundance, which has both primary and secondary nucleosynthetic components. Consequently, if a given object follows an N/O evolutionary path that differs from that of the calibrating sample, the inferred metallicity may be biased.

The DESIRED N2 calibration is based on 2007 spectra and shows the expected monotonic increase of metallicity with increasing line index, with an overall dispersion of 0.20~dex. The sampling becomes sparse at low metallicities, as \nii\ $\lambda$6584 becomes intrinsically faint in metal-poor systems, a behaviour also reflected in recent JWST studies reporting high \nii\ non-detection rates in high-redshift, low-mass galaxies \citep[e.g.][]{Shapley:23a, Sanders:23b}. At the high-metallicity end (12+log(O/H) $\gtrsim$ 8.0, corresponding to N2 $\gtrsim$ $-$0.7), the dispersion increases, consistent with the growing contribution of secondary nitrogen production in this regime. Furthermore, N2 is known to be sensitive to variations in the ionisation parameter \citep[e.g.][]{KewleyDopita:02, Nagao:06}, as reflected in the colour-coding of the data points.

The O3N2 calibration \citep{Alloin:79, PettiniPagel:04} shares with N2 the practical advantage of being largely unaffected by dust extinction, except in cases of significant differential reddening. The DESIRED O3N2 calibration, also based on 2007 spectra, displays a monotonic increase in metallicity with decreasing O3N2 index, with a slightly larger dispersion of 0.23~dex. The relation becomes increasingly degenerate at metallicities below 12+log(O/H) $\lesssim$ 8.0 (corresponding to O3N2 $\gtrsim$ 1.5). As in the case of N2, the O3N2 diagnostic inherits the same fundamental limitations associated with the behaviour of the N/O ratio and its sensitivity to ionisation conditions.

\begin{figure*}
    \centering
	\includegraphics[width=\textwidth]{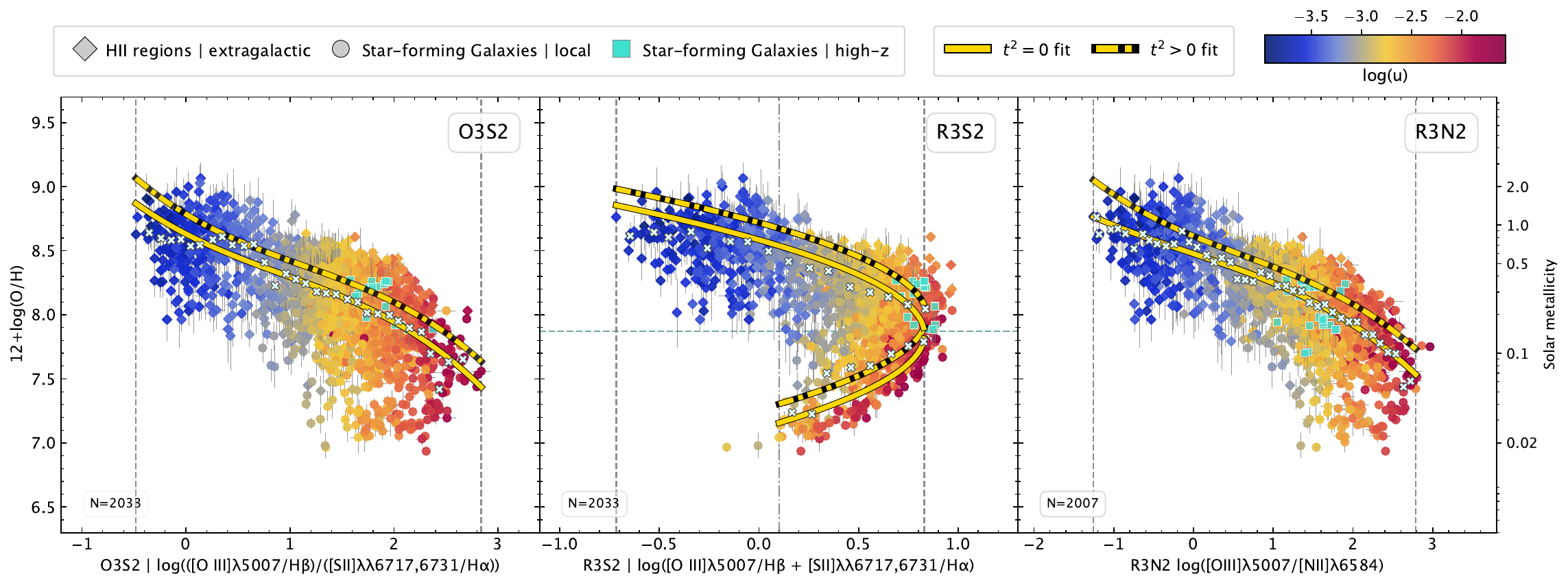}
     \caption{
     DESIRED gas-phase metallicity calibrations for the O3S2, R3S2 and R3N2 diagnostics. All symbols, definitions, and plotting conventions are identical to those described in Fig.~\ref{fig:calib_1}.
     }
    \label{fig:calib_1d}
\end{figure*}


For completeness, we also present a closely related diagnostic to O3N2, defined as R3N2 $\equiv$ log(\oiii\ $\lambda$5007/\nii\ $\lambda$6583), which omits the H$\alpha$ and H$\beta$ normalisation terms included in the standard O3N2 definition. This diagnostic was first proposed by \citet{Alloin:79} and subsequently recalibrated by \citet{Maiolino:08} and \citet{Garg:24}. The resulting relation is shown in Fig.~\ref{fig:calib_1d}. The distribution of calibrating points closely resembles that of O3N2, and the resulting dispersion is similarly 0.23~dex.

Among the nitrogen-based diagnostics, the N2O2 and N2S2 indices \citep{Jensen:76, DopitaEvans:86, KewleyDopita:02} are primarily tracers of the nitrogen-to-$\alpha$ abundance ratios, N/O and N/S, respectively \citep[e.g.][]{PerezMonteroContini:09}. Their use as metallicity indicators relies on the empirical correlation between N/$\alpha$ and O/H at intermediate and high metallicities (12+log(O/H) $\gtrsim$ 8.0), where secondary nitrogen production drives a monotonic increase of N/O with O/H. Under the commonly adopted assumption that S/O remains approximately constant with metallicity \citep[e.g.][]{Berg:20, Esteban:25}, both ratios become sensitive to N/O and thus indirectly to O/H. However, this dependence is not fundamental but mediated by the N/$\alpha$--O/H relation. 
In particular, they lose sensitivity in the low-metallicity regime where N/$\alpha$ flattens. From an ionisation standpoint, N2O2 is relatively insensitive to the ionisation parameter because \nii\ and \oii\ arise from ions with similar ionisation potentials (10.4 eV for S$^+$, 13.6 eV for O$^+$, 14.5 eV for N$^+$), making the ratio comparatively robust against variations in excitation conditions or the presence of a hard radiation field. In contrast, N2S2 retains some dependence on the ionisation parameter due to the differing ionisation properties of the species involved.

The N2O2 and N2S2 calibrations are based on 1779 and 1940 spectra, respectively. Both exhibit the expected monotonic increase of O/H with increasing line index, with overall dispersions of 0.25~dex for N2O2 and 0.24~dex for N2S2. In both cases, the dispersion increases towards the low-metallicity regime, leading to a partial degeneracy of the indices for 12+log(O/H) $\lesssim$ 8.0. Consequently, the use of N2O2 and N2S2 is better suited to intermediate and high metallicities (12+log(O/H) $\gtrsim$ 8.0), corresponding approximately to N2O2 $>-1.0$ and N2S2 $>-0.3$.

The N2S2H$\alpha$ diagnostic was introduced by \citet{Dopita:16} as a refinement of the N2S2 index. It was designed to reduce the sensitivity to ionisation parameter relative to N2 and S2, and is particularly attractive for systems where only the red spectral range is available. The DESIRED N2S2H$\alpha$ calibration, constructed from 1940 spectra, shows a monotonic increase of O/H with the line index and a slightly lower dispersion than N2S2 (0.21~dex). In addition, the low-metallicity regime appears less degenerate than in N2S2, suggesting that N2S2H$\alpha$ may be applicable over a broader metallicity range, subject to the caveats discussed above.

It is worth noting that N2O2 is sensitive to dust reddening, as its constituent emission lines lie at widely separated wavelengths across the optical spectrum. In contrast, N2S2 and N2S2H$\alpha$ benefit from their limited wavelength separation, which reduces their sensitivity to extinction effects. Note also that, as quantified in Sec.~\ref{sec:highz}, the high-redshift objects show a statistically significant offset from the full calibration sample for several N-based diagnostics (N2, N2O2, N2S2, and N2S2H$\alpha$), consistent with elevated N/O ratios at fixed O/H in high-redshift galaxies, whereas no such offset is found for O3N2.

\subsection{Sulphur-based metallicity diagnostics}
\label{sec:sulphur}

Sulphur-based calibrators benefit from the fact that sulphur is an $\alpha$-element and therefore shares the same dominant nucleosynthetic origin as oxygen, namely core-collapse supernovae. As a result, the S/O abundance ratio is expected to remain approximately constant over a wide metallicity range and to evolve much less strongly than N/O. Nevertheless, some nucleosynthesis and chemical evolution models predict a non-negligible contribution of sulphur (and argon) from Type Ia supernovae \citep[e.g.][]{Iwamoto:99, Johnson:19, Kobayashi:20}, which seems to be confirmed in some observational studies \citep{Esteban:25}. This particular behaviour could introduce secondary variations under specific evolutionary conditions. 

From a nebular physics perspective, the atomic structure of sulphur ions S$^+$ and S$^{2+}$ is similar to that of oxygen ions O$^+$ and O$^{2+}$. Their main optical emission lines: \sii\ $\lambda\lambda$6716,6731 and \siii\ $\lambda\lambda$9069,9532, are analogous to the oxygen lines that define R23. However, because of their longer wavelengths, sulphur lines contribute more efficiently to nebular cooling at relatively low temperatures and remain strong even in high-metallicity environments, extending the applicability of direct and strong-line methods toward solar abundances. The emissivities of \sii\ and \siii\ are less sensitive to electron temperature than their oxygen counterparts, implying that the turnover in their intensity--metallicity relation occurs at higher metallicities and that the relation can remain single-valued over a broader range \citep{DiazPM:00}. At low metallicities and high excitation, a non-negligible fraction of sulphur may be present as S$^{3+}$, requiring ICFs to account for usually unobserved infrared transitions (e.g. 10.4 $\mu$m). Overall, sulphur lines provide relatively robust abundance diagnostics, with moderate sensitivity to reddening due to their proximity to nearby hydrogen recombination lines.

Fig.~\ref{fig:calib_2} and \ref{fig:calib_1d} present the calibrations involving the \sii\ and \siii\ lines, as well as combinations with oxygen lines proposed in the literature. The S2, S3, and S23 diagnostics are direct analogues of the oxygen-based indicators R2, R3, and R23, replacing \oii\ and \oiii\ with \sii\ and \siii\ lines of comparable ionisation state, and using H$\alpha$ instead of H$\beta$ to reduce sensitivity to reddening corrections.

The DESIRED S2 calibration is based on 2033 spectra and exhibits a monotonic trend with the line ratio, with a relatively large dispersion of 0.27~dex that remains approximately constant across the full metallicity range. This substantial scatter is consistent with the fact that the \sii\ lines arise predominantly in partially ionised zones at the edges of \hh regions. The extent of this partially ionised zone is sensitive to the ionisation parameter, making the \sii/H$\alpha$ ratio strongly dependent on ionisation conditions and contributing to the observed dispersion. Furthermore, this ratio is particularly susceptible to contamination by DIG, which tends to enhance \sii\ emission relative to that expected from pure \hii region spectra \citep[e.g.][]{Kreckel:22}. In this context, aperture effects may also play a non-negligible role in the DESIRED sample: spatially unresolved or large-aperture observations are more prone to DIG contamination, which could introduce an additional source of scatter in S-based diagnostics.

The dependence on ionisation parameter can be mitigated by incorporating the higher-ionisation \siii\ lines in the numerator, leading to the definition of the S23 parameter, originally introduced as a sulphur abundance indicator \citep{VilchezEsteban:96} and subsequently adapted as an oxygen metallicity diagnostic \citep{DiazPM:00}. The DESIRED S23 calibration is constructed from 821 spectra, a relatively lower number than for the previous calibrations. This reflects the observational difficulty of measuring the near-infrared \siii\ $\lambda\lambda$9069,9532 lines with sufficient signal-to-noise to derive reliable $T_{\rm e}$-based abundances using the temperature-sensitive \siii\ $\lambda$6312/$\lambda$9069 ratio. Although the calibration spans a wide metallicity range 12+log(O/H) $\approx 7.1-8.7$, the number of calibrating regions decreases sharply at low metallicity, with fewer detections at 12+log(O/H) $\lesssim$ 7.7 despite the large parent compilation. Nevertheless, S23 displays a monotonic behaviour with O/H, with a dispersion of 0.23 dex. For S23 (and S3), we relaxed the statistical validity criterion from a minimum of five regions per 0.1 dex bin to five regions per 0.2 dex bin, owing to the reduced sample size. The impact of reddening can, in principle, be alleviated by using theoretical recombination ratios linking H$\alpha$ (or H$\beta$) to nearby Paschen lines close to the \siii\ features.

The S3 diagnostic, recently introduced by \citet{Sanders:26} as a calibration for high-redshift studies, is constructed from 831 spectra and exhibits a monotonic increase with metallicity, albeit with the largest overall dispersion among all the calibrations presented here ($\sigma_{\rm fit} = 0.35$~dex). The scatter is particularly pronounced at the high-metallicity end, where both the measurement uncertainties in S3 and the presence of significant outliers broaden the relation. Despite this limitation, S3 probes a wide abundance range, and together with S23 demonstrates the potential of sulphur-based diagnostics as metallicity indicators, particularly in regimes where oxygen lines are unavailable.

The S3O3 calibration, originally proposed by \citet{Stasinska:06a}, is constructed from 831 regions in the DESIRED sample. It exhibits a monotonic increase of metallicity with the line ratio and shows the smallest dispersion among the monotonic sulphur-based diagnostics, with an RMS of 0.22~dex. However, as in the case of N2O2, the dispersion increases significantly at low metallicities (12+log(O/H) $\lesssim$ 7.8), limiting its applicability in the metal-poor regime. At intermediate and high metallicities, S3O3 appears more promising, particularly given that the relevant emission lines can be detected at high redshift (up to $z \sim 4.5$ with JWST/NIRSpec). Nevertheless, caution is warranted, as photoionisation models indicate that this index may be more sensitive to variations in the ionisation parameter than to metallicity itself \citep{Kewley:19}.

The O3S2 and R3S2 calibrations, first introduced by \citet{Curti:20}, are shown in Fig.~\ref{fig:calib_1d} and are both constructed from 2033 calibrating spectra. The O3S2 diagnostic is formally analogous to O3N2, with \nii\ replaced by \sii, and exhibits a similar qualitative behaviour. The DESIRED O3S2 calibration shows a monotonic anticorrelation between the index and O/H, with a relatively large dispersion of 0.27~dex. The scatter increases at metallicities 12+log(O/H) $\lesssim$ 7.8 (O3S2 $\lesssim$ 1.0), in a manner comparable to O3N2 over the same abundance range. The relation therefore appears more reliable at intermediate and high metallicities, 12+log(O/H) $\gtrsim$ 8.0.

The R3S2 index is analogous to R23, replacing the \oii\ $\lambda$3727/H$\beta$ term with \sii\ $\lambda\lambda$6717,6731/H$\alpha$, and relies on the similar nucleosynthetic origin of sulphur and oxygen. An additional advantage is its reduced sensitivity to dust reddening due to the proximity in wavelength of the constituent lines. However, as in the case of R23, R3S2 exhibits a secondary dependence on the ionisation parameter and a bivalued structure, with a turnover at 12+log(O/H) $\approx$ 7.90. The DESIRED R3S2 calibration shows an RMS scatter of 0.18 dex (upper branch) and 0.19 dex (lower branch). As with R23, the metallicity determination becomes degenerate near the turnover of the relation, where a large fraction of the calibrating regions, including most of the high-redshift spectra, are concentrated. Its applicability therefore appears to be strongest in the intermediate- to high-metallicity regime (12+log(O/H) $\gtrsim$ 8.0).

\begin{figure*}
	\includegraphics[width=\textwidth]{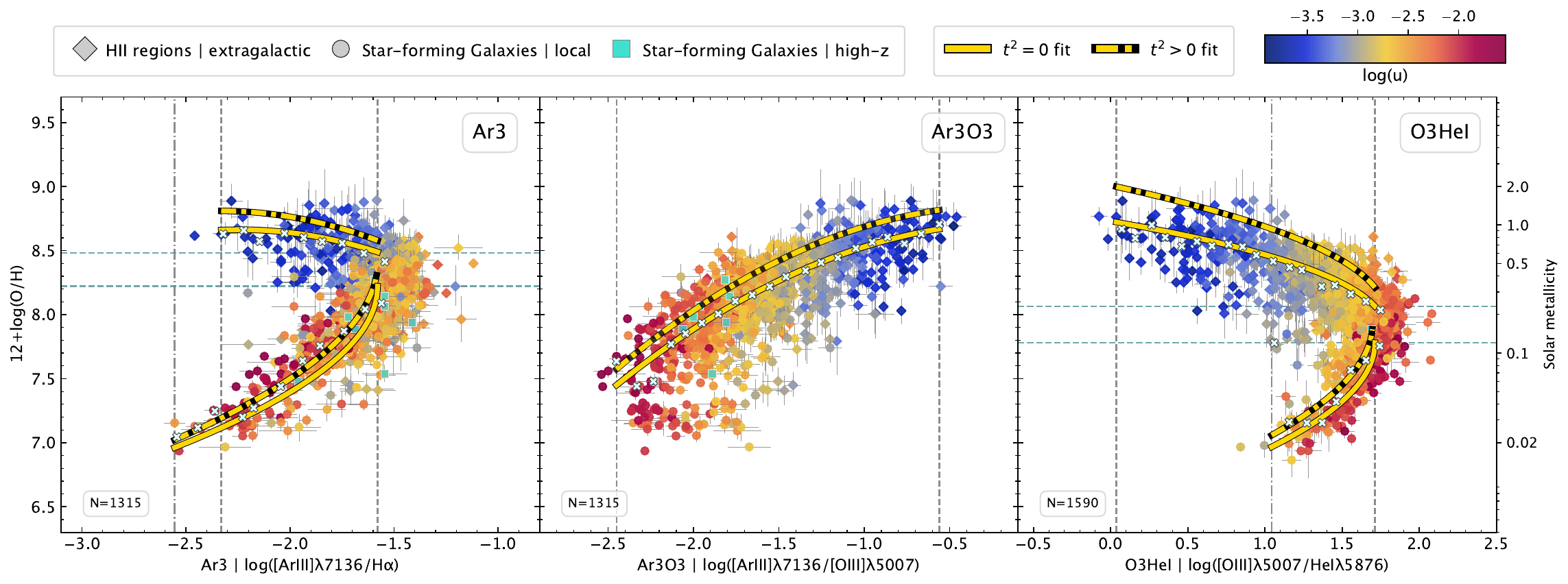}
	\includegraphics[width=\textwidth]{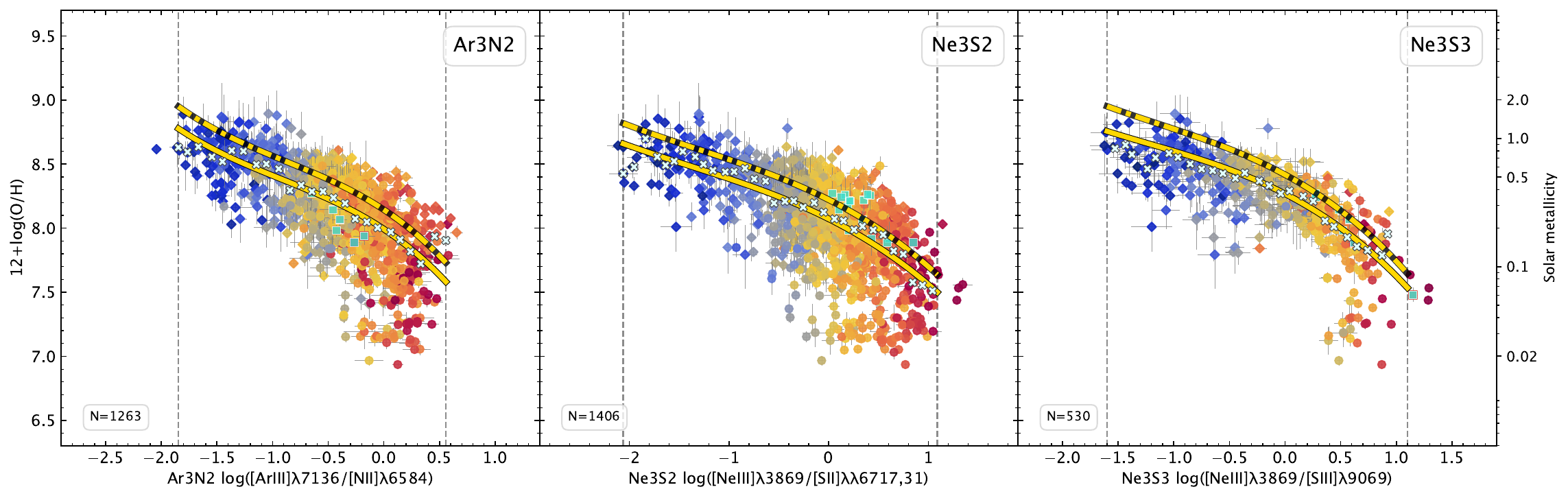}
    \caption{Gas-phase metallicity calibrations derived from the DESIRED calibration sample. Top row (left to right): Ar3, Ar3O3, O3HeI. Bottom row: Ar3N2, Ne3S2, and Ne3S3. The notation and ratio indices are defined in Table~\ref{tab:calibrations} and indicated on the x-axis of each panel. Symbols, definitions, labels and colour-coding similar to Fig.~\ref{fig:calib_1}.}
    \label{fig:calib_3}
\end{figure*}

\subsection{Argon-based metallicity diagnostics}
\label{sec:argon}

Argon-based calibrators rely on the fact that argon, like sulphur, is an $\alpha$-element synthesized primarily during oxygen burning in massive stars and released into the interstellar medium through core-collapse supernovae. In principle, Ar/O is therefore expected to remain approximately constant and to trace oxygen abundances closely. However, several observational studies have reported a systematic decrease of Ar/O with increasing O/H \citep[e.g.][]{Garnett:02, PerezMontero:07, Kojima:21, ArellanoCordova:24}, including a decline of $\sim$0.24 dex in log(Ar/O) across the full metallicity range for SFGs and \hh regions \citep{Esteban:25}. This behaviour is not straightforwardly predicted by standard nucleosynthesis and chemical evolution models and may reflect the sensitivity of Ar yields to the physical conditions in the innermost layers of the progenitor during the supernova explosion \citep{WeaverWoosley:93}. Additional complexity may arise from possible contributions of Type Ia supernovae, as suggested in some Galactic chemical evolution models \citep{Kobayashi:20} and recent observational interpretations \citep{Arnaboldi:22}. Consequently, although argon is theoretically expected to track oxygen, empirical deviations of Ar/O with metallicity introduce an additional layer of uncertainty when Ar-based strong-line diagnostics are used as metallicity tracers, specially in the high metallicity end.

Argon exhibits optical emission lines from the Ar$^{2+}$ and Ar$^{3+}$ ionisation states in \hh regions and SFGs. Ar$^{2+}$ is produced by photons with energies between 27.6 and 40.7 eV, giving rise to the \ariii\ $\lambda$7136 and $\lambda$7751 lines, while Ar$^{3+}$ requires photons in the 40.7--59.8 eV range, corresponding to the \ariv $\lambda$4711 and $\lambda$4740 transitions. It should be noted that the \ariv\ $\lambda$4711 line can blended with \hei\ $\lambda$4713 in low-resolution spectra, which can complicate the measurement of Ar$^{3+}$-based diagnostics and introduce systematic uncertainties in the derived line fluxes.

In recent years, observations of the strongest \ariii\ line at 7136 \AA (originating in an intermediate-excitation zone similar to that of \siii) have increased substantially in both \hh regions and SFGs. The large DESIRED compilation therefore enables a statistically robust reassessment of previously proposed argon-based metallicity diagnostics.

Fig.~\ref{fig:calib_3} presents the DESIRED Ar3 and Ar3O3 calibrations, both constructed from 1315 spectra. The Ar3 calibration, recently proposed by \citet{Sanders:26}, is formally analogous to R3 and therefore exhibits a bivalued behaviour, with turnover values at 12+log(O/H) $\approx$ 8.48 (upper) and $\approx$ 8.22 (lower). In contrast to other double-valued diagnostics, however, the lower-metallicity branch of Ar3 extends to substantially higher metallicities than its counterparts. The large DESIRED compilation enables a robust characterisation of the Ar3 relation in the low-metallicity regime, where the lower branch shows a dispersion of 0.19 dex. This makes Ar3 particularly suitable for determining metallicities below 12+log(O/H) $\lesssim$ 8.2, provided that the degeneracy between branches can be resolved. The few high-redshift regions populate the upper portion of the lower branch, approaching the turnover, with no coverage of the upper branch. The upper branch spans a comparatively narrow range in the Ar3 index and is populated by too few calibrating regions to exhibit the classical double-branch morphology seen in the other diagnostics; it is therefore calibrated separately using binned medians (see Appendix~\ref{app:fits}), achieving a nominally low dispersion (0.11~dex) despite the sparse sampling in this regime, and its applicability at 12+log(O/H) $\gtrsim$ 8.5 remains uncertain, as the calibration flattens at the high-metallicity end and additional observations are needed to better constrain this regime.

The Ar3O3 index was proposed by \citet{Stasinska:06a} as a metallicity indicator, motivated by the expectation that it would be less affected by DIG contamination in integrated galaxy spectra \citep{ValeAsari:19}. The DESIRED Ar3O3 calibration shows a monotonic increase in metallicity with the line ratio, with an overall dispersion of 0.23~dex, and a larger scatter at metallicities 12+log(O/H) $\lesssim$ 7.5. The behaviour of Ar3O3 is primarily driven by the metallicity dependence of the electron temperature: as metallicity increases, enhanced metal cooling lowers $T_{\rm e}$, altering the relative collisional emissivities of \ariii\ $\lambda$7136 and \oiii\ $\lambda$5007, and hence their ratio. The substantial dispersion may reflect variations in the hardness of the radiation field and/or the ionisation parameter, as well as the presence of internal temperature gradients, particularly in metal-rich \hh regions \citep{Stasinska:06a}.

In practice, the Ar3O3 diagnostic requires a relatively wide spectral coverage and a reliable reddening correction, since the involved emission lines are separated by more than 2100 \AA. Given its behaviour, it appears more suitable for metallicities 12+log(O/H) $\gtrsim$ 7.5.  For high-redshift systems, Ar3 and Ar3O3 can be observed with JWST/NIRSpec over wide redshift intervals, approximately $z \leq 6.4$ for Ar3 and $0.2 \leq z \leq 6.4$ for Ar3O3.

\subsection{New metallicity diagnostics}
\label{sec:new}

Considering the large number of emission lines available in the high-quality DESIRED sample, we explored additional line ratios for which a potentially useful correlation with metallicity could be expected, identifying four new empirical abundance relations. Fig.~\ref{fig:calib_3} presents the diagnostics O3HeI = log(\oiii\ $\lambda$5007 / \hei\ $\lambda$5876), Ar3N2 = log(\ariii\ $\lambda$7136 / \nii\ $\lambda$6584), Ne3S2 = log(\neiii\ $\lambda$3869 / \sii\ $\lambda\lambda$6717,6731), and Ne3S3 = log(\neiii\ $\lambda$3869 / \siii\ $\lambda$9069).

The O3HeI relation is analogous to R3, replacing the hydrogen recombination normalization with the nearby \hei\ $\lambda$5876 recombination line. From a physical standpoint, both H$\alpha$ and \hei\ are recombination lines; however, \hei\ emission arises from He$^+$, which requires more energetic photons than those needed to ionise hydrogen (He$^0\to$He$^+$ = 24.6 eV). In contrast, \oiii\ $\lambda$5007 traces O$^{2+}$, whose production requires even higher energies (O$^+\to$O$^{2+}$ $\approx$ 35.1 eV). The \oiii/\hei\ ratio therefore combines a high-excitation collisionally excited line with a recombination line whose presence depends on a higher ionisation threshold than hydrogen, making the diagnostic sensitive to excitation conditions in addition to its global dependence on O/H.

The O3HeI calibration was constructed from 1590 spectra and exhibits a bivalued behaviour, with turnover values at 12+log(O/H) $\approx$ 8.07 (upper) and $\approx$ 7.78 (lower). The dispersion of the upper branch is 0.16 dex, while the low-metallicity branch shows a larger dispersion of 0.20 dex. This diagnostic is particularly useful in cases where only a limited spectral range is available, or to help break degeneracies in calibrators that rely on closely spaced emission lines. In the case of NIRSpec, the indicator is observable over a broad redshift interval, $0.2 \leq z \lesssim 8$.

The Ar3N2 calibration is based on the ratio of \ariii\ $\lambda$7136, an $\alpha$-element line that is expected to follow the same dominant nucleosynthetic origin as oxygen, and \nii\ $\lambda$6584, whose properties and associated caveats were discussed in Sec.~\ref{sec:nitrogen}. This indicator is analogous to O3N2, with the \oiii\ $\lambda$5007 term replaced by \ariii\ $\lambda$7136, and similarly exhibits a monotonic behaviour, with metallicity increasing as the line ratio decreases.

From the standpoint of ionisation structure, \ariii\ traces Ar$^{2+}$, which requires photons with energies above 27.6 eV (Ar$^+\to$Ar$^{2+}$), while \nii\ $\lambda$6584 arises from N$^+$, whose parent ionisation potential is 14.5 eV (N$^0\to$N$^+$). Thus, Ar$^{2+}$ is produced in higher-excitation zones than N$^+$, although its ionisation threshold is somewhat lower than that of O$^{2+}$ (35.1 eV). The Ar3N2 ratio therefore combines a moderately high-excitation $\alpha$-element tracer with a low-excitation nitrogen line. As metallicity increases, enhanced cooling reduces the relative strength of collisionally excited lines from higher-ionisation species, while the nitrogen abundance increases due to secondary production. Consequently, one expects the \ariii/\nii\ ratio to decrease with increasing O/H, yielding a monotonic calibration, in close analogy to O3N2 but with a slightly different sensitivity to excitation conditions because of the lower ionisation potential of Ar$^{2+}$ compared to O$^{2+}$.

The Ar3N2 calibration is constructed from 1263 spectra and shows a dispersion of 0.25~dex. The scatter increases toward the low-metallicity regime, making the diagnostic particularly useful for measuring metallicities 12+log(O/H) $\gtrsim$ 7.6, corresponding to Ar3N2 $\lesssim$ $-$0.8. This ratio can be employed either as an independent metallicity indicator or to resolve the branch degeneracy of bivalued diagnostics, remaining accessible with NIRSpec up to $z \leq 6.4$.

The Ne3S2 and Ne3S3 diagnostics are analogous to the Ne3O2 indicator, with the \oii\ $\lambda$3727 term replaced by \sii\ $\lambda\lambda$6717,6731 and \siii\ $\lambda$9069, respectively. Their relevance as oxygen abundance tracers stems from the fact that neon and sulphur are $\alpha$-elements whose nucleosynthetic evolution is closely coupled to that of oxygen, as discussed in Secs.~\ref{sec:neon} and \ref{sec:sulphur}.

From the standpoint of ionisation structure, \neiii\ $\lambda$3869 traces Ne$^{2+}$, which requires photons with energies above 40.96 eV (Ne$^+\to$Ne$^{2+}$), probing the high-excitation zone of \hh regions. In contrast, \sii\ arises from S$^+$ (ionisation potential 10.36 eV from S$^0$, with S$^+\to$S$^{2+}$ at 23.3 eV), characteristic of low-ionisation zones, while \siii\ traces S$^{2+}$, produced at intermediate energies comparable to those required for O$^{2+}$ formation (O$^+\to$O$^{2+}$ = 35.1 eV). Consequently, Ne3S2 combines a high-excitation tracer with a low-excitation sulphur line, making it strongly sensitive to the ionisation state of the nebula, in close analogy to Ne3O2. Ne3S3, on the other hand, compares Ne$^{2+}$ and S$^{2+}$, both arising in relatively similar ionisation zones, and is therefore expected to be somewhat less sensitive to ionisation parameter variations than Ne3S2, though still responsive to excitation conditions. As metallicity increases, enhanced cooling reduces the strength of high-excitation lines relative to lower-excitation species, leading to a decrease in both ratios and producing a monotonic increase of metallicity toward lower line-ratio values, analogous to Ne3O2.

The Ne3S2 calibration is constructed from 1406 spectra and exhibits a dispersion of 0.26~dex, with increased scatter at low metallicities, closely resembling the behaviour of Ne3O2. It is therefore particularly useful for intermediate-to-high metallicities, 12+log(O/H) $\gtrsim$ 7.9, corresponding to Ne3S2 $\lesssim$ $-$0.2. Finally, the Ne3S3 calibration is based on 530 spectra, a smaller sample due to the observational challenge of simultaneously measuring \neiii\ $\lambda$3869 in the blue optical and \siii\ $\lambda$9069 near the red/near-infrared edge of the optical spectrum. Despite this limitation, Ne3S3 shows a relatively low dispersion of 0.21~dex, making it a useful calibration over a broad metallicity range. However, it becomes significantly degenerate at low metallicities, restricting its practical applicability to 12+log(O/H) $\gtrsim$ 7.9, corresponding to Ne3S3 $\lesssim$ 0.3. In both cases, reliable extinction correction is required, or alternatively the use of nearby hydrogen recombination lines combined with theoretical Case B ratios, to account for the large wavelength separation between the lines, particularly critical for Ne3S3. These newly proposed indicators can be employed either as independent metallicity diagnostics or to help resolve the branch degeneracy of other bivalued calibrations within a given spectral coverage.

All calibrations presented in this work are publicly available as a Python package\footnote{\url{https://github.com/frosalesortega/desired-calibrations}} that allows the user to derive 12+log(O/H) for both the $t^2=0$ and $t^2>0$ scenarios from any of the 27 strong-line indices, with full Monte Carlo error propagation.

\begin{figure*}
	\includegraphics[width=0.9\textwidth]{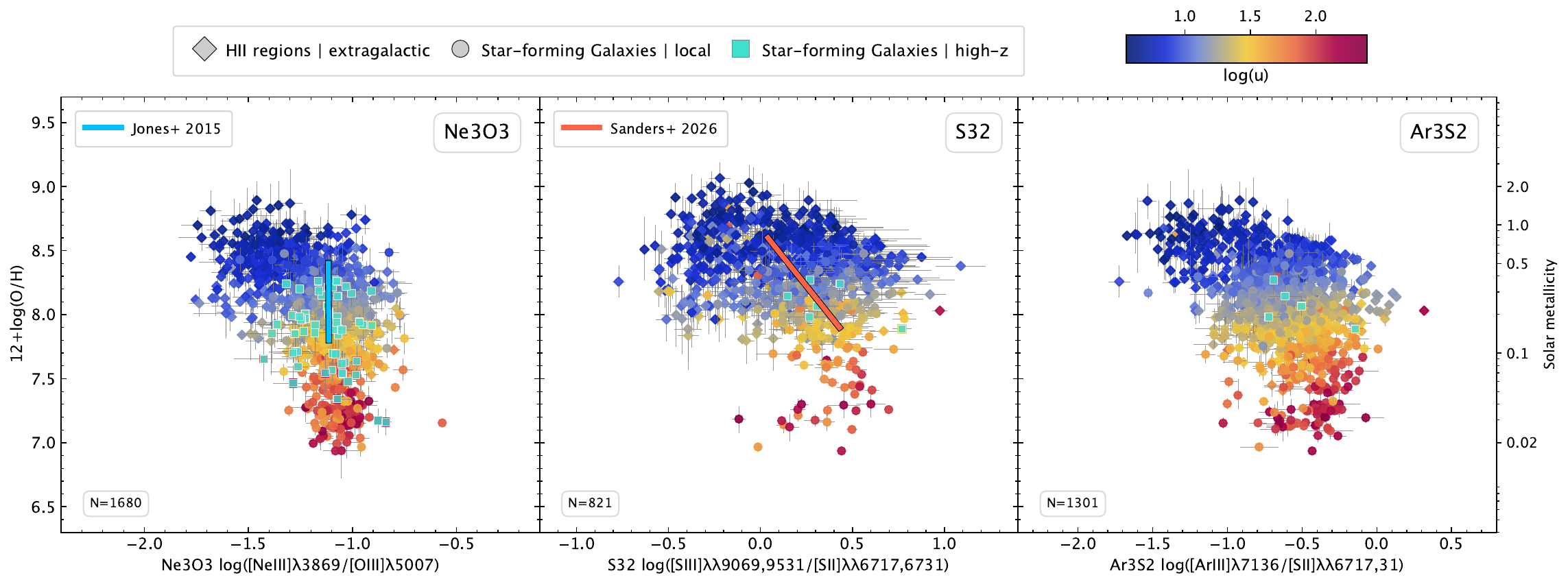}
    \caption{Strong-line diagnostics proposed in the literature found to be insensitive to metallicity. Shown is the comparison between the oxygen abundances derived from the $T_{\rm e}$-based method ($t^2=0$) and the Ne3O3, S32, and Ar3S2 indices for the DESIRED calibration sample. For NeO3, the solid blue line shows the calibration proposed by \citet{Jones:15}, while for S32 the solid red curve corresponds to the relation from \citet{Sanders:26}. Object types are indicated at the top of the figure, with high-redshift galaxies in cyan highlighted by white contours.}
    \label{fig:bad}
\end{figure*}

\subsection{Non-viable metallicity diagnostics}
\label{sec:bad}

The large statistics and high quality of the DESIRED calibration sample allow us to assess the validity of metallicity calibrations proposed in the literature, as well as other line ratios that might in principle correlate with O/H. Fig.~\ref{fig:bad} shows three relations that are found to be invalid. The Ne3O3 ratio, defined as \neiii\ $\lambda$3869/\oiii\ $\lambda$5007, was presented by \citet{Jones:15}, who already noted its weak sensitivity to metallicity and suggested that its primary usefulness might lie as a consistency check for nebular reddening. Using 1680 regions from the DESIRED calibration sample, we confirm that Ne3O3 is almost completely insensitive to metallicity: most regions cluster around Ne3O3 $\approx$ $-$1.2 over the wide range 12+log(O/H) $\approx$ 7.0--8.8, with only the highest-metallicity objects bending toward slightly lower values (Ne3O3 $\lesssim$ $-$1.4). A Pearson correlation test confirms a formally significant, but weak, correlation between Ne3O3 and metallicity ($r=-0.459$, $p=3.0\times10^{-88}$).

From a physical standpoint, this behaviour is expected. Both \neiii\ $\lambda$3869 and \oiii\ $\lambda$5007 arise from doubly ionised species, Ne$^{2+}$ and O$^{2+}$, whose formation requires relatively high-energy photons (O$^+\to$O$^{2+}$ = 35.1 eV; Ne$^+\to$Ne$^{2+}$ = 40.96 eV). These ions therefore occupy similar high-excitation zones within \hh regions. As a result, the Ne3O3 ratio primarily reflects the relative emissivities of two ions that coexist in similar physical zones and whose elemental abundances scale together. Variations in electron temperature with metallicity affect both collisionally excited lines in a broadly similar manner, further reducing the sensitivity of their ratio to O/H. The slight bending at the highest metallicities likely reflects the stronger temperature dependence of high-excitation lines as enhanced cooling lowers $T_{\rm e}$, but this effect is insufficient to generate a monotonic or dynamically useful metallicity trend.

For comparison, we include the calibration provided by \citet{Jones:15}, which in the $y$ = 12+log(O/H) versus $x$ = Ne3O3 plane corresponds almost to a vertical relation at Ne3O3 $\approx$ $-$1.1 over a narrow metallicity interval, reinforcing the conclusion that Ne3O3 does not constitute a viable metallicity diagnostic.

The S32 ratio, defined as log(\siii\ $\lambda\lambda$9069,9531 / \sii\ $\lambda\lambda$6717,6731) by \citet{Sanders:26}, was proposed as analogous to O32 and Ne3O2, with a metallicity dependence driven by the anticorrelation between ionisation parameter and O/H. They constructed this relation using 47 high-redshift regions from the AURORA survey. Using the DESIRED calibration sample, we computed S32 for 821 regions and find instead a clustering of points centred at S32 $\approx$ 0.3 and 12+log(O/H) $\approx$ 8.4, with only a few outliers. A Pearson correlation test detects a formally significant correlation between S32 and metallicity ($r=-0.305$, $p=4.1\times10^{-19}$); however, the large dispersion around the \citet{Sanders:26} relation visible in Fig.~\ref{fig:bad} indicates that S32 provides little practical constraining power on metallicity despite the detectable overall trend. Although the sample contains relatively few objects in the interval 12+log(O/H) $\approx$ 7.3--7.8, the handful of regions around 12+log(O/H) $\approx$ 7.0 already indicate that the relation would remain degenerate over the full DESIRED metallicity range.

From a physical perspective, this behaviour can be understood by examining the ionisation structure traced by sulphur lines. The S32 ratio compares S$^{2+}$ and S$^+$. The ionisation potential for S$^+\to$S$^{2+}$ is 23.3 eV, significantly lower than that required to produce O$^{2+}$ (35.1 eV) or Ne$^{2+}$ (40.96 eV). Consequently, both S$^+$ and S$^{2+}$ arise in relatively extended and overlapping zones within \hh regions, probing lower-to-intermediate excitation conditions compared to O32 or Ne3O2. Because sulphur is an $\alpha$-element expected to track oxygen closely in abundance, the S$^{2+}$/S$^+$ ratio is primarily sensitive to the ionisation parameter rather than to elemental abundance variations \citep{Kewley:19}. 

In addition, changes in electron temperature with metallicity affect both ions in a similar manner, further reducing any differential metallicity dependence. As a result, S32 predominantly traces the degree of ionisation rather than metallicity, and without a strong, systematic anticorrelation between ionisation parameter and O/H across the sample, no monotonic metallicity trend emerges. For comparison, we include the calibration proposed by \citet{Sanders:26}, which in the $y$ = 12+log(O/H) versus $x$ = S32 plane corresponds to a straight line with negative slope crossing the clustered DESIRED data over a limited metallicity interval.

Finally, we explored the Ar3S2 index, defined as log(\ariii\ $\lambda$7136 / \sii\ $\lambda\lambda$6717,6731), motivated by the expectation that it might display a behaviour analogous to that observed in the O/H versus Ar3N2 plane. We constructed Ar3S2 using 1301 regions from the DESIRED calibration sample; a Pearson correlation test detects a formally significant correlation between Ar3S2 and metallicity ($r=-0.386$, $p=2.3\times10^{-47}$); however, as for Ne3O3, the index remains largely degenerate with metallicity across most of the sample's dynamic range. Over a wide range of Ar3S2 values, the metallicity spans 12+log(O/H) $\approx$ 7.0--8.5 with large dispersion, while the highest-metallicity regions show a mild bending toward lower Ar3S2 values ($\lesssim$ $-$1.2), suggestive of a weak residual trend, consistent with the modest overall correlation quantified above.

From the standpoint of ionisation structure, Ar3S2 compares Ar$^{2+}$ and S$^+$. The production of Ar$^{2+}$ requires photons with energies above 27.6 eV (Ar$^+\to$Ar$^{2+}$), tracing intermediate-to-high excitation zones, whereas S$^+$ is associated with much lower ionisation energies (S$^0\to$S$^+$ = 10.36 eV), characteristic of the outer, low-excitation regions of \hh regions. The ratio therefore strongly reflects the ionisation stratification of the nebula rather than elemental abundance variations. Moreover, Ar/S should not vary dramatically with O/H in standard chemical evolution scenarios. As a consequence, the Ar3S2 ratio is primarily sensitive to the ionisation parameter and to the detailed geometry of the ionised region, while its dependence on metallicity is indirect and weak. Changes in electron temperature with metallicity affect both collisionally excited lines, but not in a manner sufficient to generate a monotonic metallicity sequence across the sample. Overall, we conclude that Ar3S2 does not provide a reliable strong-line metallicity diagnostic, as its behaviour is dominated by the degree of ionisation rather than by systematic variations with O/H.

These examples illustrate the importance of a large, high-quality calibration sample such as DESIRED to robustly assess the validity of strong-line diagnostics across the full metallicity range, and highlight the risks associated with low-number statistics in deriving empirical metallicity relations. We stress that formal statistical significance is not equivalent to practical diagnostic power, which must instead be assessed against the dispersion of already-established calibrations.

\section{Discussion}
\label{sec:4}

\subsection{Comparison of calibrations with $T_{\rm e}$-based metallicities}
\label{sec:sigma}

\begin{figure*}
    \centering
	\includegraphics[width=\textwidth]{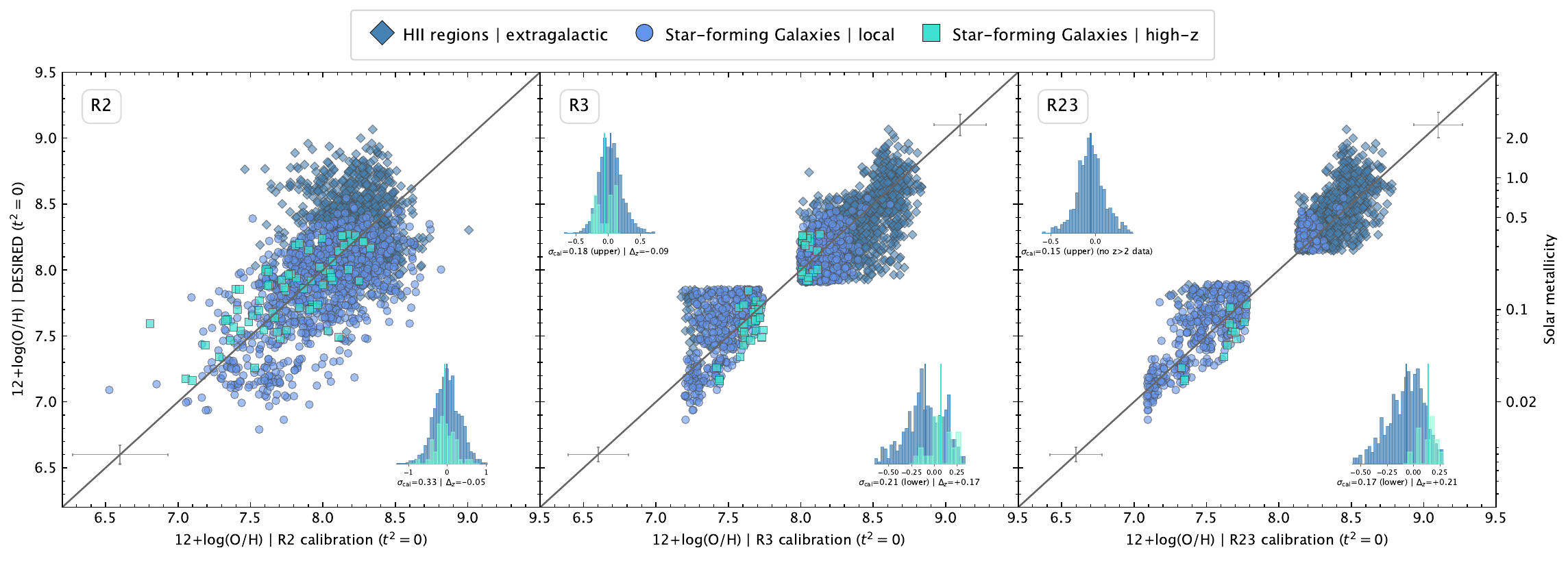}
	\includegraphics[width=\textwidth]{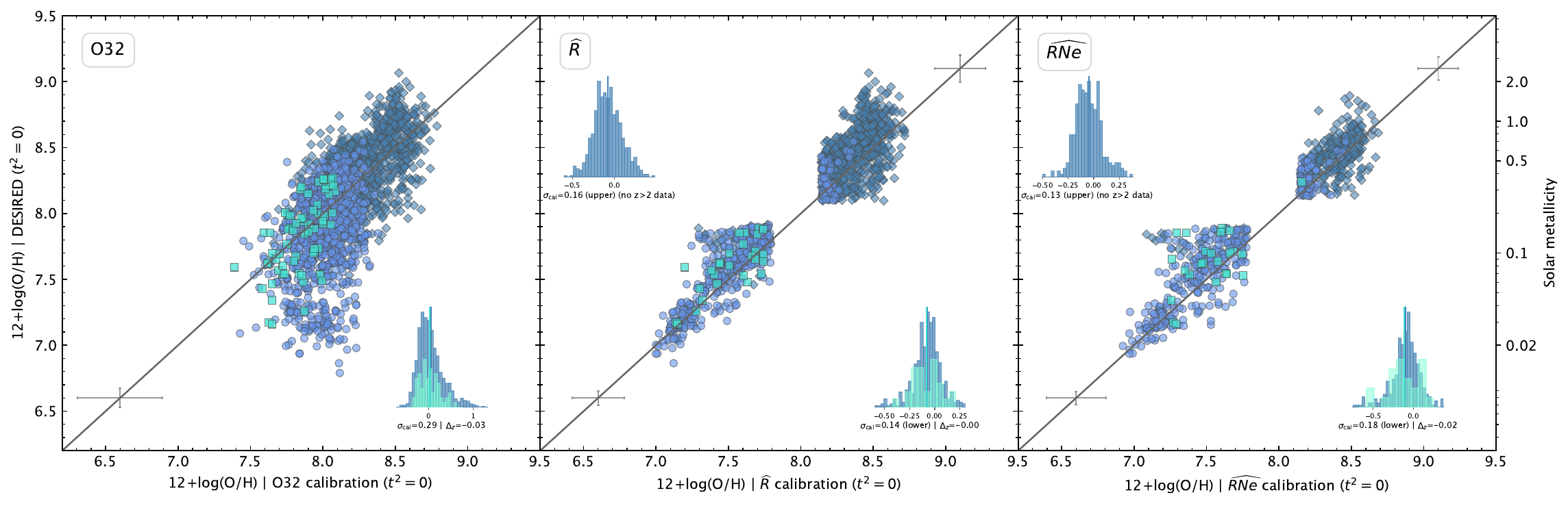}
	\includegraphics[width=\textwidth]{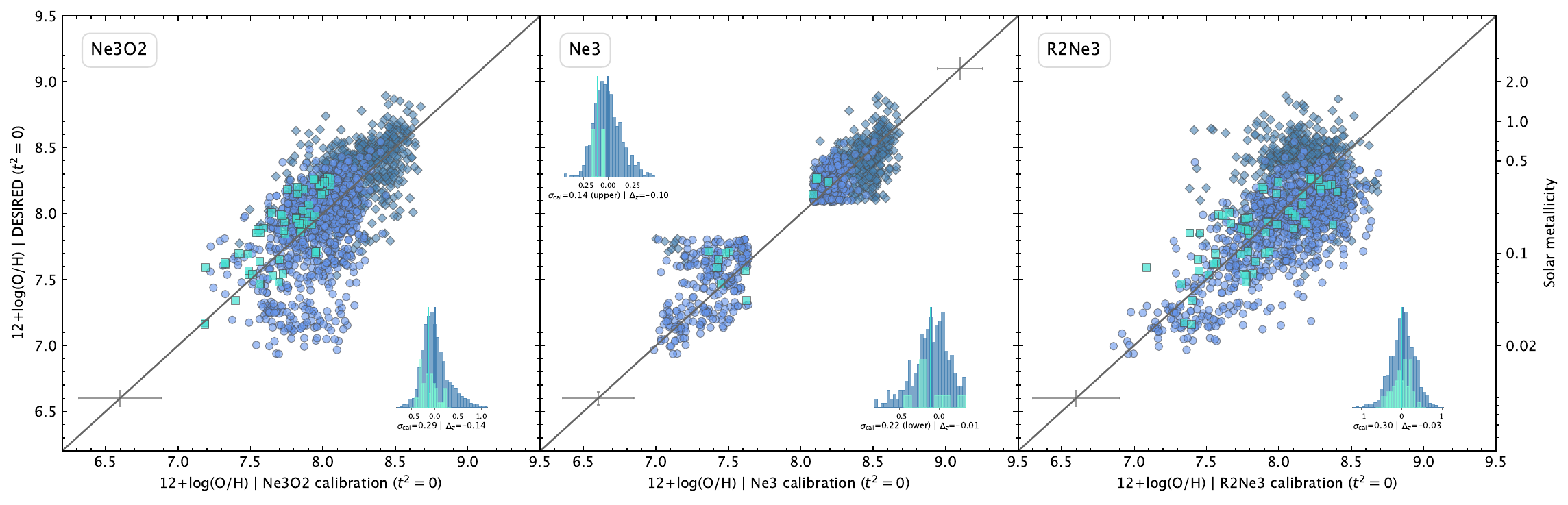}
    \caption{Comparison between the oxygen abundances derived from the calibrations shown in Fig.~\ref{fig:calib_1} and the direct ($T_{\rm e}$-based, $t^2=0$) metallicities for the DESIRED calibration sample, restricted to objects whose measured line ratio falls within the formally valid index range of the corresponding calibration and, for bivalued diagnostics, excluding the ambiguous region between the two branch turnovers as defined in Sec. \ref{sec:calibrations}. The solid line denotes the one-to-one relation. Error bars at the corners correspond to the mean 1$\sigma$ uncertainties for the $T_{e}$-based metallicities, and the root-mean-square scatter of the residuals for the best-fitting solution for each calibration. The inset panels show the distributions of residuals, defined as $\Delta$(O/H) = (O/H)$*{\rm calibration}$ $-$ (O/H)$*{T_{\rm e}}$, for the full calibration sample (blue) and, overlaid for comparison, for the $z>2$ subsample alone (turquoise). The vertical lines mark the mean of each distribution; the $z>2$ mean is omitted, and the panel labelled accordingly, when fewer than four $z>2$ objects satisfy these criteria. Here, $\sigma_{\rm cal}$ denotes the dispersion of the residual distribution for the full calibration sample. For the bivalued calibrations, separate histograms are shown for the upper and lower branches, with $\Delta_{z}$ (the difference between the $z>2$ and full-sample means) reported for each. Object types are indicated at the top of the figure.}
    \label{fig:sigma_1}
\end{figure*}

To evaluate the performance of the derived relations and to quantify their empirical dispersion (i.e., the intrinsic predictive scatter in 12+log(O/H)), we compare the oxygen abundances inferred from each calibration presented in Sec.~\ref{sec:3} with the corresponding $T_{\rm e}$-based metallicities of the calibration sample.


Fig.~\ref{fig:sigma_1} shows illustrative examples for the oxygen- and neon-based diagnostics presented in Fig.~\ref{fig:calib_1d}, including both monotonic (e.g. R2, O32, Ne3O2) and bivalued calibrations (e.g. R3, R23, \Rhat), restricted to objects whose measured line ratio falls within the formally valid index range of the corresponding calibration and, for bivalued diagnostics, excluding the ambiguous region between the two branch turnovers. In all cases, both the $T_{\rm e}$-based metallicity ($y$-axis) and the calibration-inferred metallicity ($x$-axis) are computed under the $t^2=0$ assumption. The inset panels display the distributions of residuals, defined as $\Delta$(O/H) = (O/H)$_{\rm calibration}$ $-$ (O/H)$_{T_{\rm e}}$, for the full calibration sample; overlaid for comparison, in the same panels, are the residuals of the $z>2$ subsample alone. A vertical line marks the mean of each distribution, omitted for the $z>2$ subsample when fewer than four objects satisfy these criteria. A quantitative assessment of these $z>2$ residuals is presented separately in Sect.~\ref{sec:highz}.

We define $\sigma_{\rm cal}$ as the standard deviation of these residuals, which provides a quantitative measure of the empirical predictive scatter of each calibration. A tight concentration of points around the one-to-one relation indicates that the calibration reproduces the direct abundances with small systematic offset and limited intrinsic dispersion. From these representative examples, which include some of the most widely used diagnostics in the literature (e.g. R3, R23) as well as recently introduced indices designed to minimise dependence on ionisation conditions (e.g. \Rhat, \RNe), it is possible to infer the strengths and limitations of different diagnostics as a function of metallicity range and of the physical sensitivities inherent to each index.

For instance, in the case of R2, the calibration departs significantly from the one-to-one relation at both ends of the metallicity range: at 12+log(O/H) $\gtrsim$ 8.5, a population of objects is systematically displaced above the one-to-one line, indicating that the calibration overestimates the direct metallicity, while at 12+log(O/H) $\lesssim$ 7.5 the opposite trend is apparent. This behaviour is consistent with the strong sensitivity of R2 to the ionisation parameter, which introduces a metallicity-dependent bias that is not fully absorbed by the calibration. Despite these systematic departures at the extremes, the residual distribution is approximately Gaussian with broad wings, yielding an overall $\sigma_{\rm cal} = 0.33$ dex. The high-redshift objects are distributed in the intermediate-to-low metallicity range and do not show a statistically significant offset from the full-sample distribution.

For the double-valued R3 and R23 calibrations, the comparison is performed separately for the upper and lower branches, selecting the appropriate branch according to the $T_{\rm e}$-based metallicity. Both calibrations follow the one-to-one relation reasonably well at the extremes of the metallicity range, but the scatter increases substantially near the turnover of each relation, where the degeneracy between branches is most severe. The residual distributions for the upper branches are close to Gaussian, with $\sigma_{\rm cal} = 0.18$ and $0.15$ dex for R3 and R23, respectively; for R23, however, no $z>2$ objects fall within the formally valid index range of the upper branch. In the lower branch, R3 shows a broader and more asymmetric distribution ($\sigma_{\rm cal} = 0.21$ dex), reflecting the stronger degeneracy at low metallicities, whereas R23 retains a more symmetric profile ($\sigma_{\rm cal} = 0.17$ dex). For both lower branches, the $z>2$ residual distribution is measurably offset towards higher values relative to the full sample; as discussed in Sec.~\ref{sec:highz}.

In other cases, the calibrations perform satisfactorily over a restricted abundance range but become unreliable elsewhere. This behaviour is exemplified by Ne3O2 and O32, which reproduce the direct metallicities reasonably well at 12+log(O/H) $\gtrsim 8.0$, but deviate markedly from the one-to-one relation at low metallicities, where the calibration-inferred abundances are systematically overestimated. This systematic trend is reflected in strongly skewed residual distributions with pronounced positive tails, yielding overall dispersions of 0.29 dex for both Ne3O2 and O32. The $z>2$ residuals show a negative offset of $0.14$,dex for Ne3O2, compared with a negligible offset of $-0.03$,dex for O32.

The Ne3 calibration presents a double-valued behaviour analogous to that of R3 and R23, and its performance is accordingly branch-dependent. The upper branch ($\sigma_{\rm cal} = 0.14$ dex) follows the one-to-one relation closely, with a residual distribution comparable to that of the upper branches of R3 and R23. The lower branch shows a broader but approximately symmetric distribution ($\sigma_{\rm cal} = 0.22$ dex), with the scatter increasing near the turnover as expected from the inherent degeneracy of the index. The number of $z>2$ objects within the formally valid range of either branch is too small (fewer than ten in each case) to support a statistically robust assessment of the applicability of this diagnostic at high redshift.


By contrast, \Rhat\ and \RNe\ provide the closest overall agreement with the direct abundances among all the diagnostics shown here, and the residual distributions are symmetric and well-behaved in both branches. The upper branches yield dispersions of 0.16 and 0.13 dex for \Rhat\ and \RNe, respectively; neither branch, however, retains any $z>2$ object within its formally valid index range, so high-redshift applicability cannot be assessed for these two branches with the current sample (see Sec.~\ref{sec:highz}). The lower branches (12+log(O/H) $\lesssim$ 8.0) are equally well-constrained, with dispersions of 0.14 and 0.17 dex for \Rhat\ and \RNe\ respectively, and show a negligible offset for the $z>2$ objects that do populate this regime, representing a tight branch-level agreement.

Finally, R2Ne3 follows the one-to-one relation reasonably well across most of the metallicity range covered by the calibration sample, with a residual distribution that is close to Gaussian and only a modest departure at the lowest metallicities. The overall dispersion, $\sigma_{\rm cal} = 0.30$ dex, is comparable to that of R2 and reflects the combined sensitivity of this index to both excitation and ionisation conditions. The $z>2$ objects are well distributed across the covered metallicity range and do not show a statistically significant offset from the full-sample distribution.

The calibration–$T_{\rm e}$ comparisons for the full set of relations are shown in Appendix~\ref{app:sigma} (Figs.~\ref{fig:sigma_2} to \ref{fig:sigma_t2_3}), for both the $t^2=0$ and  $t^2>0$ cases. The corresponding values of $\sigma_{\rm cal}$, i.e. the standard deviation of the metallicity residuals relative to the $T_{\rm e}$ abundances, are reported in Tables~\ref{tab:coeff_t2eq0} and \ref{tab:coeff_t2ge0}, for $t^2=0$ and  $t^2>0$, respectively.

\begin{figure*}
	\includegraphics[width=\textwidth]{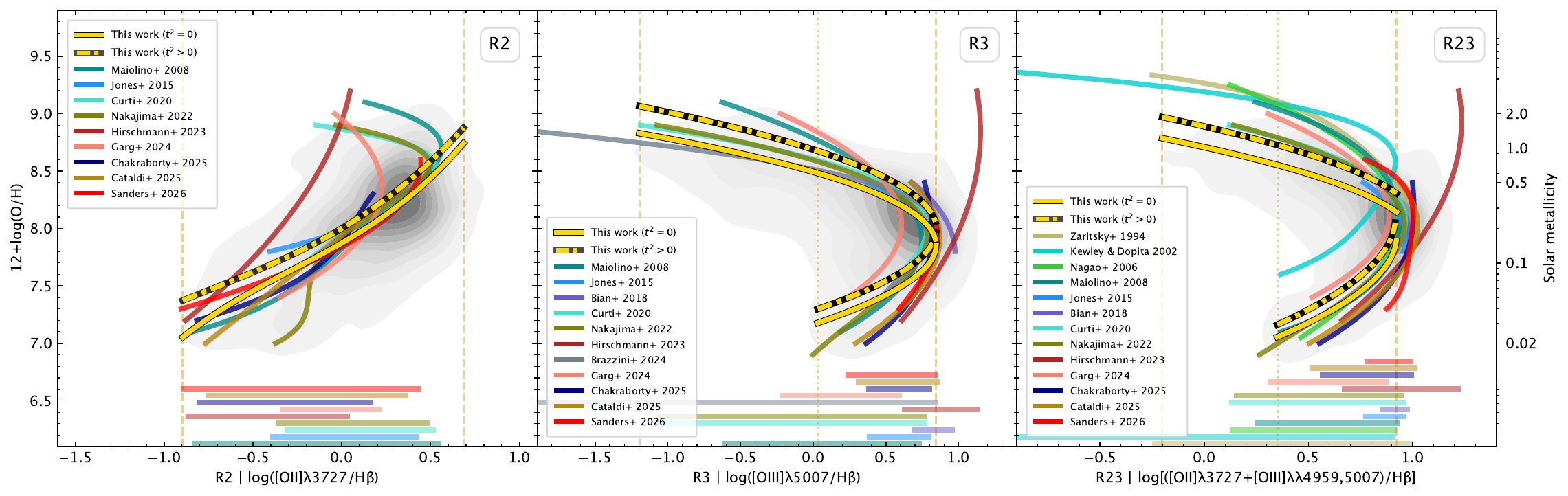}
	\includegraphics[width=\textwidth]{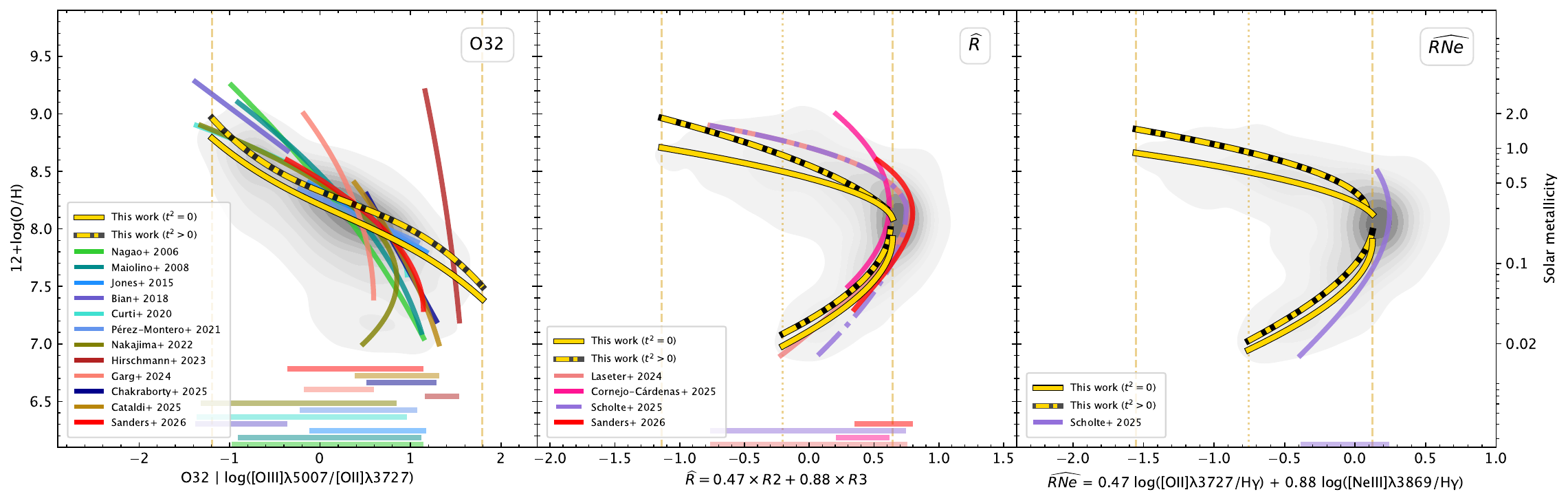}
	\includegraphics[width=\textwidth]{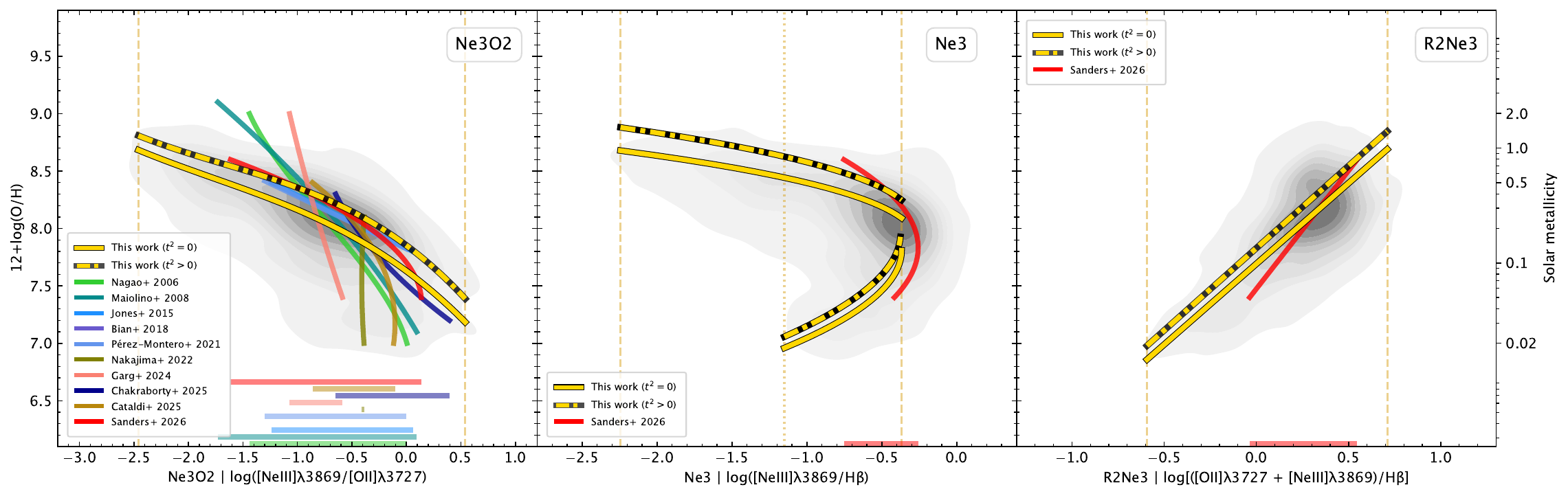}
      \caption{Comparison of the DESIRED metallicity calibrations shown in Fig.~\ref{fig:calib_1} for the oxygen- and neon-based indicators with previous determinations from the literature. Grey contours represent the density distribution of the calibration regions, with darker shades indicating higher source densities in the metallicity–index plane for the $t^2=0$ case. Ten contours are plotted at equally spaced probability levels. The yellow solid line and yellow–black dashed line correspond to the $t^2=0$ and $t^2>0$ calibrations derived in this work, respectively. Vertical dashed yellow lines mark the validity range of the DESIRED calibrations, while dotted yellow lines mark the minimum index value for the lower branch in the bivalued relations. Previous calibrations are shown as solid colours, with the corresponding reference in the legends. The horizontal coloured bars beneath each panel indicate the validity ranges of the corresponding literature calibrations.}
    \label{fig:compar_1}
\end{figure*}

\subsection{Comparison with the literature}
\label{sec:comparison}

The large size of the DESIRED calibration sample ––spanning the full metallicity range and the complete dynamic range of the diagnostic indices–– places us in a favourable position to carry out a meaningful and comprehensive comparison with strong-line calibrations previously reported in the literature. We have aimed to be as inclusive as possible, ensuring that the principal diagnostics are represented, whether by chronological precedence or by their established impact and widespread adoption. While some relations may not have been included, we are confident that the most influential calibrations in the field are covered.

\begin{figure*}
	\includegraphics[width=\textwidth]{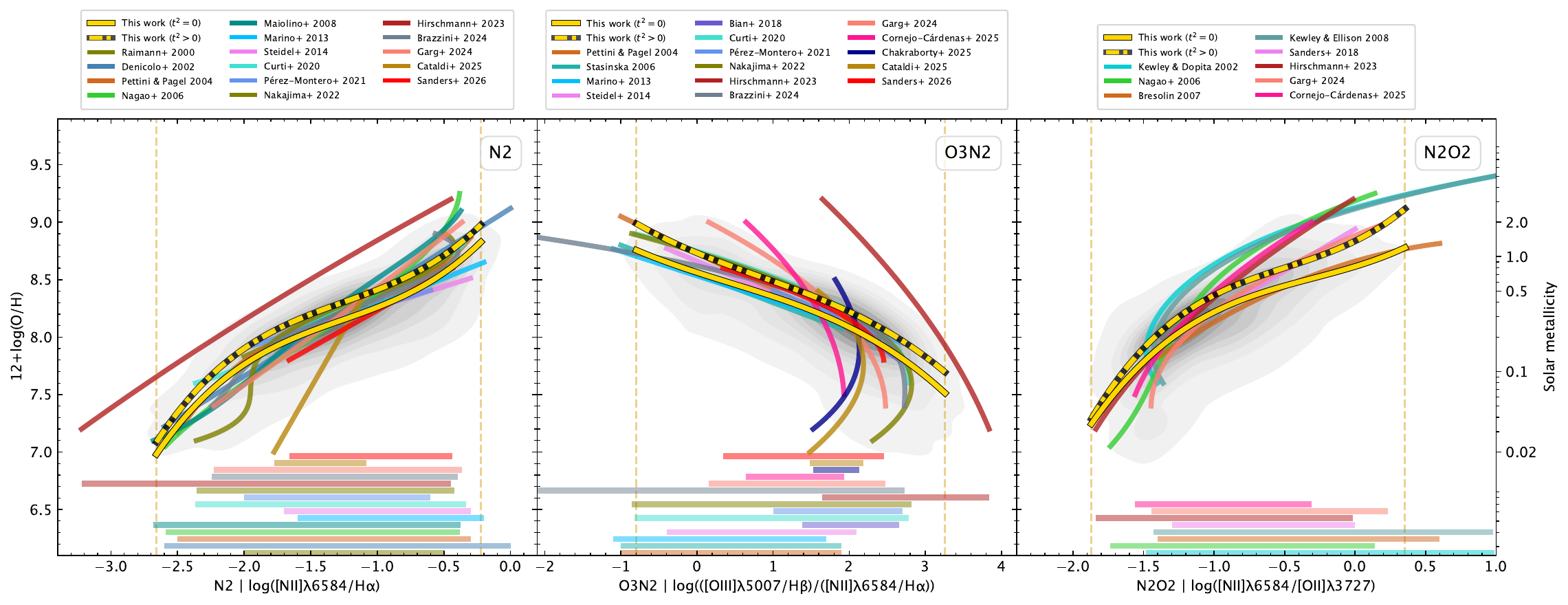}
	\includegraphics[width=\textwidth]{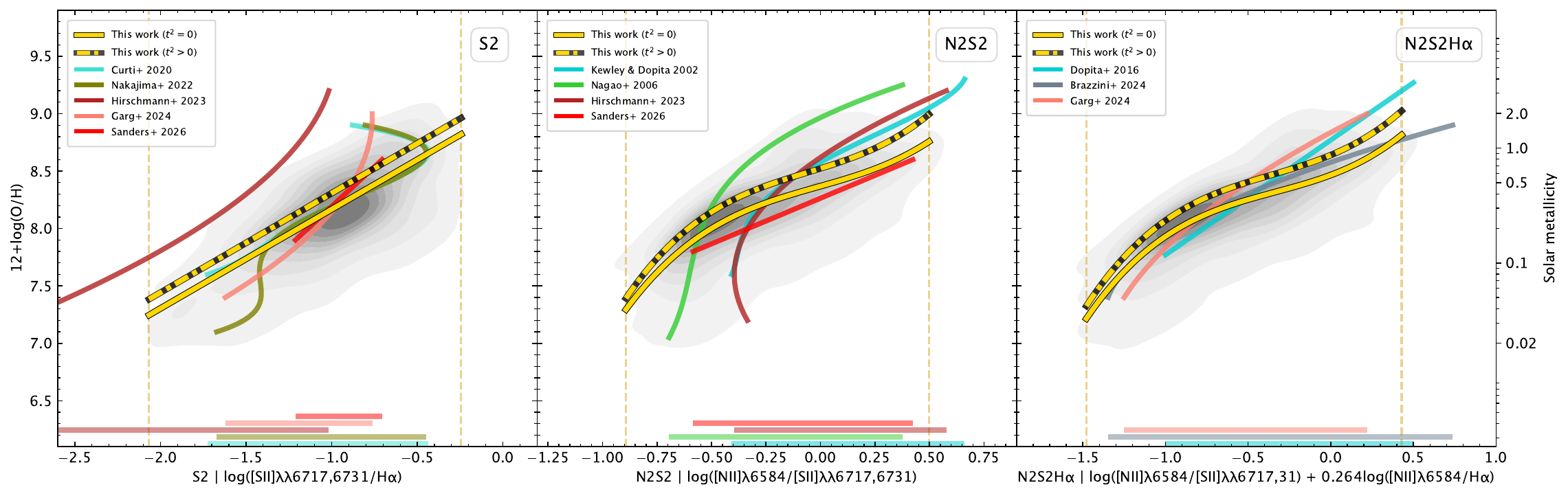}
	\includegraphics[width=\textwidth]{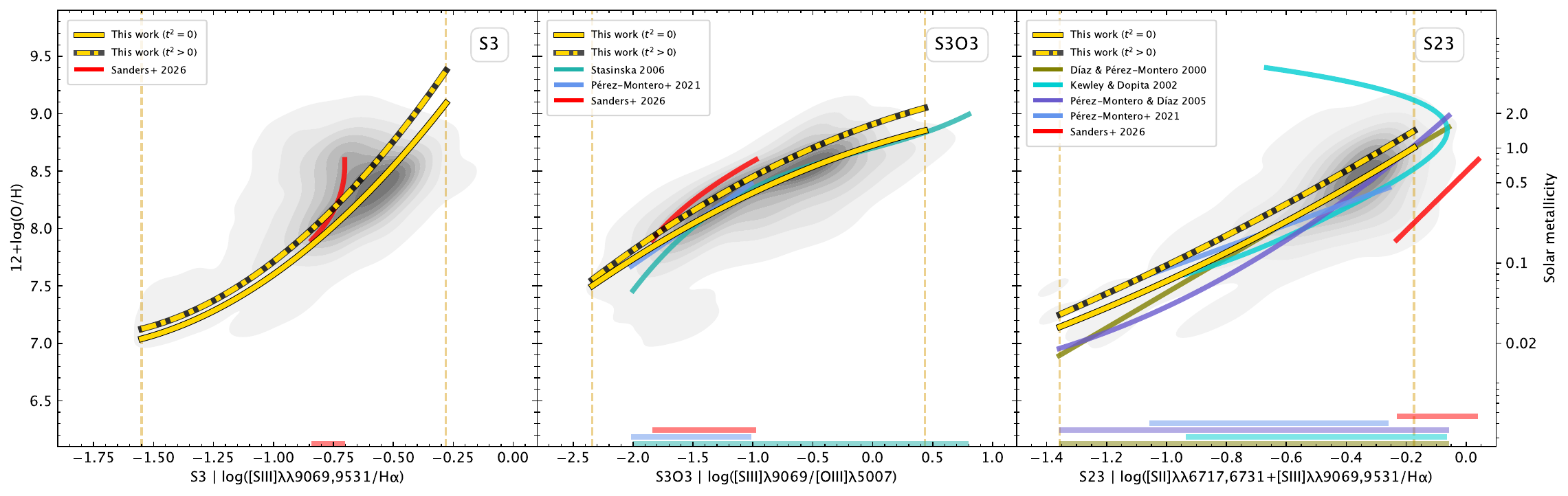}
    \caption{Comparison of the DESIRED metallicity calibrations shown in Fig.~\ref{fig:calib_2} for the nitrogen- and sulphur-based indicators with previous determinations from the literature. All definitions and plotting conventions as in Fig.~\ref{fig:compar_1}.}
    \label{fig:compar_2}
\end{figure*}

Figs.~\ref{fig:compar_1} to \ref{fig:compar_3} present this comparison. The grey contours represent the density distribution of the calibration regions, obtained through a kernel density estimation (KDE), and illustrate both the global morphology and the relative concentration of sources in the metallicity--index plane for the $t^2=0$ case. The yellow solid and yellow--black dashed curves represent the $t^2=0$ and $t^2>0$ calibrations derived in this work, respectively. Previous calibrations are shown as solid coloured curves, with the corresponding references indicated in the legends. The horizontal coloured bars beneath each panel denote the validity ranges reported for the literature calibrations. For a consistent and physically meaningful comparison throughout this section, literature calibrations should be contrasted with the DESIRED $t^2=0$ relations (yellow solid curves), as most published strong-line calibrations are anchored to the classical direct method under the assumption of a homogeneous temperature structure. The $t^2>0$ calibrations, however, provide additional insight, since the systematic offset in metallicity between the two sets of relations reflects the impact of temperature inhomogeneities across the parameter space.

\subsubsection{Oxygen-- and neon--based calibrations}

Fig.~\ref{fig:compar_1} shows the oxygen- and neon-based calibrations. In the case of R2, the literature exhibits a wide variety of proposed functional forms. In some cases, bivalued relations have been suggested \citep[e.g.][]{Maiolino:08, Curti:20, Nakajima:22, Garg:24}, possibly reflecting the limited size of earlier calibration samples, the inverse fitting approach adopted in several works (see Appendix~\ref{app:fits}), or the intrinsic nature of the R2 diagnostic itself, which displays a large dispersion (see Sec.~\ref{sec:sigma}).

In the case of R2, the literature exhibits a wide variety of proposed functional forms. In some cases, bivalued relations have been suggested \citep[e.g.][]{Maiolino:08, Curti:20, Nakajima:22, Garg:24}, consistent with the flattening of R2 towards solar and super-solar metallicities expected from photoionisation models. This may equally reflect the limited size and high-metallicity coverage of earlier calibration samples, or the intrinsic nature of the R2 diagnostic itself, which displays a large dispersion (see Sec.~\ref{sec:sigma}). The DESIRED calibration is consistent, within the intrinsic scatter and in the region of highest source density, with several previous determinations \citep[e.g.][]{Maiolino:08, Curti:20, Nakajima:22, Cataldi:25}, and agrees over a substantial R2 index range with \citet{Jones:15} and \citet{Sanders:26}. However, clear discrepancies arise at the high-metallicity end, where the relations from \citet{Maiolino:08}, \citet{Curti:20}, \citet{Nakajima:22}, and \citet{Garg:24} become bivalued, in contrast with the monotonic behaviour recovered by DESIRED. While the DESIRED sample does show a genuine saturation of the sensitivity of R2 towards these high metallicities, we do not find evidence for R2 becoming double-valued within our calibration sample, nor for a flattening sharp enough to constitute a turn analogous to that reported in these literature relations. At the low-metallicity end, the calibrations of \citet{Maiolino:08}, \citet{Nakajima:22}, \citet{Garg:24}, \citet{Chakraborty:25}, and \citet{Cataldi:25} predict systematically lower metallicities than those inferred from DESIRED. Note that the validity interval of the DESIRED calibration (vertical yellow dashed lines) spans a broader index range than that covered by any of the previous relations.

In the case of R3, all literature calibrations for this diagnostic adopt a bivalued functional form, although with markedly different validity intervals, as illustrated by the varying lengths of the horizontal coloured bars at the bottom of the panel. For the upper-metallicity branch, the DESIRED $t^2=0$ calibration shows agreement only with \citet{Brazzini:24}, who, however, define an index validity interval extending to values of R3 that are not sampled by the calibration data. All other relations systematically overestimate the metallicity in this regime, with those of \citet{Curti:20} and \citet{Nakajima:22} lying between the DESIRED $t^2=0$ and $t^2>0$ relations. The prescriptions by \citet{Maiolino:08} and \citet{Garg:24} exhibit substantial deviations from our calibration along this branch. In the lower-metallicity branch, the DESIRED $t^2=0$ calibration agrees with those of \citet{Maiolino:08}, \citet{Curti:20}, and \citet{Nakajima:22} over a very narrow range around R3 $\sim0.5$, whereas the remaining literature relations are shifted to lower R3 values (higher $\log u$), predicting systematically lower abundances, with the exception of \citet{Garg:24}, whose calibration overlaps with the $t^2>0$ relation at the lowest-metallicity end. A dedicated comparison with the two-dimensional $P$-method calibration of \citet{PilyuginThuan:05}, which parametrises R3 as a function of the excitation parameter $P$, is presented in Appendix~\ref{app:pt05}.

For the R23 diagnostic, notably none of the previous calibrations reproduce the DESIRED relation for the upper branch. All literature prescriptions systematically overestimate the metallicity in this regime, in several cases yielding abundances even higher than those implied by the DESIRED $t^2>0$ calibration. 
We attribute this discrepancy primarily to the fitting methodology commonly adopted in previous work. As discussed in Appendix~\ref{app:fits}, several of these calibrations are constructed using a single function fitted to medians binned in O/H. While this approach reproduces the morphology of the lower branch, the fit is poorly constrained in the sparser high-metallicity regime of the upper branch and consequently diverges. By contrast, fitting each branch of the bivalued diagnostics independently ensures that the precision of neither branch is compromised by the sampling density or morphology of the other.
The validity interval of the DESIRED upper branch is the broadest among the most recent R23 calibrations. Although the interval reported by \citet{Zaritsky:94} extends to lower index values than ours, their relation lies outside the locus of calibrating regions in the upper branch. A comparable behaviour is seen in the \citet{KewleyDopita:02} calibration, based on photoionisation models, which is overall shifted towards higher metallicities and defines a validity range for the upper branch that extends beyond the domain populated by observed regions. For the lower branch, the DESIRED calibration is in good agreement with several previous empirical relations, sampling comparably validity ranges (the yellow dotted lines mark the minimum index threshold adopted for the lower branch in bivalued diagnostics). Notable exceptions are the recent calibrations derived from high-redshift samples by \citet{Cataldi:25}, \citet{Chakraborty:25}, and \citet{Sanders:26}, which were established over comparatively narrow validity intervals. As for R3, the comparison with the \citet{PilyuginThuan:05} $P$-method calibration for R23 is discussed in Appendix~\ref{app:pt05}, where we show that the DESIRED calibration sample covers nearly the full parameter space originally defined by excitation parameter $P$, and that the offset between the solutions can be interpreted as a systematic shift in effective excitation within the $P$-method framework.

For O32 and Ne3O2, previous calibrations exhibit markedly different functional forms, both in shape and slope, likely reflecting limitations in the size and parameter-space coverage of their respective calibration samples. For both diagnostics, all literature prescriptions systematically overestimate the metallicity with respect to the DESIRED calibration at the high-metallicity end. In the case of O32, none of the previous calibrations reproduce the DESIRED $t^2=0$ relation in this regime. The relation of \citet{Jones:15} overlaps with DESIRED only within its own low-metallicity range, crossing and reaching the DESIRED $t^2>0$ calibration towards the edge of its validity interval, while \citet{PerezMontero:21} is the only literature relation that closely follows the DESIRED $t^2>0$ calibration across essentially its full validity range. A similar behaviour is observed for Ne3O2, where the relations of \citet{Jones:15}, \citet{PerezMontero:21}, and \citet{Sanders:26} follow the DESIRED $t^2>0$ calibration for 12+log(O/H) $\gtrsim$ 8.0, suggesting that their functional forms effectively trace a metallicity scale that is offset relative to the classical homogeneous-temperature assumption.

The relations presented by \citet{Hirschmann:23}, derived from emission-line catalogues of simulated galaxies at $z = 0$--8 from the IllustrisTNG simulation, departs significantly from all previous calibrations and the locus of the empirical data for the R2, R3, R23 and O32 relations, especially at the high-metallicity end. For O32 and Ne3O2, some recent calibrations sample the regime where the index shows large intrinsic dispersion and becomes weakly sensitive to metallicity (12+log(O/H) < 8.0), leading to partially degenerate functional forms (e.g. \citealt{Cataldi:25, Chakraborty:25}). Notably, the DESIRED calibrations span the widest validity interval among all O32 and Ne3O2 prescriptions considered here.

The \Rhat\ and \RNe\ diagnostics have been introduced only recently, so there are relatively few literature calibrations available for comparison. In the case of \Rhat, similarly to previously proposed bivalued relations, the upper branch of the DESIRED calibration lies systematically below the literature prescriptions. The relations of \citet{Laseter:24} and \citet{Scholte:25} share the same functional form along the upper branch (for clarity, the \citet{Scholte:25} relation is shown as a dash-dotted line), and both lie above to the DESIRED $t^2>0$ calibration. For the lower branch, the DESIRED calibration shows very good agreement with \citet{Laseter:24}, while it departs from the relation of \citet{Scholte:25}, which systematically underestimate the metallicity in this regime. The calibration by \citet{Cornejo:25}, derived from synthetic spectra of simulated galaxies using the PRISMA model, lies above the DESIRED $t^2>0$ calibration along the lower-metallicity branch, but deviates from all previous prescriptions at the high-metallicity end. The relation by \citet{Sanders:26} agrees with the DESIRED $t^2=0$ calibration over the low-metallicity range covered by its validity interval, but departs from it at metallicities above the lower-branch turnover.
For \RNe, we provide calibrations for both the upper and lower metallicity branches. The only available comparison in the literature is the relation by \citet{Scholte:25}, which predominantly samples the lower branch and does not reproduce the DESIRED \RNe\ calibration. In both diagnostics, the DESIRED relations cover by far the widest validity interval for both the upper and lower branches.

For the recently proposed Ne3 and R2Ne3 diagnostics, \citet{Sanders:26} provide the only available point of comparison in the literature. The Ne3 calibration by \citet{Sanders:26} was derived from 111 high-redshift galaxies in the AURORA sample, whereas the DESIRED Ne3 calibration is constructed from 1680 spectra in the DESIRED compilation. The \citet{Sanders:26} relation primarily samples the high-density locus at the transition between the upper and lower metallicity branches, and is defined over a narrow index interval. In contrast, the substantially larger and statistically more complete DESIRED sample enables the derivation of calibrations that robustly cover both the upper and lower metallicity branches across a significantly broader index validity range. The \citet{Sanders:26} calibration shows only limited agreement with the DESIRED $t^2>0$ relation just above the upper-branch turnover, while on the lower branch it lies at lower index values, with no overlap with either DESIRED calibration.
Similarly, the DESIRED R2Ne3 calibration spans a substantially wider range of the monotonic line ratio. It shows good agreement, within the intrinsic dispersion, with the \citet{Sanders:26} relation in the region of highest source density, although with a slightly different slope.

\subsubsection{Nitrogen-- and sulphur--based calibrations}

The comparisons for the nitrogen- and sulphur-based calibrations are presented in Fig.~\ref{fig:compar_2}. The N2 and O3N2 diagnostics are among the most widely used strong-line metallicity indicators, and consequently numerous empirical calibrations are available in the literature for these ratios. In both cases, most literature prescriptions lie close to or between the DESIRED $t^2=0$ and $t^2>0$ calibrations within the region of highest source density and across their respective validity intervals. This includes the seminal relations proposed by \citet{Raimann:00}, \citet{Denicolo:02}, and \citet{PettiniPagel:04}, as well as subsequent calibrations by \citet{Steidel:14} and \citet{Bian:18}.

For N2, noticeable discrepancies arise at the high-metallicity end, for 12+log(O/H) $\gtrsim$ 8.5, where the relations by \citet{Denicolo:02}, \citet{Nagao:06}, \citet{Maiolino:08}, and \citet{Garg:24} overestimate the oxygen abundance by up to 0.5~dex relative to DESIRED. The calibrations by \citet{Marino:13}, \citet{Curti:20}, \citet{Brazzini:24}, and \citet{Sanders:26} show very good agreement with the DESIRED relations for both N2 and O3N2, within the quoted uncertainties, although all of them sample a narrower index range than that covered by DESIRED.

Exceptions include the recent relations by \citet{Cataldi:25} for both N2 and O3N2, as well as the \citet{Garg:24}, \citet{Cornejo:25}, and \citet{Chakraborty:25} calibrations for O3N2, which exhibit markedly divergent trends over relatively restricted index intervals, particularly in the low-metallicity, high-dispersion regime of O3N2. The \citet{Hirschmann:23} relations for N2, O3N2, and S2 show the same critical deviations discussed for the oxygen-based diagnostics, lying in regions characterised by unphysical line ratios and outside the locus populated by observed calibration regions.

For the N2O2 diagnostic, literature prescriptions display significant differences in both curvature and slope, especially towards the high-metallicity regime. The relations by \citet{KewleyDopita:02}, \citet{Nagao:06}, \citet{KewleyEllison:08}, \citet{Hirschmann:23}, and \citet{Cornejo:25} are clearly offset towards higher metallicities with respect to both the DESIRED $t^2=0$ and $t^2>0$ calibrations. In contrast, the empirical relation by \citet{Bresolin:07a} closely follows the $t^2=0$ DESIRED locus over the metallicity range 12+log(O/H) > 8.0, while \citet{Sanders:18} and \citet{Garg:24} follow the $t^2>0$ calibration at higher metallicities.

In the case of the S2 diagnostic, the relations by \citet{Curti:20} and \citet{Nakajima:22} lie very close to the DESIRED $t^2=0$ calibration within the regime most densely populated by the calibration regions (S2 $\in$ [$-$1.4,$-$0.5]), although systematic offsets appear toward both extremes of the metallicity range. For \citet{Nakajima:22}, these deviations arise from an unusual functional form at low metallicities and from the adoption of a bivalued parametrisation at high metallicities. The \citet{Garg:24} and \citet{Sanders:26} calibrations also follow the general behaviour of the data within the high-density region, but are defined over a much narrower index interval and display a noticeably steeper slope, crossing both the DESIRED $t^2=0$ and $t^2>0$ relations.

For N2S2, the calibrations by \citet{KewleyDopita:02}, \citet{Nagao:06}, and \citet{Hirschmann:23} show substantial differences in functional form with respect to DESIRED. The \citet{Nagao:06} relation departs significantly from the empirical locus, while the \citet{KewleyDopita:02} and \citet{Hirschmann:23} prescriptions follow the locus traced by DESIRED but are more consistent with the $t^2>0$ calibration at metallicities 12+log(O/H) $\gtrsim$ 8.5. Notably, the \citet{Sanders:26} relation provides the closest match, being nearly parallel to the DESIRED $t^2=0$ calibration, although defined over a narrower validity interval and offset toward lower metallicities by $\sim$0.1 dex.

\begin{figure*}
    \includegraphics[width=\textwidth]{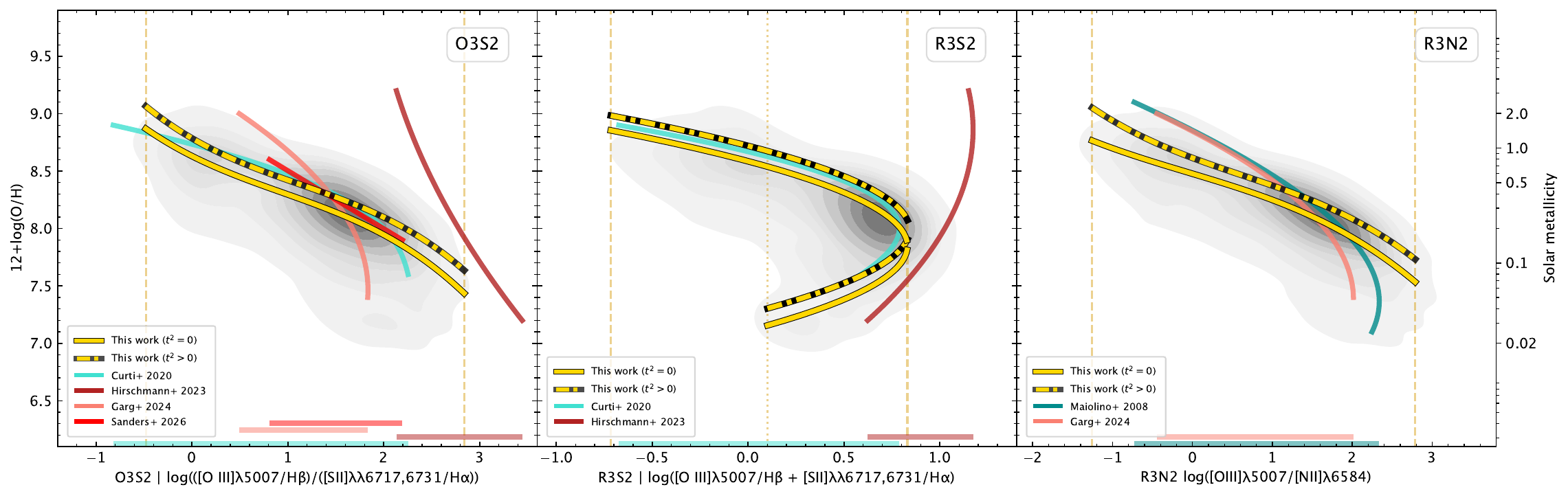}
	\includegraphics[width=0.69\textwidth]{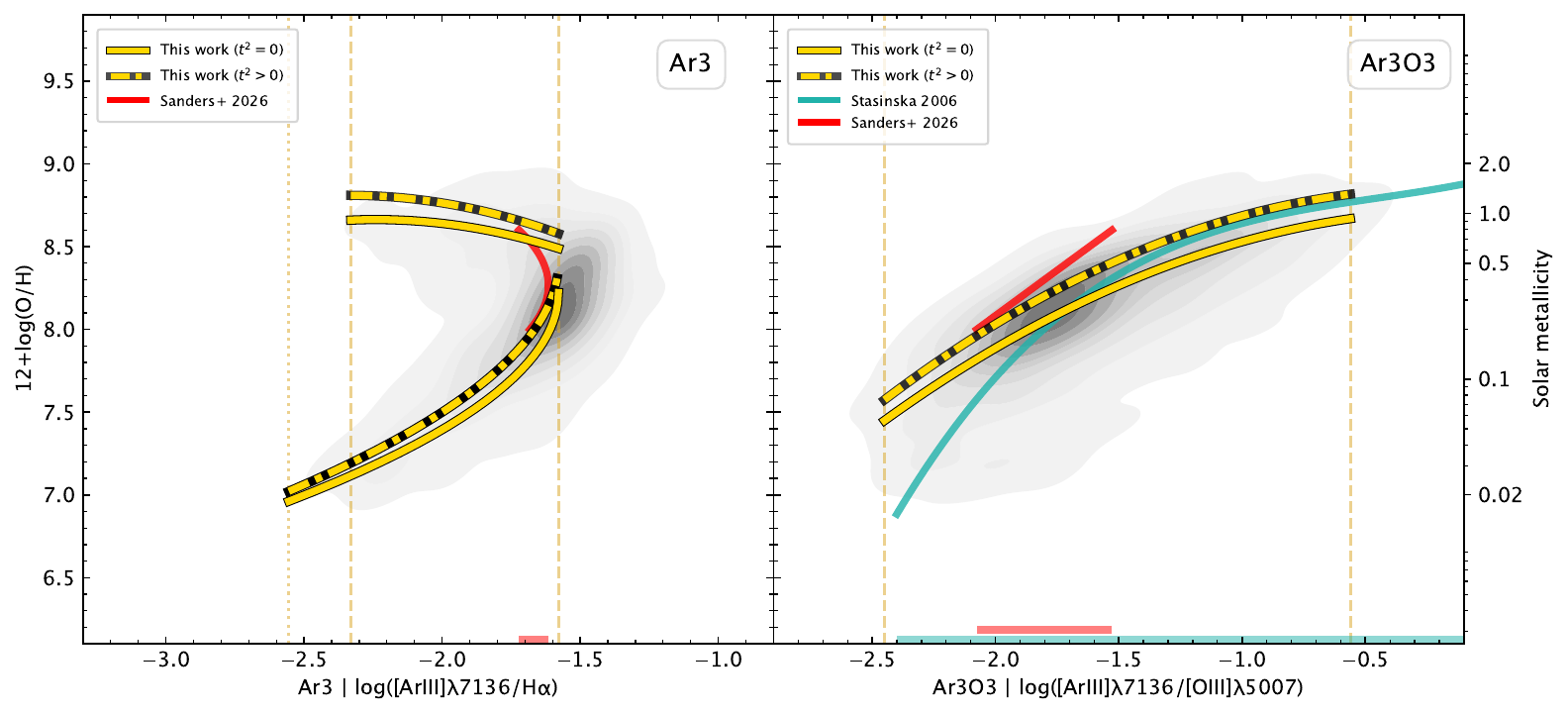}
    \caption{Comparison of the DESIRED metallicity calibrations shown in Figs.~\ref{fig:calib_1d} and \ref{fig:calib_3} for the sulphur- and argon-based indicators with previous determinations from the literature. All definitions and plotting conventions as in Fig.~\ref{fig:compar_1}.}
    \label{fig:compar_3}
\end{figure*}

For the N2S2H$\alpha$ diagnostic, the empirical calibration by \citet{Brazzini:24} shows the closes agreement with the DESIRED relation across a substantial fraction of the index range, lying between the $t^2=0$ and $t^2>0$ calibrations at the high-metallicity end, although extending toward index values not sampled by the calibration regions. The \citet{Dopita:16} and \citet{Garg:24} relations broadly follow the overall trend of the data and lie near the DESIRED locus, but exhibit a steeper behaviour and depart from the DESIRED relations, particularly toward the high-metallicity end. For R3N2 (Fig.~\ref{fig:compar_3}), the relations by \citet{Maiolino:08} and \citet{Garg:24} overestimate the metallicity relative to both the DESIRED $t^2=0$ and $t^2>0$ relations at 12+log(O/H) $\gtrsim8.2$, while underestimating it at 12+log(O/H) $\lesssim8.2$, where the index becomes less sensitive to metallicity.

As in the case of Ne3, for the S3 diagnostic the only literature calibration available for comparison is that of \citet{Sanders:26}, which was derived from a high-redshift sample and therefore spans a very narrow index interval concentrated around the highest-density locus of the DESIRED distribution. Within the overlapping region, the \citet{Sanders:26} relation lies between the DESIRED $t^2=0$ and $t^2>0$ calibrations, although it rapidly departs and becomes degenerate at S3 $\approx -0.75$. In contrast, the DESIRED sample enables a calibration of the S3 relation over a substantially broader validity interval, covering the full dynamic range of the index sampled by the calibration regions, albeit with lower statistics toward the low-metallicity end.

For S3O3, the relations by \citet{Stasinska:06a} closely follows the DESIRED calibration within its quoted validity range. 
The prescription by \citet{PerezMontero:21}, constructed from a sample of extreme emission-line galaxies (EELGs) at $z < 0.49$ selected from the Sloan Digital Sky Survey (SDSS), also tracks the DESIRED $t^2=0$ calibration well in the region of highest source density, but is defined over a more restricted index interval than DESIRED. The \citet{Sanders:26} relation is defined over an even narrower index interval, and lies above both the DESIRED $t^2=0$ and $t^2>0$ calibrations throughout this range.

The S23 diagnostic has been calibrated by early empirical prescriptions \citep[e.g.][]{DiazPM:00, PerezMonteroDiaz:05}, model-based relations \citep[e.g.][]{KewleyDopita:02}, and more recent empirical recalibrations \citep[e.g.][]{PerezMontero:21, Sanders:26}. Although the qualitative trend is broadly similar among the different prescriptions, they differ markedly in curvature and absolute abundance scale. Most relations lie below the DESIRED calibration over a large fraction of the validity range, approaching and crossing the DESIRED locus in the region most densely populated by the calibration sample. 
The \citet{Sanders:26} relation, however, exhibits a steeper positive slope, lies well outside the locus of the DESIRED calibration sample, and is defined over a comparatively narrower index interval.

The upper panels of Fig.~\ref{fig:compar_3} show the comparisons for the O3S2, R3S2 and R3N2 calibrations. For the O3S2 diagnostic, the relation by \citet{Curti:20} overpredicts the metallicity by $\sim$0.2 dex relative to the DESIRED $t^2=0$ calibration over a substantial fraction of the index range, effectively tracing instead the DESIRED $t^2>0$ relation. The \citet{Sanders:26} calibration lies close to the region of highest source density and broadly reproduces the trend of the DESIRED relation, although it also overestimates the metallicity by $\sim$0.2 dex and exhibits a steeper slope. The \citet{Garg:24} prescription shows a noticeably different slope compared to the previous references, leading to increasing deviations across the index range.
For the bivalued R3S2 diagnostic, the upper branch of the \citet{Curti:20} calibration closely follows the functional form of both the DESIRED $t^2=0$ and $t^2>0$ calibrations, lying between them over the full validity interval of the upper branch. Below the turnover, however, the low-metallicity branch of \citet{Curti:20} follows the DESIRED $t^2>0$ relation only over a very narrow validity interval. For O3S2 and both branches of R3S2, the \citet{Hirschmann:23} prescriptions deviate significantly from the empirical locus, extending into index--metallicity combinations that are not populated by the observed calibration regions.

Overall, the DESIRED calibrations for the nitrogen- and sulphur-based diagnostics span broader index validity intervals than the majority of previous prescriptions, providing improved coverage while remaining anchored to the empirical distribution of the calibration sample.

\subsubsection{Argon--based calibrations}

The lower panels of Fig.~\ref{fig:compar_3} present the comparisons for the argon-based diagnostics. For Ar3, the \citet{Sanders:26} calibration is defined over an extremely narrow index interval, concentrated around the highest-density locus near the turnover between the two metallicity branches, where the diagnostic has minimal sensitivity to abundance variations. In contrast, the DESIRED calibration traces the full double-valued structure of the Ar3 relation, covering a substantially broader index range, particularly along the lower branch.

For Ar3O3, the prescription by \citet{Stasinska:06a} deviates from the DESIRED calibrations at metallicities 12+log(O/H) $\lesssim$ 8.0, but converges towards the DESIRED $t^2>0$ calibration at the high-metallicity end (12+log(O/H) $\gtrsim$ 8.2), where it extends to Ar3O3 values beyond the range sampled by our calibrating spectra. Meanwhile, the \citet{Sanders:26} relation is found closely to the high-density locus, but predicts systematically higher abundances than both the DESIRED $t^2=0$ and $t^2>0$ calibrations, and is defined over a comparatively restricted index interval.

Overall, the comparisons presented in this section show that the DESIRED calibrations provide a consistent and statistically robust reference across essentially all commonly used strong-line diagnostics, while spanning the broadest validity intervals in index space and metallicity among the relations considered. Where literature prescriptions disagree, the discrepancies can typically be traced to limited calibration samples, restricted parameter-space coverage, or the adoption of functional forms that extrapolate beyond the locus populated by the empirical data. In several cases, published relations converge toward the DESIRED $t^2>0$ solutions at the high-metallicity end, highlighting the potential impact of temperature inhomogeneities on the absolute abundance scale. In the next section, we build on these results to discuss the applicability of the DESIRED calibrations at high redshift.

Overall, the comparisons presented in this section show that the DESIRED calibrations provide a consistent and statistically robust reference across essentially all commonly used strong-line diagnostics, while spanning the broadest validity intervals in index space and metallicity among the relations considered. Where literature prescriptions disagree, the discrepancies can typically be traced to limited calibration samples, restricted parameter-space coverage, or the adoption of functional forms that extrapolate beyond the locus populated by the empirical data. In several cases, published relations converge toward the DESIRED $t^2>0$ solutions at the high-metallicity end, highlighting the potential impact of temperature inhomogeneities on the absolute abundance scale. In other cases, however, such discrepancies may instead reflect genuine physical differences in the average ISM conditions of the calibrating samples rather than methodological limitations. This is particularly relevant when comparing DESIRED with calibrations based on qualitatively distinct populations, such as high-redshift JWST samples or local extreme emission-line galaxies (EELGs), whose average ionisation conditions and abundance properties may differ systematically from those of the broader DESIRED compilation.  In the next section, we build on these results to discuss the applicability of the DESIRED calibrations at high redshift.

\subsection{Applicability of DESIRED calibrations at high redshift}
\label{sec:highz}


The question of whether strong-line metallicity calibrations derived from local-Universe samples can be reliably applied to high-redshift galaxies has become one of the central topics in extragalactic chemical abundance studies. It is now well established that the ionised ISM at $z > 2$ exhibits systematically different conditions compared to typical $z \sim 0$ SFGs, including higher electron densities, harder ionising spectra driven by $\alpha$-enhanced massive stars, and potentially enhanced N/O abundance ratios at fixed O/H \citep{Steidel:14, Shapley:15, Sanders:16a, Strom:17, Strom:18, Topping:20b, Cullen:21, Stanton:24, Stanton:25}. These differences have led several authors to argue that calibrations anchored to representative $z \sim 0$ samples will not reliably yield accurate metallicities when applied to high-redshift galaxies \citep{Steidel:14, Sanders:24, Sanders:26, Cataldi:25}. In particular, \citet{Sanders:26} conclude that calibrations based on typical $z \sim 0$ \hh regions or SFGs would return systematically biased metallicities when applied to high-redshift samples, with offsets most often in excess of 0.1~dex in O/H at fixed line ratio. As shown below, a direct statistical test of this claim using the DESIRED sample supports it only for a subset of diagnostics, rather than for the sample as a whole.

Before the advent of JWST, the first direct test of strong-line calibration validity at cosmological distances was carried out by \citet{Jones:15}, who used $T_{\rm e}$-based metallicities for a sample of galaxies at $z \simeq 0.8$ drawn from the DEEP2 survey. Their analysis demonstrated that the relation between metallicity and strong emission lines of oxygen, hydrogen, and neon does not evolve between $z = 0$ and $z \simeq 0.8$, to within a measurement precision of 0.03~dex. Crucially, they found that galaxies at all redshifts populate a constant locus of $\alpha$-element and hydrogen line ratios, with negligible offset (0.01~dex in the \oiii/H$\beta$ versus \neiii/\oii\ diagnostic diagram) and very low scatter ($\sim$0.04~dex). \citet{Jones:15} concluded that evolution in the \nii-BPT diagram at high redshift is driven by the \nii/H$\alpha$ ratio--and therefore by systematically higher N/O ratios--rather than by changes in the $\alpha$-element emission-line properties at fixed O/H.


This conclusion has been further supported by JWST-based studies that probe the redshift evolution of calibrations among high-redshift galaxies themselves. \citet{Chakraborty:25} divided a sample of 67 galaxies at $3 < z < 10$ into two redshift bins ($3 < z < 5$ and $5 < z < 10$) and found that the R23 calibration shows metallicity differences of less than 0.02~dex between the two bins, indicating negligible internal evolution. This result is consistent with cosmological simulations \citep{Hirschmann:23, Garg:24} and with the analysis of \citet{Sanders:24}, who examined the residuals around the best-fit R23 calibrations as a function of redshift and found no significant trends up to $z \sim 8$. In the same vein, \citet{Sanders:26} find that none of the O-, Ne-, and N-based line ratios show a statistically significant ($>3\sigma$) offset from zero in the binned medians out to $z \approx 6$, and conclude that they do not find evidence for strong evolution of metallicity calibrations between $z \sim 2$ and $z \sim 10$, noting that if any evolution exists between $z \sim 0$ and $z \sim 2$, it may plausibly slow down or halt at $z \gtrsim 2$ as galaxies approach the maximum $\alpha$/Fe ratio attainable from pure core-collapse supernovae enrichment. 
We independently test this within the DESIRED $z>2$ subsample itself. For each calibration/branch with a sufficient number of $z>2$ objects within its formally valid range, we compute the Spearman rank correlation between the calibration residuals and redshift. Of the 26 tests, 5 yield a nominal $p<0.05$, consistent with the $\sim$1--2 false positives expected by chance alone at this significance level. However, none remains statistically significant after applying the Benjamini--Hochberg correction for multiple testing \citep{Benjamini:95} at $\alpha=0.05$. We therefore find no statistically robust evidence for a residual trend with redshift in any DESIRED calibration within our $z>2$ subsample, consistent with the negligible internal evolution reported previously.

In this context, the DESIRED calibration sample offers a fundamentally new perspective on this debate. Previous $z \sim 0$ calibration samples--such as those employed by \citet{Maiolino:08}, \citet{Curti:20}, or \citet{Sanders:21}--are based either on representative populations of local galaxies or on stacked spectra and were built with the goal of being ``typical'' of the $z \sim 0$ star-forming population. However, they were not designed to encompass the full diversity of ionisation conditions present across cosmic time. As a result, they may not include a sufficient population of compact, high-excitation, high-ionisation-parameter systems analogous to those commonly observed at high redshift. The DESIRED calibration sample overcomes this limitation: by assembling 2392 high-quality spectra from a heterogeneous compilation of \hh regions and SFGs--including individual \hh regions, nearby dwarfs, blue compact dwarfs, extreme emission-line galaxies, and high-redshift galaxies spanning from $z \sim 0$ up to $z \sim 10$--the parameter space coverage is vastly expanded relative to any single previous calibration effort.

\begin{figure}
    \centering
	\includegraphics[width=\columnwidth]{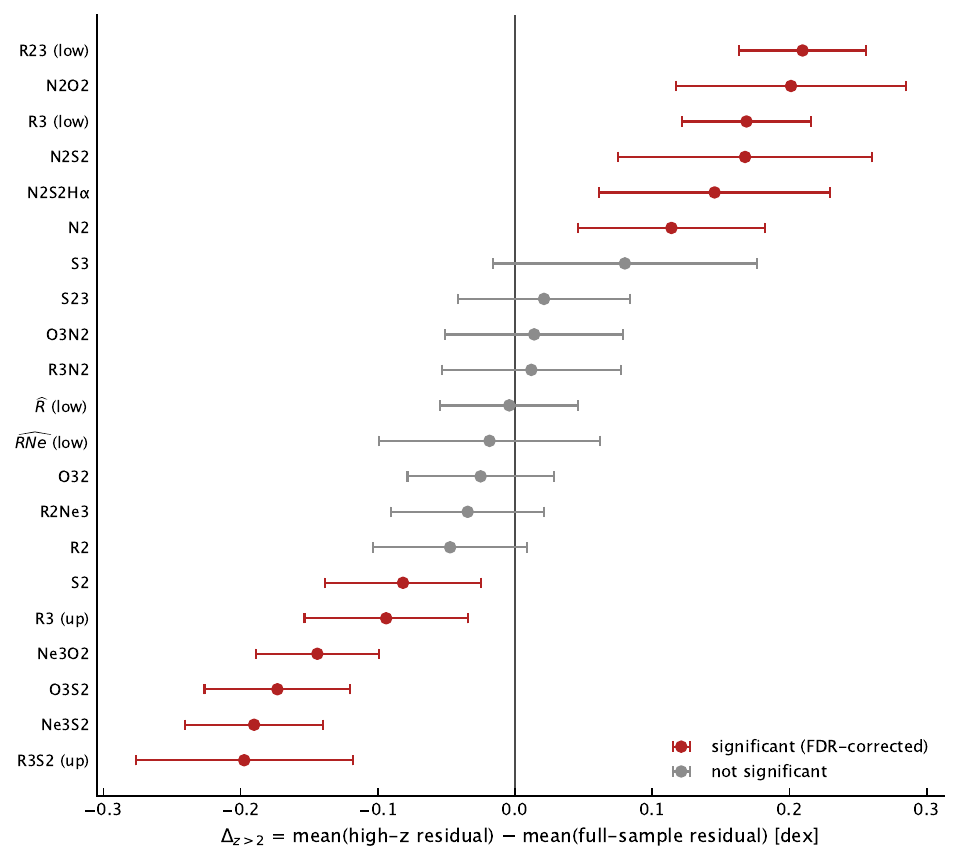}
     \caption{High-redshift ($z>2$) versus full-sample residual offset, $\Delta_{z>2} = \langle\Delta\rangle_{z>2} - \langle\Delta\rangle_{\rm full}$, where $\Delta = ({\rm O/H})_{\rm calibration} - ({\rm O/H})_{T_{\rm e}}$, for each DESIRED calibration/branch ($t^2=0$) with at least five $z>2$ objects within its formally valid line-ratio range. Error bars indicate the approximate 95 per cent confidence interval, given by $\pm1.96\times$ the standard error of the mean, for $\Delta_{z>2}$. Points are colour-coded according to whether the offset remains statistically significant after applying the Benjamini--Hochberg correction for multiple testing, which controls the false discovery rate (FDR) at $\alpha=0.05$ across all diagnostics/branches shown here. Diagnostics are ordered by the magnitude of $\Delta_{z>2}$. Branches without sufficient $z>2$ coverage within their valid range (e.g. the upper branches of R23, $\widehat{R}$, and RNe) are omitted.}
    \label{fig:highz_offset_forest}
\end{figure}

The residual distributions shown in Fig. \ref{fig:sigma_1} for the $t^2=0$ case, and for the full set of diagnostics under both temperature-fluctuation treatments in Figs.~\ref{fig:sigma_2}--\ref{fig:sigma_t2_3} (Appendix~\ref{app:sigma}), already illustrate, panel by panel, how the $z>2$ residuals compare with those of the full calibration sample. For each diagnostic and branch, we compare the mean residual of the $z>2$ objects with that of the full calibration sample using a one-sample $t$-test and account for multiple comparisons using the Benjamini--Hochberg procedure, controlling the false discovery rate (FDR) at $\alpha=0.05$. Fig.~\ref{fig:highz_offset_forest} shows the resulting offset and its uncertainty for each diagnostic and branch, distinguishing those that remain statistically significant after correction from those that do not. (i) For a number of diagnostics, including R2, O32, R2Ne3, O3N2, R3N2, S3, S23, and the lower branches of \Rhat\ and \RNe, the mean residual of the $z>2$ objects is statistically indistinguishable from that of the full calibration sample. (ii) A second group, comprising the nitrogen-based diagnostics N2, N2O2, N2S2, and N2S2H$\alpha$, together with the low-metallicity branches of R3 and R23, shows a statistically significant offset, with the calibrations under-predicting O/H for the $z>2$ objects. This behaviour is consistent with the elevated N/O ratios at fixed O/H established for high-redshift galaxies (see below). (iii) A third group, comprising diagnostics more directly sensitive to the ionisation parameter, Ne3O2, O3S2, Ne3S2, S2, and the high-metallicity branches of R3 and R3S2, shows a significant offset in the opposite direction, consistent with the harder ionising spectra reported for $z>2$ star-forming galaxies. (iv) Finally, for the upper branches of R23, $\widehat{R}$, and $\widehat{RNe}$, and for the lower branch of Ar3, the current $z>2$ subsample provides insufficient coverage within the formally valid index range of the corresponding calibration to allow a meaningful statistical assessment. For these branches, we are therefore unable to assess their applicability at high redshift with the current data, rather than being able to claim either agreement or disagreement.

\begin{figure}
    \centering
	\includegraphics[width=\columnwidth]{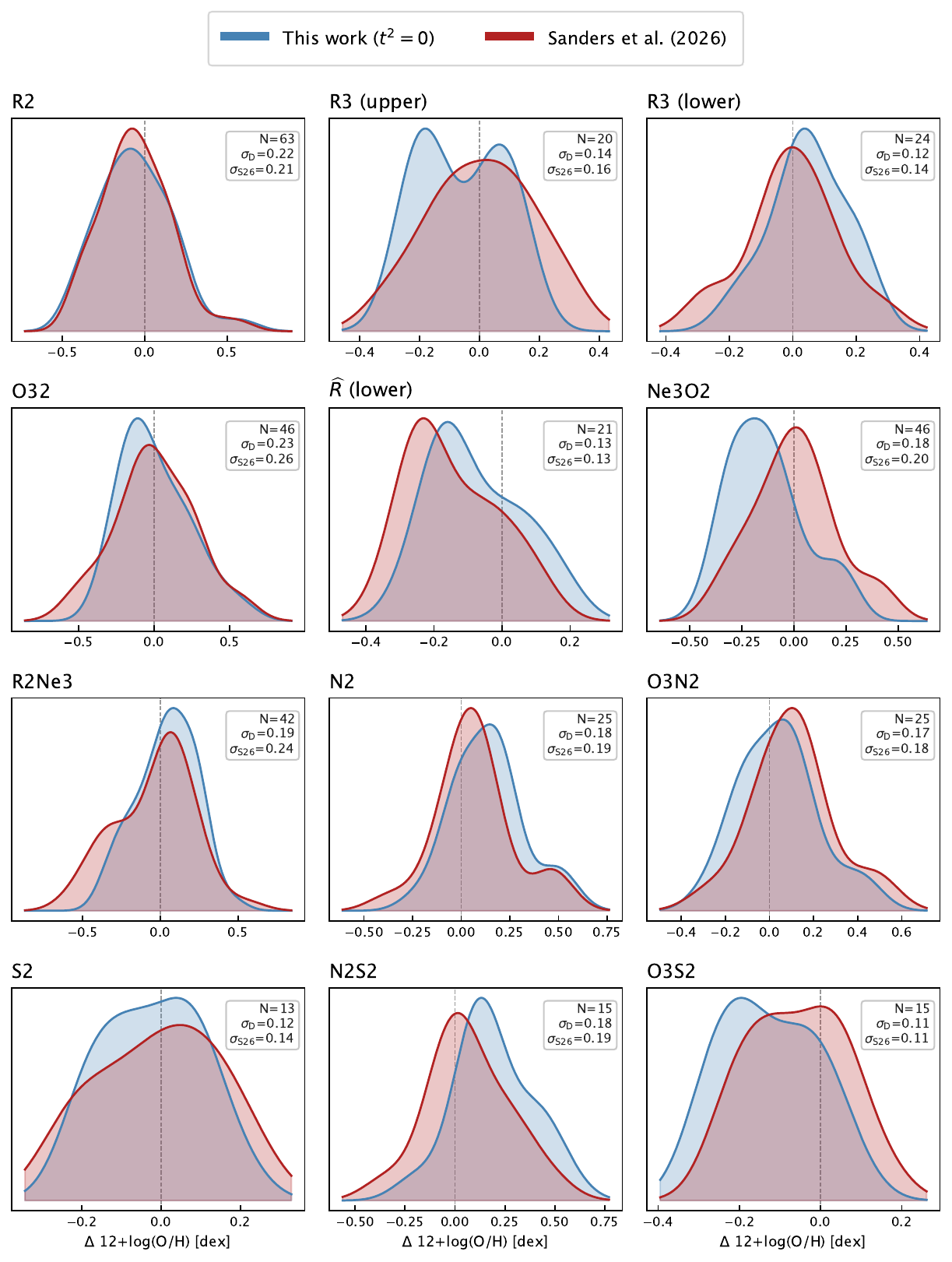}
     \caption{Comparison of the DESIRED ($t^2=0$; blue) and \citet{Sanders:26} (red) calibrations for the twelve diagnostics with a sufficient number ($N\geq10$) of common $z>2$ objects within the formal validity ranges of both calibration sets. Each panel shows the Gaussian kernel density estimate of the residual $\Delta = ({\rm O/H})_{\rm calibration} - ({\rm O/H})_{T_{\rm e}}$ for the same objects, computed independently for each calibration. $N$ denotes the number of objects common to the two calibration sets, while $\sigma_D$ and $\sigma_{S26}$ denote the standard deviations of $\Delta$ for the DESIRED and \citet{Sanders:26} calibrations, respectively. See Table~\ref{tab:sanders_comparison} for the full statistical comparison.}
    \label{fig:sanders26}
\end{figure}

Furthermore, we perform a direct, paired comparison between the DESIRED calibrations and a set of calibrations derived exclusively from $z>2$ galaxies \citep{Sanders:26}, evaluated on the same $z>2$ objects and restricted to the intersection of both calibrations' formal validity ranges. For each of the twelve diagnostics with a sufficient number ($N\geq10$) of common, valid objects, we build the residual distribution of each calibration relative to the direct ($T_{\rm e}$-based) abundance and estimate its density with a Gaussian kernel. Fig.~\ref{fig:sanders26} shows these paired density estimates for each diagnostic, and Table~\ref{tab:sanders_comparison} lists, for the same objects, the dispersion, the mean absolute residual, and the fraction of objects for which the DESIRED residual is smaller in magnitude. A paired Wilcoxon signed-rank test \citep{Wilcoxon:45} on the absolute residuals shows no statistically significant difference between the two calibration sets after correction for multiple testing (Benjamini--Hochberg, $\alpha=0.05$); by the raw dispersion $\sigma$, DESIRED matches or improves on the \citet{Sanders:26} calibrations in ten of the twelve cases. This result indicates that, for the diagnostics in common use, DESIRED's typical accuracy at $z>2$ is not inferior to that of a calibration set purpose-built for this redshift range. This agreement is particularly noteworthy because the DESIRED sample was not specifically constructed for high-redshift applications, yet it recovers comparable empirical performance to calibrations built exclusively from $z>2$ galaxies. For the subset of diagnostics discussed above where a genuine offset is detected (categories ii and iii), this offset is itself of comparable magnitude to what is found when applying a dedicated $z>2$ calibration set to the same objects, so it more plausibly reflects a real, physically-driven dependence of the line-ratio--metallicity relation on N/O and ionisation conditions than a shortcoming specific to the DESIRED sample selection.


\begin{table}
\centering
\caption{Comparison of the residuals of the DESIRED and \citet{Sanders:26} calibrations for the high-$z$ ($z>2$) subsample, restricted to the intersection of their respective formal validity ranges. $N$ is the number of galaxies included in each comparison. $\sigma_{\rm D}$ and $\sigma_{\rm S26}$ are the standard deviations, and $|\Delta|_{\rm D}$ and $|\Delta|_{\rm S26}$ are the mean absolute residuals (dex), with respect to the direct ($T_{\rm e}$-based) oxygen abundances, for the DESIRED and \citet{Sanders:26} calibrations, respectively. $p$ is the $p$-value of the paired Wilcoxon signed-rank test applied to the absolute residuals. None of the twelve comparisons remains statistically significant after applying the Benjamini--Hochberg correction for multiple testing; all adjusted $p$-values satisfy $p_{\rm adj}>0.11$.}
\label{tab:sanders_comparison}
\begin{tabular}{lcccccc}
\hline
Calibration & $N$ & $\sigma_{\rm D}$ & $\sigma_{\rm S26}$ & $|\Delta|_{\rm D}$ & $|\Delta|_{\rm S26}$ & $p$ \\
\hline
R2 & 63 & 0.22 & 0.21 & 0.19 & 0.17 & 0.13 \\
R3 (upper) & 20 & 0.14 & 0.16 & 0.13 & 0.14 & 0.65 \\
R3 (lower) & 24 & 0.12 & 0.14 & 0.10 & 0.11 & 0.55 \\
O32 & 46 & 0.23 & 0.26 & 0.19 & 0.20 & 0.59 \\
$\widehat{R}$ (lower) & 21 & 0.13 & 0.13 & 0.13 & 0.17 & 0.04 \\
Ne3O2 & 46 & 0.18 & 0.20 & 0.19 & 0.15 & 0.02 \\
R2Ne3 & 42 & 0.19 & 0.24 & 0.16 & 0.19 & 0.06 \\
N2 & 25 & 0.18 & 0.19 & 0.17 & 0.15 & 0.09 \\
O3N2 & 25 & 0.17 & 0.18 & 0.14 & 0.16 & 0.26 \\
S2 & 13 & 0.12 & 0.14 & 0.10 & 0.12 & 0.09 \\
N2S2 & 15 & 0.18 & 0.19 & 0.22 & 0.15 & 0.03 \\
O3S2 & 15 & 0.11 & 0.11 & 0.14 & 0.11 & 0.04 \\
\hline
\end{tabular}
\end{table}


A notable exception to the general consistency between local and high-redshift data arises for diagnostics involving the S3O3 and Ar3O3 ratios, for which \citet{Sanders:26} find systematic offsets that they attribute to sub-solar S/O and Ar/O at high redshift, potentially reflecting a delayed contribution from Type Ia supernovae to sulphur and argon enrichment in young galaxies at $z \gtrsim 2$ \citep{Kobayashi:20}. Our own $z>2$ subsample for these two diagnostics is too small ($N<10$) to independently confirm or refute this offset. Some tension also persists in the O32 diagnostic, where hints of increased scatter and mild systematic shifts have been reported at the lowest metallicities in the high-redshift regime \citep{Cataldi:25, Sanders:26}; we do not detect a statistically significant offset for O32 in the DESIRED $z>2$ subsample, though we note that the objects populating the lowest-metallicity end of this diagnostic remain comparatively few. Additionally, the high dispersion in N/O at fixed O/H observed among high-redshift sources \citep{Sanders:26} implies that N-based indicators (N2, O3N2, N2S2) are less reliable tracers of O/H at $z \gtrsim 2$, reinforcing the long-standing conclusion that nitrogen-based diagnostics are systematically affected by N/O variations at high redshift \citep{Jones:15, Masters:14, Shapley:15}, consistent with category (ii) above.


Overall, the evidence suggests that the question is not whether a calibration was derived at $z \sim 0$ or at $z > 2$, but whether the calibrating sample adequately spans the multi-dimensional parameter space--in metallicity, ionisation conditions, excitation, density, and chemical abundance patterns--that is relevant for the galaxy population under study, and whether a given diagnostic is applied within its formally valid range and with due regard for its known sensitivity to secondary parameters such as N/O. A sufficiently large and diverse sample, such as DESIRED, naturally encompasses much of the physical conditions found in high-redshift galaxies without requiring a dedicated high-redshift calibrating set, though it does not eliminate the physically-driven residual dependence on ionisation conditions and N/O discussed above. Conversely, calibrations derived from small high-redshift samples ($\sim$20--140 galaxies) are currently limited to a restricted metallicity range \citep[typically $12+\log({\rm O/H}) < 8.4$;][]{Laseter:24, Chakraborty:25, Cataldi:25} and lack the statistical leverage needed to constrain the full shape of the calibration relations, especially near the turnover region of bivalued diagnostics and at the high-metallicity end where $\sim 10^{10}$~M$_\odot$ galaxies at $z \sim 2$--4 are expected to lie \citep{Sanders:21, Sanders:26}.


A comprehensive assessment of the validity of the DESIRED strong-line abundance calibrations across cosmic time, and a detailed investigation of the redshift evolution of the metallicity calibrations--including multi-variate approaches that combine strong-line ratios with an independent tracer of the ionisation parameter or excitation conditions--is beyond the scope of this paper and will be addressed in a forthcoming paper of this series.

\begin{figure*}
    \centering
	\includegraphics[width=\textwidth]{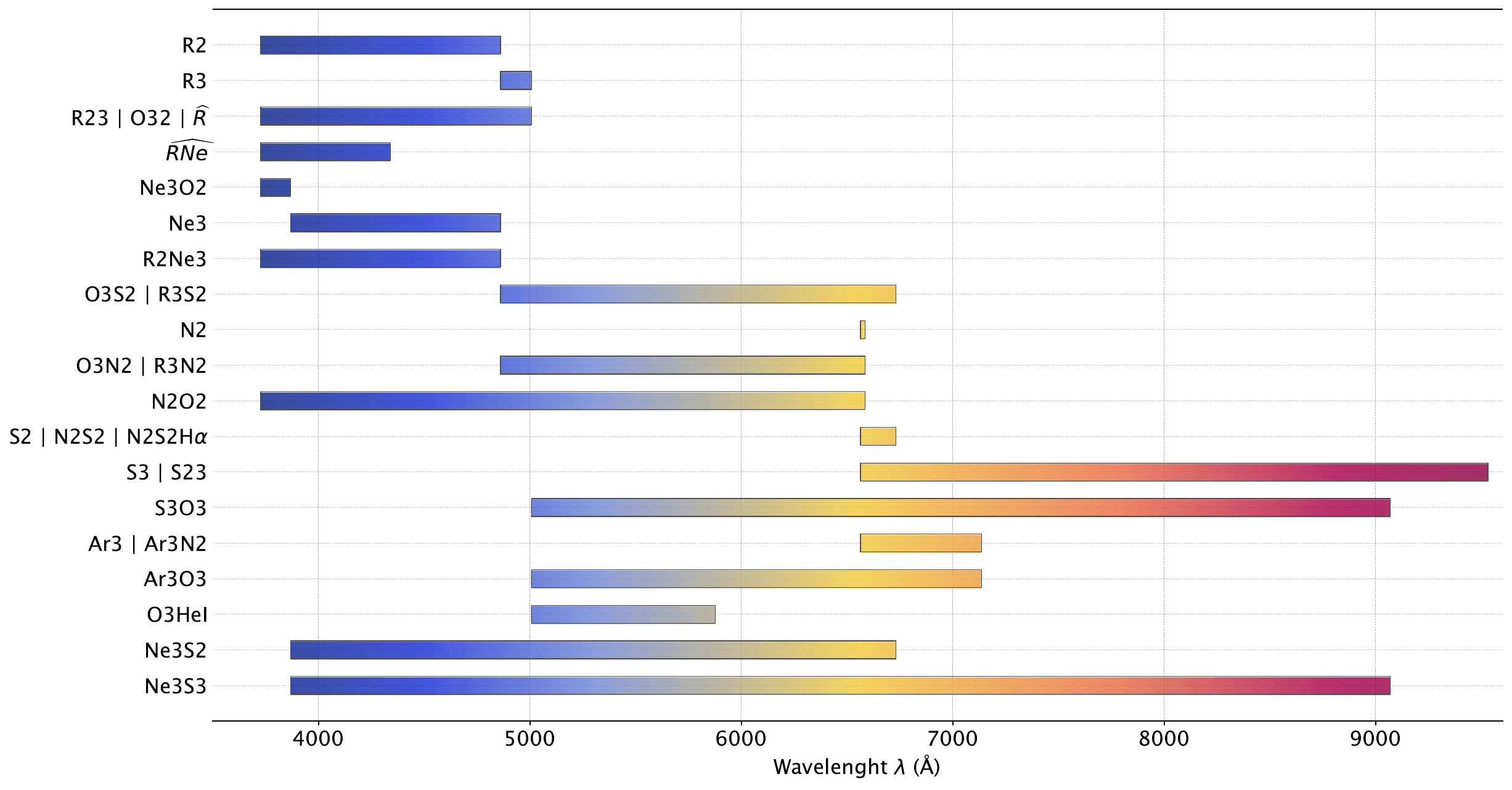}
     \caption{Minimum rest-frame spectral coverage required to apply each of the strong-line metallicity calibrators considered in this work. Each bar spans the wavelength range between the shortest and longest emission lines involved in a given diagnostic. Calibrations sharing an identical observational requirements are grouped on a single row.}
    \label{fig:redshift}
\end{figure*}

\subsection{Guidelines for the application of the DESIRED calibrations}
\label{sec:guide}

In this section, we provide practical guidance for the application of the DESIRED strong-line metallicity calibrations to spectroscopic observations of star-forming regions. Ideally, a strong-line metallicity indicator should be single-valued with respect to O/H, dominated by a well-understood physical dependence on metallicity, unaffected by the presence of DIG, and independent of chemical evolution effects beyond the oxygen abundance itself \citep{Stasinska:05, Stasinska:10, MaiolinoMannucci:19}. In practice, however, no single diagnostic satisfies all of these conditions simultaneously, and each metallicity index presents a complex behaviour across different metallicity regimes, with distinct validity ranges and sensitivities to the physical properties of the ionised gas. The following recommendations are therefore intended to aid the user in selecting, applying, and interpreting the DESIRED calibrations in a robust and informed manner.

The DESIRED calibrations are designed for, and should only be applied to, sources in which the line emission is predominantly powered by star formation. A key advantage of the DESIRED calibration sample is that it encompasses a large and diverse range of ionised ISM properties, including variations in ionisation parameter, electron density, excitation, and chemical abundance patterns. An equally important advantage is the fact that, for the first time, strong-line metallicity estimates can be derived while explicitly accounting for temperature inhomogeneities in the ionised gas. Both the $t^2=0$ and $t^2>0$ calibrations are provided for all 27 diagnostics (Tables~\ref{tab:coeff_t2eq0} and \ref{tab:coeff_t2ge0}). The choice between these two sets depends on the stance regarding the role of temperature fluctuations in nebular abundance determinations: the $t^2=0$ calibrations are anchored to the classical direct-method abundance scale, whereas the $t^2>0$ calibrations incorporate the empirical correction for temperature inhomogeneities following \citet{MendezDelgado:23a}, yielding systematically higher abundances by $\sim$0.13~dex at fixed line ratio. We recommend that the metallicity scale adopted in any analysis be explicitly reported and, where possible, that results be provided for both assumptions to facilitate comparison across studies. It is important to note that stellar oxygen abundances have been found to be more consistent with those derived assuming $t^2>0$, whereas those obtained using the ``direct method'' typically show a systematic offset \citep{MartinezHernandez:2026}. This is particularly relevant when comparing stellar and nebular abundances. Note that these calibrations may prove unreliable for sources with particularly extreme or anomalous physical conditions not well represented by the calibration sample, such as very high Lyman continuum escape fractions or electron densities ($n_{\rm e} \gtrsim 10^5$~cm$^{-3}$), where the collisional de-excitation of the emission lines used in the diagnostics may significantly alter the observed line ratios.

It is strongly recommended that these calibrations only be used within the metallicity ranges where the comparison with the $T_{\rm e}$-based abundances shows tight clustering around the one-to-one relation, indicating negligible systematic offsets and limited intrinsic scatter. The validity range of each calibration---in both the line-ratio index and the corresponding metallicity interval---is provided in Tables~\ref{tab:coeff_t2eq0} and \ref{tab:coeff_t2ge0}, for the $t^2=0$ and $t^2>0$ cases, respectively. Extrapolating beyond these ranges may yield unreliable results and lead to incorrect physical interpretations of the observed regions, including erroneous conclusions about their chemical enrichment or ionisation state; therefore, such extrapolations should be treated with great caution.

To assess the suitability of a particular calibration for a given science application, we recommend examining the calibration--$T_{\rm e}$ comparison plots presented in Sec.~\ref{sec:sigma} (Fig.~\ref{fig:sigma_1} and Figs.~\ref{fig:sigma_2} to \ref{fig:sigma_t2_3} in Appendix~\ref{app:sigma}), which directly reveal the metallicity regimes where each diagnostic performs well and where systematic deviations from the one-to-one relation become important. Individual calibrations should be applied within the metallicity ranges where these comparisons show tight agreement with the direct abundances. For instance, diagnostics such as \Rhat, \RNe, and Ar3 exhibit particularly tight clustering around the one-to-one relation along their respective lower branches (12+log(O/H) $\lesssim$ 8.0), with dispersions as low as 0.14~dex, making them excellent choices for the low-metallicity regime. Conversely, Ne3O2 performs well at intermediate-to-high metallicities but becomes increasingly degenerate below 12+log(O/H) $\sim$ 8.0. The N2 and N2S2H$\alpha$ diagnostics---which have sometimes been regarded with scepticism owing to the well-known secondary nucleosynthetic origin of nitrogen---exhibit relatively low scatter and good agreement with the direct abundances over a broad range of intermediate and low metallicities, but diverge at 12+log(O/H) $\gtrsim$ 8.5, where the sensitivity of \nii\ to variations in N/O at fixed O/H introduces additional scatter \citep{PerezMonteroContini:09, Scholte:25, Sanders:26}.

In general, we recommend the use of at least two metallicity calibrations, preferably based on different line ratios, to verify the robustness of the derived metallicities and to assess how the scientific conclusions depend on the choice of diagnostic. The simultaneous use of multiple diagnostics is not only desirable but required when employing bivalued indicators (R3, R23, \Rhat, \RNe, R3S2, Ne3, O3HeI, Ar3), for which the branch degeneracy must be resolved. In such cases, a monotonic calibration such as N2, O3N2, Ne3O2, S2, Ar3N2, or another single-valued ratio available for the source, should be used to determine whether the object lies on the upper or lower metallicity branch. More generally, simultaneously fitting multiple line ratios to derive the metallicity is desirable whenever sufficient measurements are available, as this approach reduces the magnitude of systematic uncertainties and helps to mitigate the sensitivity of individual diagnostics to secondary physical parameters \citep{Curti:17, Curti:20, Brazzini:24}. The intrinsic scatter estimates ($\sigma_{\rm cal}$) reported in Tables~\ref{tab:coeff_t2eq0} and \ref{tab:coeff_t2ge0} can be used as weights in any such simultaneous fitting procedure to account for the different predictive power of each diagnostic. 

In this context, the $\sigma_{\rm cal}$ value, defined in Sec.~\ref{sec:sigma} as the standard deviation of the metallicity residuals relative to the $T_{\rm e}$-based abundances, plays a central role: it quantifies the intrinsic predictive scatter of each calibration and represents the minimum 1$\sigma$ uncertainty that should be assigned to any individual metallicity estimate derived from these strong-line methods. This scatter arises from the fact that strong-line ratios depend not only on O/H, but also on other physical parameters of the ionised gas, such as the ionisation parameter, electron density, and the shape of the ionising spectrum, that vary at fixed metallicity. Since this source of systematic uncertainty is inherent to the method and cannot be reduced by improving the signal-to-noise ratio of the observations, the total uncertainty in a strong-line metallicity estimate is given by the quadrature sum of $\sigma_{\rm cal}$ and the measurement error ($\sigma_{\rm meas}$) propagated through the calibration relation. Consequently, claims of individual strong-line metallicity uncertainties significantly below $\sigma_{\rm cal}$ (i.e., below $\sim$0.1--0.2~dex for most diagnostics) are not supported by the empirical scatter of the calibrating data \citep[see also][]{Jones:15, Curti:17, Sanders:26}. It is worth noting, however, that while the precision of individual metallicity determinations is fundamentally limited by $\sigma_{\rm cal}$, sample-averaged metallicities---obtained by averaging over many objects---can achieve considerably higher precision, as the random component of the calibration scatter is reduced by $\sqrt{N}$ for $N$ independent measurements.

In addition to the global $\sigma_{\rm cal}$, we provide a quadratic parametrisation of the calibration uncertainty as a function of the emission-line index, $\sigma(x)$ (see Eq.~\eqref{eq:sigma_profile}), whose coefficients are listed alongside the calibration coefficients in Tables~\ref{tab:coeff_t2eq0} and \ref{tab:coeff_t2ge0}. 
This index-dependent dispersion profile provides a more estimate of the local calibration uncertainty than the single $\sigma_{\rm fit}$ value, which may underestimate the scatter in sparsely sampled or degenerate regimes and overestimate it where the calibration is best constrained. In practice, we recommend adopting $\max[\sigma_{\rm cal},\, \sigma(x)]$ as the calibration uncertainty for individual metallicity estimates, where $\sigma(x)$ is evaluated at the observed index value.

Bivalued diagnostics warrant a further, specific caveat in the vicinity of the turnover metallicity separating their two branches. By construction, the line ratio becomes increasingly insensitive to metallicity as the turnover is approached, such that a given observed value can correspond to a wide range of oxygen abundances. This represents an intrinsic degeneracy of the diagnostic itself, rather than a statistical artefact of the fit. Moreover, the polynomial approximations tabulated in Tables~\ref{tab:coeff_t2eq0} and \ref{tab:coeff_t2ge0} do not reproduce the vertical-tangent behaviour of the exact branch-inversion relation at the turnover (Appendix~\ref{app:fits}). Consequently, neither $\sigma_{\rm cal}$ nor $\sigma(x)$ fully captures the true rate at which the metallicity uncertainty increases in this regime. Metallicities inferred from line-ratio values close to the turnover of a bivalued calibration should therefore be interpreted with particular caution.

Regarding the nitrogen-based diagnostics (N2, O3N2, N2O2, N2S2, N2S2H$\alpha$, Ar3N2), although their calibration scatters are quantitatively comparable to those of O- and Ne-based ratios within the respective validity ranges, their systematic sensitivity to N/O variations argues against their use when alternative line ratios that do not include \nii\ are available, particularly at metallicities 12+log(O/H) $\gtrsim$ 8.5 or in environments where the N/O--O/H relation may deviate from the mean trend of the calibrating sample. As discussed in Sec.~\ref{sec:highz}, the enhanced dispersion in N/O at fixed O/H observed at high redshift further limits the reliability of these indicators at $z \gtrsim 2$. An important related caveat concerns studies of the nitrogen-to-oxygen abundance ratio as a function of global metallicity: in such analyses, none of the strong-line diagnostics involving nitrogen lines should be used to measure O/H, as both axes would then essentially trace the nitrogen abundance rather than providing an independent constraint on O/H. For these studies, the global metallicity should be measured using only nitrogen-free diagnostics \citep{Florido:22, PerezMontero:25}. On a practical level, N-based indicators will also have limited utility for metal-poor high-redshift galaxies due to the intrinsic faintness of the \nii\ $\lambda$6585 line, for which there are high non-detection rates even with deep JWST spectroscopy \citep[e.g.,][]{Cataldi:25}.

It should also be noted that the present calibrations are derived from individual \hh regions, integrated spectra of nearby SFGs, and a number of high-redshift SFGs, without an explicit correction for the contribution of DIG emission. The DIG---characterised by enhanced \sii/H$\alpha$ and, to a lesser extent, \nii/H$\alpha$ ratios relative to classical \hh regions \citep{Zurita:00, Oey:07, Florido:22, PerezMontero:25, GonzalezDiaz:24a}—can bias metallicity estimates based on diagnostics involving low-ionisation species, especially in spatially integrated or fibre-based spectra of local galaxies \citep{ValeAsari:19, Kumari:19, Kreckel:22}. High-redshift galaxies are expected to be less affected by DIG contamination owing to their higher star-formation rate surface densities and more compact morphologies \citep{Sanders:17, Florido:22}. 
Similarly, recent studies have shown that electron-density inhomogeneities can bias abundance determinations through their effects on \oii\ and \sii\ emission-line ratios, while having a comparatively minor impact on \nii\ diagnostics \citep{MendezDelgado:23b, Cataldi:25, Martinez:25}. The effects of DIG and density inhomogeneities on the DESIRED calibrations will be investigated in detail in a subsequent paper in this series.

Finally, as demonstrated in Sec.~\ref{sec:highz}, the applicability of the DESIRED calibrations at high redshift is diagnostic-dependent rather than uniform: for roughly a third of the diagnostics testable with the current $z>2$ subsample, no statistically significant offset is found between the local and high-redshift populations, while the remainder show a modest (typically $0.08-0.21$\,dex) but statistically robust offset, attributable to elevated N/O ratios or harder ionising conditions at $z>2$. Where such an offset exists, it does not increase with redshift within the $z>2$ range probed here, and is comparable in magnitude to that found when applying a set of calibrations derived exclusively from high-redshift galaxies to the same objects. Users should therefore consult the categorisation and Fig.~\ref{fig:highz_offset_forest} in Sec.~\ref{sec:highz} when selecting a diagnostic for high-redshift work, favouring category-(i) diagnostics where the science case allows, or explicitly accounting for the expected sign and magnitude of the offset for diagnostics in categories (ii) and (iii).

In terms of observational accessibility with JWST/NIRSpec, the O- and Ne-based diagnostics (\oii\ $\lambda$3727, \neiii\ $\lambda$3869, H$\delta$, H$\beta$, \oiii\ $\lambda\lambda$4959,5007) can be observed across a very broad redshift range ($z \lesssim 8$--10 for O-based ratios, up to $z \approx 11.7$ for Ne-based ratios that rely exclusively on lines at $\lambda_{\rm rest} \lesssim 4100$~\AA). In particular, the Ne-based indicators (Ne3, Ne3O2, R2Ne3) will be of special utility at $z > 9.7$, where \oiii\ $\lambda\lambda$4960,5008 and H$\beta$ shift redward of NIRSpec's wavelength coverage, and thus represent the most promising diagnostics for probing chemical abundances in the very early Universe. The S- and Ar-based diagnostics, which require \sii, \siii, and \ariii\ lines at $\lambda_{\rm rest} \sim 6700$--9100~\AA, are observable at somewhat lower redshifts ($z \lesssim 6$--7 depending on the specific lines involved). N-based diagnostics involving \nii\ $\lambda$6584 are restricted to $z \lesssim 6.6$ and, as noted above, become practically inaccessible for metal-poor galaxies at $z > 6$ due to the intrinsic faintness of the \nii\ line. Overall, the combined set of DESIRED calibrations provides metallicity diagnostics accessible across the entire redshift range currently probed by JWST, with O- and Ne-based ratios offering the broadest coverage and the strongest prospects for application in the epoch of reionisation. Fig.~\ref{fig:redshift} shows the minimum rest-frame spectral coverage required to apply each of the calibrators presented in this work, where each bar spans the wavelength range between the shortest- and longest-wavelength emission lines involved in a given diagnostic. This figure can be used as a practical reference to assess whether a given calibrator is accessible at a particular redshift with a given instrument or spectral setup.

\section{Summary and conclusions}
\label{sec:5}

In this paper, the first of a series, we have presented a new generation of empirical optical strong-line metallicity calibrations based on the DESIRED (DEep Spectra of Ionised REgions Database) project. From the full DESIRED compilation, we constructed a high-quality calibration sample of 2392 objects---including 1029 extragalactic \hh regions, 1296 local star-forming galaxies, and 67 high-redshift ($z > 2$) galaxies---drawn from 201 independent literature references, after applying stringent selection criteria on auroral-line uncertainties, emission-line ratio consistency, and derived abundance quality. This constitutes the most extensive $T_{\rm e}$-based calibration dataset ever assembled, nearly four times larger than previous heterogeneous compilations, spanning an unprecedented metallicity range of $12+\log({\rm O/H}) \in [6.79, 9.07]$ for $t^2=0$ and $[6.87, 9.32]$ for $t^2>0$.

All physical conditions and chemical abundances were derived homogeneously using PyNeb with up-to-date atomic data and the most recent DESIRED $T_{\rm e}$--$T_{\rm e}$ relations \citep{OrteGarcia:26}, ensuring full internal consistency. For the first time in a systematic effort of this kind, the calibrations have been derived under two complementary scenarios: the classical homogeneous temperature assumption ($t^2=0$), and a scenario accounting for temperature inhomogeneities ($t^2>0$), directly addressing the abundance discrepancy problem \citep{Peimbert:67, MendezDelgado:23a}. The $t^2>0$ calibrations yield systematically higher metallicities by $\sim$0.13~dex at fixed line ratio, providing the community with a quantitative measure of the systematic uncertainty introduced by the treatment of the thermal structure of the ionised gas.

We have presented a total of 27 strong-line metallicity calibrations---23 corresponding to diagnostics previously introduced in the literature and 4 new indices proposed in this work---spanning the full suite of commonly used optical emission-line ratios, organised into four families: oxygen-based (R2, R3, R23, O32, \Rhat, \RNe, Ne3O2, R2Ne3, Ne3), nitrogen-based (N2, O3N2, R3N2, N2O2, N2S2, N2S2H$\alpha$), sulphur-based (S2, S3, S23, S3O3, O3S2, R3S2), and argon- and neon-based diagnostics (Ar3, Ar3O3, O3HeI, Ar3N2, Ne3S2, Ne3S3). Each calibration is parametrised as a polynomial function with best-fitting coefficients, validity ranges, and dispersions reported for both the $t^2=0$ and $t^2>0$ cases. Typical intrinsic dispersions range from $\sim$0.15~dex for the tightest diagnostics (\RNe, Ne3) to $\sim$0.35~dex for those most sensitive to the ionisation parameter (R2, S3).

The broad parameter-space coverage of the DESIRED sample also enabled a critical assessment of diagnostics proposed in the literature. We demonstrated that the Ne3O3, S32, and Ar3S2 line ratios do not constitute reliable metallicity indicators, being dominated by ionisation-structure effects rather than systematic variations with O/H.
A detailed comparison with previous calibrations revealed that the DESIRED relations span the broadest validity intervals in both index and metallicity space among all prescriptions considered, while remaining anchored to the empirical locus of the data. Where literature calibrations disagree, the discrepancies can typically be traced to limited calibration samples, restricted parameter-space coverage, or functional forms that extrapolate beyond the observed data. Notably, several widely used calibrations converge toward the DESIRED $t^2>0$ solutions at the high-metallicity end, reinforcing the need for calibrations that explicitly account for temperature inhomogeneities.


Regarding the applicability at high redshift, the DESIRED calibration sample ––by virtue of its large size, heterogeneous composition, and inclusion of compact, high-excitation systems such as blue compact dwarfs and extreme emission-line galaxies–– broadly encompasses the physical conditions found in high-redshift galaxies without requiring a dedicated high-redshift calibrating set. We find, however, that this applicability is diagnostic-dependent rather than uniform: for a fraction of the diagnostics that can be robustly tested with our $z>2$ subsample, no statistically significant offset is found between the local and high-redshift populations, whereas others show modest but significant offsets associated with elevated N/O ratios or harder ionising conditions at $z>2$. The recently proposed high-redshift, JWST-based calibrations from the AURORA, JADES, PRIMAL, and MARTA surveys are nonetheless statistically indistinguishable from our relations, and the offsets reported between earlier ``typical $z \sim 0$'' calibrations and high-redshift data can largely be attributed to the limited parameter-space coverage of those reference samples, rather than indicating a universal redshift dependence of the strong-line diagnostics themselves. Careful selection of the appropriate diagnostic and validity range is therefore key to reliable abundance determinations across cosmic time.

The importance of these calibrations for modern astrophysics cannot be overstated. Strong-line methods remain the only practical means of measuring gas-phase metallicities for the vast majority of star-forming regions and galaxies observed across cosmic time. The DESIRED calibrations represent a substantial advance in several key respects: the unprecedented size and diversity of the calibration sample, which mitigates sample-selection biases that have limited earlier relations; the fully homogeneous analysis with up-to-date atomic data, which eliminates methodological inconsistencies; the simultaneous provision of $t^2=0$ and $t^2>0$ calibrations, which enables users to bracket the plausible range of metallicities; the broad coverage in metallicity and ionisation conditions, making the calibrations applicable from the most metal-poor local dwarfs to massive galaxies at cosmic noon and beyond; and the critical evaluation of both viable and non-viable diagnostics, providing the community with a well-defined set of recommended relations together with explicit guidance on their limitations. In Sec.~\ref{sec:guide}, we described the recommended guidelines for the practical application of the DESIRED calibrations, including the selection of appropriate diagnostics, the treatment of bivalued relations, and the expected impact of adopting $t^2=0$ versus $t^2>0$.

The calibrations presented here open several avenues for future work. The most immediate application is the derivation of metallicity scaling relations and radial abundance gradients in galaxies of the local Universe and at high redshift, which will allow us to characterise how the DESIRED calibrations perform across different galaxy populations and to trace the redshift evolution of the strong-line diagnostics themselves. Building on this observational foundation, the unique statistical leverage of the DESIRED sample makes it particularly well suited for the development of multi-dimensional calibrations in the spirit of \citet{PilyuginGrebel:16}, where multiple line ratios are combined simultaneously to reduce residual dependences on the ionisation parameter and to improve abundance accuracy across the full metallicity range. Finally, the same empirical framework can be extended beyond oxygen to the nitrogen-to-oxygen abundance ratio, providing N/O calibrations anchored to the DESIRED sample and, for the first time, accounting for the effects of temperature inhomogeneities. Together, these O/H and N/O calibrations will constitute a self-consistent framework for characterising chemical abundance patterns in star-forming galaxies across cosmic time.

\section*{Acknowledgements}

We are grateful to the referee for their thoughtful and constructive comments, which led to significant improvements in the final version of this paper.
CE, JGR, MOG and ERR acknowledge financial support from the Agencia Estatal de Investigación of the Ministerio de Ciencia e Innovación  y Universidades (AEI-MCIU) under grant ``The internal structure of ionised nebulae and its effects in the determination of the chemical composition of the interstellar medium and the Universe'' with reference PID-2023-151648NB-I00 (DOI:10.13039/5011000110339). 
JEMD, AZLA, CM and SFS thank the support by UNAM/DGAPA/PAPIIT/IA103326 project ``DESIRED (DEep Spectra of ionised Regions Database): de las emisiones más sutiles a la física fundamental del universo’’ .
JEMD, AZLA, CM and SFS acknowledge support from the Secretaría de Ciencia, Humanidades, Tecnología e Innovación (SECIHTI) project CBF-2025-I-2048, ``Resolviendo la Física Interna de las Galaxias: De las Escalas Locales a la Estructura Global con el SDSS-V Local Volume Mapper''.
JGR also acknowledges support from the AEI-MCIU and from the European Regional Development Fund (ERDF) under grant ``Planetary nebulae as the key to understanding binary stellar evolution'' with reference PID2022-136653NA-I00 (DOI:10.13039/501100011033). 
KZAC acknowledges support from a UKRI Frontier Research Guarantee Grant (PI Cullen; grant reference: EP/X021025/1).
KK gratefully acknowledges funding from the Deutsche Forschungsgemeinschaft (DFG, German Research Foundation) in the form of an Emmy Noether Research Group (grant number KR4598/2-1, PI Kreckel) and the European Research Council’s starting grant ERC StG-101077573 (“ISM-METALS").
OE acknowledges funding from the Deutsche Forschungsgemeinschaft (DFG, German Research Foundation) -- project-ID 541068876.
SFS acknowledges the support of the Mexican Federal Government's Secretaría de Ciencia, Humanidades, Tecnología e Innovación (SECIHTI) through project CBF-2025-I-236, as well as grant PID2022-136598NB-C31 (ESTALLIDOS) from the Spanish Ministry of Science and Innovation (MCINN).
This work has been supported by grant PID2022-136598NB-C33, funded by MCIN/AEI/10.13039/501100011033 and by ``ERDF A way of making Europe''.
IAZ acknowledges funding from the Deutsche Forschungsgemeinschaft (DFG; German Research Foundation)---project-ID 550945879.
JMV acknowledges financial support from the Severo Ochoa grant 
CEX2021-001131-S, funded by MICIU/AEI/10.13039/501100011033 from the research grant PID2022-136598NB-C32 (``Estallidos8'').

\section*{Data Availability}

All data used in this study are publicly available in the original publications from which the emission-line fluxes and/or intensities were compiled. Appendix~\ref{app:references} provides the complete list of references corresponding to the spectroscopic data comprising the DESIRED calibration sample. The full DESIRED calibration catalogue—including object identification, derived physical conditions, ionic and total oxygen abundances, and the corresponding bibliographic reference for each spectrum—is made available online as Supplementary Material (see Appendix~\ref{app:sample}). The calibrations and high-resolution plots presented in this work are publicly available at \url{https://github.com/frosalesortega/desired-calibrations}.
Software: 
\texttt{PyNeb} \citep{Luridiana:15, Morisset:20}, 
\texttt{numpy} \citep{numpy}, 
\texttt{scipy} \citep{scipy}.
\texttt{astropy} \citep{astropy:18, astropy:22}, 
\texttt{pandas} \citep{pandas:10,pandas:20},
\texttt{matplotlib} \citep{matplotlib},
\texttt{seaborn} \citep{seaborn},
\texttt{jupyter} \citep{jupyter}.



\bibliographystyle{mnras}
\bibliography{biblio} 



\appendix

\section{Atomic data}

The adopted atomic data are listed in Table~\ref{tab:atomic_data}, selected following previous analyses of the physical conditions and chemical properties of star-forming regions \citep[e.g.,][]{MendezDelgado:23b, Esteban:25}.

\begin{table}
\centering
\footnotesize
\caption{Atomic data used in this work.}
\label{tab:atomic_data}
\begin{tabular}{lll}
\toprule
Ion & Transition probabilities ($A_{ij}$) & Collision strengths ($\Upsilon_{ij}$) \\
\midrule
O$^{+}$   & \citet{FroeseFischerTachiev:04} & \citet{Kisielius:09} \\
O$^{2+}$  & \citet{Wiese:96},   & \citet{AggarwalKeenan:99} \\    
 & \citet{StoreyZeippen:00} \\
S$^{+}$   &  \citet{Irimia:05} & \citet{Tayal:10}\\
S$^{2+}$  &  \citet{FroeseFischer:06} & \citet{Grieve:14}\\
Cl$^{2+}$ &  \citet{Fritzsche:99} & \citet{Butler:89}\\
Ar$^{3+}$ &   \citet{Mendoza:82b}  & \citet{Ramsbottom:97}\\
Fe$^{2+}$ & \citet{Deb:09} & \citet{Zhang:96}\\
& \citet{Mendoza:23} & \citet{Mendoza:23}\\
N$^{+}$   & \citet{FroeseFischerTachiev:04}  & \citet{Tayal:11}  \\
N$^{2+}$   & \citet{Galavis:98}     & \citet{BlumPradhan:92} \\
\bottomrule
\end{tabular}
\end{table}

\label{app:atomic}

\section{References of the DESIRED calibration data}

Here we provide the complete list of references corresponding to the spectroscopic data comprising the DESIRED calibration sample used in this work, ordered alphabetically. The number given after each reference indicates the number of spectra extracted from that study and included in the present analysis. The complete catalogue of spectroscopic data is presented in Appendix~\ref{app:sample}.\\

\noindent
References: 
(1) \citet{Annibali:15}: 9;
(2) \citet{Annibali:17}: 6;
(3) \citet{Annibali:19}: 3;
(4) \citet{ArellanoCordova:16}: 12;
(5) \citet{Berg:12}: 42;
(6) \citet{Berg:13}: 21;
(7) \citet{Berg:15}: 43;
(8) \citet{Berg:16}: 7;
(9) \citet{Berg:20}: 17;
(10) \citet{Berg:21}: 2;
(11) \citet{Bhattacharya:25}: 1;
(12) \citet{Bresolin:04}: 10;
(13) \citet{Bresolin:05}: 11;
(14) \citet{Bresolin:07a}: 2;
(15) \citet{Bresolin:07b}: 2;
(16) \citet{Bresolin:09}: 4;
(17) \citet{Bresolin:09b}: 28;
(18) \citet{Bresolin:10}: 3;
(19) \citet{Bresolin:11a}: 4;
(20) \citet{Bresolin:11b}: 8;
(21) \citet{Bresolin:12}: 18;
(22) \citet{Brown:14}: 4;
(23) \citet{Buckalew:05}: 17;
(24) \citet{Campbell:86}: 29;
(25) \citet{Castellanos:02}: 4;
(26) \citet{Crockett:06}: 5;
(27) \citet{Croxall:09}: 27;
(28) \citet{Croxall:15}: 17;
(29) \citet{Croxall:16}: 67;
(30) \citet{Curti:23}: 2;
(31) \citet{Diaz:00}: 1;
(32) \citet{Diaz:87}: 1;
(33) \citet{DominguezGuzman:22}: 8;
(34) \citet{Edmunds:84}: 3;
(35) \citet{Egorova:21}: 1;
(36) \citet{Esteban:09}: 13;
(37) \citet{Esteban:14}: 9;
(38) \citet{Esteban:20}: 17;
(39) \citet{Fernandez:18}: 25;
(40) \citet{Fernandez:22}: 3;
(41) \citet{Fricke:01}: 1;
(42) \citet{Garcia-Benito:10}: 1;
(43) \citet{Garnett:04}: 1;
(44) \citet{Garnett:94}: 1;
(45) \citet{Garnett:97}: 3;
(46) \citet{Goddard:11}: 1;
(47) \citet{GomezGonzalez:24}: 1;
(48) \citet{Gonzalez-Delgado:94}: 15;
(49) \citet{Guseva:00}: 3;
(50) \citet{Guseva:01}: 2;
(51) \citet{Guseva:03a}: 2;
(52) \citet{Guseva:03b}: 1;
(53) \citet{Guseva:03c}: 2;
(54) \citet{Guseva:04}: 2;
(55) \citet{Guseva:09}: 29;
(56) \citet{Guseva:11}: 79;
(57) \citet{Guseva:12}: 1;
(58) \citet{Guseva:13}: 30;
(59) \citet{Guseva:20}: 5;
(60) \citet{Guseva:24}: 2;
(61) \citet{Hadfield:07}: 6;
(62) \citet{Hagele:06}: 3;
(63) \citet{Hagele:08}: 7;
(64) \citet{Hagele:11}: 3;
(65) \citet{Hagele:12}: 3;
(66) \citet{Harikane:25}: 2;
(67) \citet{Hirschauer:15}: 10;
(68) \citet{Hirschauer:16}: 2;
(69) \citet{Hodge:95}: 1;
(70) \citet{Hsyu:18}: 2;
(71) \citet{Hsyu:20}: 5;
(72) \citet{Isobe:22}: 9;
(73) \citet{Izotov:01}: 2;
(74) \citet{Izotov:04}: 30;
(75) \citet{Izotov:04b}: 3;
(76) \citet{Izotov:06}: 253;
(77) \citet{Izotov:07}: 152;
(78) \citet{Izotov:09}: 14;
(79) \citet{Izotov:11}: 1;
(80) \citet{Izotov:12}: 12;
(81) \citet{Izotov:16a}: 4;
(82) \citet{Izotov:17b}: 2;
(83) \citet{Izotov:18a}: 1;
(84) \citet{Izotov:18c}: 5;
(85) \citet{Izotov:19}: 1;
(86) \citet{Izotov:20a}: 7;
(87) \citet{Izotov:21a}: 8;
(88) \citet{Izotov:21b}: 5;
(89) \citet{Izotov:22}: 7;
(90) \citet{Izotov:24a}: 7;
(91) \citet{Izotov:94}: 9;
(92) \citet{Izotov:97a}: 1;
(93) \citet{Izotov:97b}: 28;
(94) \citet{Izotov:98a}: 4;
(95) \citet{Izotov:98b}: 18;
(96) \citet{Izotov:99}: 11;
(97) \citet{Izotov:99a}: 3;
(98) \citet{Kehrig:11}: 2;
(99) \citet{Kehrig:20}: 3;
(100) \citet{Kennicutt:03}: 18;
(101) \citet{Kniazev:00}: 1;
(102) \citet{Kniazev:05}: 6;
(103) \citet{Kniazev:18}: 11;
(104) \citet{Kobulnicky:96}: 2;
(105) \citet{Kobulnicky:97}: 3;
(106) \citet{Kobulnicky:97b}: 4;
(107) \citet{Kobulnicky:98}: 1;
(108) \citet{Kojima:21}: 8;
(109) \citet{Kunth:83}: 13;
(110) \citet{Laseter:24}: 5;
(111) \citet{Lee:03a}: 3;
(112) \citet{Lee:03b}: 1;
(113) \citet{Lee:04}: 14;
(114) \citet{Lee:04a}: 5;
(115) \citet{Lee:05}: 4;
(116) \citet{Lee:06}: 2;
(117) \citet{Li:13}: 10;
(118) \citet{Lin:17}: 38;
(119) \citet{Lopez-Sanchez:09}: 18;
(120) \citet{LopezSanchez:07}: 4;
(121) \citet{LopezSanchez:15}: 4;
(122) \citet{Luridiana:02}: 2;
(123) \citet{Magrini:09}: 3;
(124) \citet{Magrini:10}: 24;
(125) \citet{Magrini:17}: 2;
(126) \citet{Melbourne:04}: 12;
(127) \citet{Mendez:99}: 3;
(128) \citet{Miller:96}: 1;
(129) \citet{Miralles-Caballero:14}: 16;
(130) \citet{Nakajima:22}: 10;
(131) \citet{Nakajima:23}: 5;
(132) \citet{Nicholls:14}: 16;
(133) \citet{Nishigaki:23}: 10;
(134) \citet{Noeske:00}: 1;
(135) \citet{Pagel:80}: 2;
(136) \citet{Pastoriza:93}: 2;
(137) \citet{Patterson:12}: 5;
(138) \citet{Peimbert:05}: 2;
(139) \citet{Peimbert:12}: 4;
(140) \citet{Peimbert:86}: 1;
(141) \citet{Pena:07}: 8;
(142) \citet{PenaGuerrero:12}: 4;
(143) \citet{PerezMontero:09}: 2;
(144) \citet{Pilyugin:07}: 44;
(145) \citet{Pollock:25}: 3;
(146) \citet{Popescu:00}: 22;
(147) \citet{Pustilnik:02}: 2;
(148) \citet{Pustilnik:03a}: 3;
(149) \citet{Pustilnik:03b}: 1;
(150) \citet{Pustilnik:05}: 6;
(151) \citet{Pustilnik:06}: 1;
(152) \citet{Pustilnik:10}: 1;
(153) \citet{Pustilnik:11}: 15;
(154) \citet{Pustilnik:16}: 7;
(155) \citet{Pustilnik:20}: 2;
(156) \citet{Pustilnik:21}: 2;
(157) \citet{Pustilnik:24}: 9;
(158) \citet{Rickards-Vaught:24}: 209;
(159) \citet{Rogers:21}: 27;
(160) \citet{Rogers:22}: 57;
(161) \citet{Rosolowsky:08}: 62;
(162) \citet{Sanders:20}: 2;
(163) \citet{Sanders:24}: 23;
(164) \citet{Saviane:08}: 5;
(165) \citet{Schaerer:24a}: 1;
(166) \citet{Scholte:25}: 18;
(167) \citet{Scholtz:25}: 1;
(168) \citet{Skillman:03}: 6;
(169) \citet{Skillman:13}: 2;
(170) \citet{Skillman:85}: 5;
(171) \citet{Skillman:93}: 2;
(172) \citet{Skillman:94}: 1;
(173) \citet{Stanghellini:10}: 7;
(174) \citet{Stanghellini:14}: 3;
(175) \citet{Stasinska:13}: 6;
(176) \citet{Stiavelli:25}: 1;
(177) \citet{Tang:25}: 2;
(178) \citet{Thuan:05}: 30;
(179) \citet{Thuan:95}: 5;
(180) \citet{Thuan:99}: 2;
(181) \citet{Toribio:16}: 14;
(182) \citet{TorresPeimbert:89}: 3;
(183) \citet{Tullmann:03}: 1;
(184) \citet{Valerdi:19}: 1;
(185) \citet{Valerdi:21}: 6;
(186) \citet{Vilchez:03}: 7;
(187) \citet{Vilchez:88b}: 2;
(188) \citet{Vilchez:88a}: 1;
(189) \citet{Vilchez:98}: 13;
(190) \citet{Watanabe:24}: 3;
(191) \citet{Welch:25}: 1;
(192) \citet{Westmoquette:13}: 6;
(193) \citet{Zahid:11}: 3;
(194) \citet{Zamora:23}: 9;
(195) \citet{Zinchenko:24}: 16;
(196) \citet{Zurita:12}: 11;
(197) \citet{vanZee:00}: 2;
(198) \citet{vanZee:06a}: 25;
(199) \citet{vanZee:06b}: 2;
(200) \citet{vanZee:97}: 13;
(201) \citet{vanZee:98}: 14.




\label{app:references}

\section{Calibration fitting methodology}

This section describes the methodology employed to derive the strong-line calibrations presented in Sec.~\ref{sec:calibrations}. Given a dataset of $N$ calibration spectra with $T_e$-based oxygen abundances ($y = 12 + \log(\mathrm{O/H})$) and emission-line ratios ($x$), two natural regression directions exist. In Direction~A (direct), the metallicity is modelled as a function of the line ratio, $y = f(x)$, minimising residuals in O/H. This is the classical approach adopted in calibration studies \citep[e.g.][]{PettiniPagel:04, Marino:13, PilyuginGrebel:16}. In Direction~B (inverse), the line ratio is modelled as a function of metallicity, $x = g(y)$, and the resulting polynomial is numerically inverted to recover $y = g^{-1}(x)$ for practical application. This approach has been developed by \citet{Curti:17, Curti:20}, \citet{Brazzini:24}, \citet{Sanders:26}, and \citet{Cataldi:25}, among others.

Direction~B is motivated by the argument that, at fixed metallicity, the line ratio varies due to secondary physical parameters such as ionisation parameter, electron density, and nebular geometry, placing the dominant scatter along the index axis. However, the converse also holds: at a fixed value of the emission-line ratio, metallicity varies due to genuine abundance differences between objects, and this scatter is physically informative. Since the end-user of a strong-line calibration possesses a measured line ratio and seeks to infer a metallicity, providing the calibration as $12+\log(\mathrm{O/H}) = f(\mathrm{index})$ is the most natural formulation. We explored both directions for every diagnostic and adopted the solution that best traces the underlying trend. All final calibrations are reported in the form of Eq.~\eqref{eq:polynomial}, regardless of the direction in which the fit was originally performed. For Direction~B fits, the coefficients of the inverted relation are obtained by evaluating the forward polynomial on a fine grid in $12+\log(\mathrm{O/H})$, numerically inverting it, and re-fitting a polynomial to the resulting curve, with approximation residuals verified to be $\lesssim 0.02$~dex in all cases.

Three regression methods were considered for each direction. The first is ordinary polynomial regression (unweighted least-squares), which serves as the baseline approach and has been widely used in the calibration literature \citep[e.g.][]{PettiniPagel:04, Curti:20, Nakajima:22}. The second is orthogonal distance regression \citep[ODR;][]{BoggsRogers:90}, which minimises the sum of squared orthogonal distances from the data to the fitted curve, accounting for measurement uncertainties in both variables. ODR is particularly appropriate when both the line ratio and the abundance carry measurement errors, and has been adopted in recent high-redshift calibration work \citep[e.g.,][]{Sanders:24, Sanders:26}. The third is density-equalised regression, a weighted least-squares approach in which each data point receives a weight inversely proportional to the local data density, estimated from bin counts along the fitting axis. This prevents the fit from being dominated by the metallicity regime with the most data points, and is useful for diagnostics with uneven coverage across the metallicity range. Polynomial degrees from 1 to 3 were tested for each combination of method and direction.

Two data configurations were explored for each fit: individual data points and binned medians. For the individual configuration, all $T_e$-based measurements passing quality criteria are included, subject to the condition that each 0.1~dex bin along the line-ratio axis contains at least $N_\mathrm{min}$ objects (typically $N_\mathrm{min} = 5$, relaxed to 3 for diagnostics with fewer available measurements). Data falling outside this range are excluded from the fit. For the binned configuration, the median $x$ and $y$ within bins of width 0.1~dex are computed (retaining only bins with $N \geq N_\mathrm{min}$), and the fit is performed on these medians. A similar approach is employed by \citet{Curti:17}, \citet{Brazzini:24}, and \citet{Sanders:26}, who use binned medians as a visual guide or directly as fitting data. Binning was performed along both the index and O/H axes; the axis producing the lower dispersion was adopted.

Model selection among the candidate fits is based on three metrics. The total fit dispersion, $\sigma_\mathrm{fit}$, is the standard deviation of the residuals $y_i - f(x_i)$ evaluated on the individual (unbinned) data within the validity range. The intrinsic dispersion, $\sigma_\mathrm{int}$, is obtained by subtracting in quadrature the contribution of measurement uncertainties from the total dispersion:

\begin{equation}
        \sigma_\mathrm{int}^2 = \sigma_\mathrm{fit}^2 - \left\langle \sigma_{y,i}^2 + \left(\frac{\partial f}{\partial x}\bigg|_{x_i}\right)^2 \sigma_{x,i}^2 \right\rangle,
        \label{eq:sigma_int}
\end{equation}

\noindent
where $\sigma_{x,i}$ and $\sigma_{y,i}$ are the measurement errors of the $i$-th data point and the derivative term propagates the index errors into the O/H residual space. This formulation follows the approach of \citet{Sanders:26}, who define the intrinsic scatter as the value that yields a reduced $\chi^2$ of unity. Finally, the Bayesian Information Criterion \citep[BIC;][]{Schwarz:78} is used to penalise model complexity:

\begin{equation}
        \mathrm{BIC} = N \ln\left(\frac{\mathrm{RSS}}{N}\right) + k \ln N,
        \label{eq:BIC}
\end{equation}

\noindent
where RSS is the residual sum of squares and $k$ is the number of free parameters. A difference $\Delta\mathrm{BIC} > 10$ is taken as strong evidence against the higher-BIC model \citep{KassRaftery:95}. Among all models within $\Delta\mathrm{BIC} < 10$ of the best-fit value, the one with the lowest $\sigma_\mathrm{fit}$ at the lowest polynomial degree is adopted. All selected fits were additionally verified by visual inspection against binned medians to ensure the model traces the data without oscillations or extrapolation artefacts.

\begin{table*}
\centering
\caption{Bivalued diagnostics: best-fitting polynomial coefficients of the underlying Direction-B relations. For each diagnostic, the corresponding upper- and lower-branch solutions are given separately, with coefficients listed for both the homogeneous ($t^2=0$) and inhomogeneous ($t^2>0$) temperature structure cases. The branch calibrations listed in Tables~\ref{tab:coeff_t2eq0} and \ref{tab:coeff_t2ge0} are obtained as the exact numerical inverses of the relations tabulated here. The $12+\log(\mathrm{O/H})$ columns list the minimum and maximum metallicities over which each branch fit is valid.
}
\label{tab:coeff_B}


\begin{tabular}{l SSSS SS SSSS SS}
\toprule
& \multicolumn{4}{c}{Coefficients | $t^2=0$}
& \multicolumn{2}{c}{12+log(O/H)}
&  \multicolumn{4}{c}{Coefficients | $t^2>0$}
& \multicolumn{2}{c}{12+log(O/H)} \\
\cmidrule(lr){2-5} \cmidrule(lr){6-7} \cmidrule(lr){8-11} \cmidrule(lr){12-13}
Calibration
  & {$a_0^{\prime}$} & {$a_1^{\prime}$} & {$a_2^{\prime}$} & {$a_3^{\prime}$}
  & {Min} & {Max}
  & {$a_0^{\prime}$} & {$a_1^{\prime}$} & {$a_2^{\prime}$} & {$a_3^{\prime}$}
  & {Min} & {Max} \\
\midrule

R3 ~{\scriptsize upper} & -0.63 & -3.69 & -2.50 & -0.23 & \NA & 8.83 & -0.00 & -2.48 & -1.84 & -0.06 & \NA & 9.07 \\
\phantom{R3} ~{\scriptsize lower} & -0.63 & -3.69 & -2.50 & -0.23 & 7.18& \NA & -0.00 & -2.48 & -1.84 & -0.06 & 7.30& \NA \\
R23 ~{\scriptsize upper} & 0.05 & -2.43 & -1.48 & 0.05 & 8.14 & 8.79 & 0.42 & -1.84 & -1.37 & 0.06 & 8.31 & 8.97 \\
\phantom{R23} ~{\scriptsize lower} & 0.50 & -1.11 & -0.74 & \NA & 7.06 & 7.89 & 0.68 & -0.78 & -0.66 & \NA & 7.16 & 8.05 \\
\Rhat ~{\scriptsize upper} & -1.06 & -5.12 & -3.36 & 0.76 & 8.09 & 8.71 & -0.33 & -2.55 & -1.47 & \NA & 8.18 & 8.96 \\
\phantom{\Rhat} ~{\scriptsize lower} & -0.12 & -2.06 & -1.53 & -0.17 & 6.98 & 7.92 & 0.22 & -1.23 & -0.94 & -0.01 & 7.09 & 8.00 \\
\RNe ~{\scriptsize upper} & -1.73 & -5.84 & -4.54 & \NA & 8.13 & 8.66 & -0.84 & -3.48 & -3.28 & \NA & 8.19 & 8.87 \\
\phantom{\RNe} ~{\scriptsize lower} & -0.69 & -2.15 & -1.55 & -0.17 & 6.95 & 7.89 & -0.31 & -1.48 & -1.34 & -0.17 & 7.03 & 7.98 \\
R3S2 ~{\scriptsize upper} & -0.25 & -2.57 & -1.53 & \NA & \NA & 8.85 & 0.07 & -2.21 & -1.58 & \NA & \NA & 8.98 \\
\phantom{R3S2} ~{\scriptsize lower} & -0.25 & -2.57 & -1.53 & \NA & 7.16& \NA & 0.07 & -2.21 & -1.58 & \NA & 7.30& \NA \\
Ar3 ~{\scriptsize upper} & -2.75 & -15.95 & -85.69 & -173.02 & 8.48 & 8.66 & -1.75 & -2.32 & -13.05 & -95.73 & 8.60 & 8.80 \\
\phantom{Ar3} ~{\scriptsize lower} & -1.71 & -0.58 & -0.61 & \NA & 6.96 & 8.22 & -1.62 & -0.28 & -0.50 & \NA & 7.02 & 8.31 \\
Ne3 ~{\scriptsize upper} & -2.31 & -5.91 & -5.23 & -1.30 & 8.09 & 8.68 & -1.38 & -3.79 & -3.84 & -0.99 & 8.25 & 8.88 \\
\phantom{Ne3} ~{\scriptsize lower} & -1.21 & -1.91 & -1.08 & \NA & 6.96 & 7.81 & -0.89 & -1.40 & -0.96 & \NA & 7.06 & 7.93 \\
O3HeI ~{\scriptsize upper} & 0.19 & -4.47 & -3.79 & -0.88 & 8.07 & 8.72 & 0.99 & -2.42 & -2.08 & -0.34 & 8.21 & 9.00 \\
\phantom{O3HeI} ~{\scriptsize lower} & 0.88 & -1.84 & -1.01 & \NA & 6.97 & 7.78 & 1.20 & -1.31 & -0.86 & \NA & 7.06 & 7.89 \\

\bottomrule
\end{tabular}
\end{table*}

The coefficients of the adopted fit were refined using a bootstrap procedure: in each of 1000 realisations, the data are perturbed according to their measurement uncertainties ($x_i' = x_i + \mathcal{N}(0, \sigma_{x,i})$, $y_i' = y_i + \mathcal{N}(0, \sigma_{y,i})$), and the fit is repeated using the same method and polynomial degree. The final coefficients are obtained by fitting the median of the bootstrap realisations, and the coefficient uncertainties are computed as the standard deviation of the distribution across realisations. In all cases, the bootstrap coefficients are consistent with the direct fit within the uncertainties, confirming the robustness of the adopted solutions.

For bivalued diagnostics, a single polynomial $12+\log(\mathrm{O/H}) = f(x)$ cannot capture the full relation, as a given value of the index corresponds to two possible metallicities. In these cases, the upper and lower branches are each calibrated by fitting a polynomial $x = g(12+\log(\mathrm{O/H}))$ in Direction~B, following the same model-selection procedure described above. For each branch, the turnover metallicity is determined analytically as the value of $12+\log(\mathrm{O/H})$ at which $\mathrm{d}g/\mathrm{d}(12+\log(\mathrm{O/H})) = 0$, which for a polynomial of degree $n$ reduces to finding the real roots of a polynomial of degree $n-1$ within the appropriate metallicity range. For most bivalued diagnostics, the model-selection criteria described above favour independent Direction-B fits for the upper and lower branches, yielding two closely-spaced turnover values with a narrow intervening metallicity interval excluded from both calibrations; for the R3 and R3S2 diagnostics, a single Direction-B polynomial fitted to the full calibration sample reproduces both branches equally well, yielding a single turnover value. Both cases are reported in Tables~\ref{tab:coeff_t2eq0} and \ref{tab:coeff_t2ge0} for each bivalued diagnostic. The calibration relations listed in these tables, in the form of Eq.~\eqref{eq:polynomial}, are obtained for each branch as the exact numerical inverse of its corresponding Direction-B fit, evaluated over the branch's validity range up to its turnover value. For Ar3, whose upper branch is populated by too few calibrating regions and does not exhibit the classical double-branch morphology observed in the other diagnostics, only the lower branch follows the procedure described below, whereas the upper branch is calibrated using a direct Direction-A fit to binned medians. For consistency, we adopt an upper-branch turnover 0.26 dex above the analytically defined lower-branch turnover, corresponding to the mean excluded interval found across the other affected diagnostics.

For completeness, the underlying Direction-B fits from which the branch calibrations above are derived are reported separately in Table~\ref{tab:coeff_B}, following the relation:

\begin{equation}
    \log(R) = \sum_{i} a_{i}^{\prime}\, x^{i},
    \label{eq:polynomial_B}
\end{equation}

\noindent
where $R$ is the bivalued emission-line ratio, $x \equiv 12+\log(\mathrm{O/H}) - 8.70$ is the metallicity expressed relative to the solar value \citep{AmarsiGrevesse:26}, and $a_{i}^{\prime}$ are the corresponding best-fitting coefficients. Users should employ an auxiliary monotonic diagnostic to determine which branch is appropriate for their objects.

In addition to the global $\sigma_\mathrm{fit}$, we provide a quadratic parametrisation of the dispersion as a function of the line ratio, $\sigma(x)$ (Eq.~\eqref{eq:sigma_profile}). The coefficients are obtained by fitting a second-order polynomial to the binned dispersions $\sigma_k$, computed from the median absolute deviation (MAD) of the residuals within each bin (converted to equivalent Gaussian $\sigma$ via the factor 1.4826), weighted by $\sqrt{N_k}$ to give higher weight to better-sampled bins. No correction for measurement uncertainties is applied at this stage, so $\sigma(x)$ reflects the total dispersion of the residuals around the fit at each value of the line ratio, rather than the intrinsic dispersion $\sigma_{\rm int}$. The polynomial is evaluated only within the range covered by bins with sufficient statistics, and held constant at the boundary value outside this range. This parametrisation allows the user to estimate the calibration uncertainty at any value of the emission-line ratio, accounting for the fact that the dispersion may vary substantially across the metallicity range. 

A Python implementation for evaluating $\sigma(x)$ from the tabulated coefficients is publicly available at: \\
{\small  \url{https://github.com/frosalesortega/desired-calibrations}}.

\label{app:fits}


\section{Rotation angle of the \texorpdfstring{\Rhat}{Rhat} and
\texorpdfstring{\RNe}{RNe} diagnostics}

The \Rhat\ \citep{Laseter:24} and \RNe\ \citep{Scholte:25} calibrations are defined as a linear combination of the R2--R3 (\oii/H$\gamma$--Ne3/H$\gamma$) plane around the O/H axis, rotated by an angle $\phi$ chosen to minimise the scatter of the resulting index at fixed metallicity.

The calibration samples involved in defining this rotation differ substantially from one another. \citet{Laseter:24} combined 465 individual low-metallicity objects with direct ($T_{\rm e}$-based) oxygen abundances with \emph{stacked} SDSS spectra from \citet{Curti:17}, binned in 12+log(O/H), for the metal-rich (12+log(O/H)$\gtrsim8.2$) branch; stacking thousands of galaxies per bin substantially suppresses both measurement noise and object-to-object scatter in this high-metallicity regime, at the cost of a comparatively narrow overall metallicity range (12+log(O/H)$\lesssim8.4$). \citet{Scholte:25} instead calibrated both diagnostics using 782 individual, unstacked local galaxies with direct metallicities from the DESI Early Data Release, extending to 12+log(O/H)$\sim8.5$; no stacking was used, and their calibration was validated against a separate, held-out sample of 22 high-redshift
JWST/EXCELS galaxies, which was not itself used in the fit. The DESIRED calibration sample used here is more than four times larger than that of \citet{Laseter:24}, and 1.7--2.7 times larger than the local DESI EDR sample of \citet{Scholte:25} (1338 objects for \RNe\ and 2111 for \Rhat,
respectively), and spans a substantially wider metallicity range, extending to 12+log(O/H)$\sim9.1$, deep into the declining branch of R3 and Ne3/H$\gamma$. Following the same approach already adopted elsewhere in this paper for similarly behaved bivalued diagnostics, we fit the low-metallicity branch using the individual calibration objects directly, and the high-metallicity branch using data binned by the index itself, which plays a
role analogous to spectral stacking in suppressing scatter in this regime.

Given these differences in sample size, construction, and metallicity coverage, the rotation that minimises the scatter could plausibly differ considerably between the three calibrations: R3 and Ne3/H$\gamma$ turn over and decline at high metallicity, so a rotation optimised over a narrow, high-precision, low-metallicity-biased sample need not minimise the scatter over the full dynamic range spanned by a large, heterogeneous compiled sample such as DESIRED's, and vice versa. We test this directly by repeating the rotation optimisation independently over the DESIRED calibration sample for both diagnostics, sweeping $\phi$ between 0 and 90 degrees and evaluating, at each point, the total dispersion produced by the same fitting methodology
used throughout this paper (Appendix~\ref{app:fits}), refining the search with a fine grid around the minimum.

\begin{figure}
    \centering
    \includegraphics[width=0.9\linewidth]{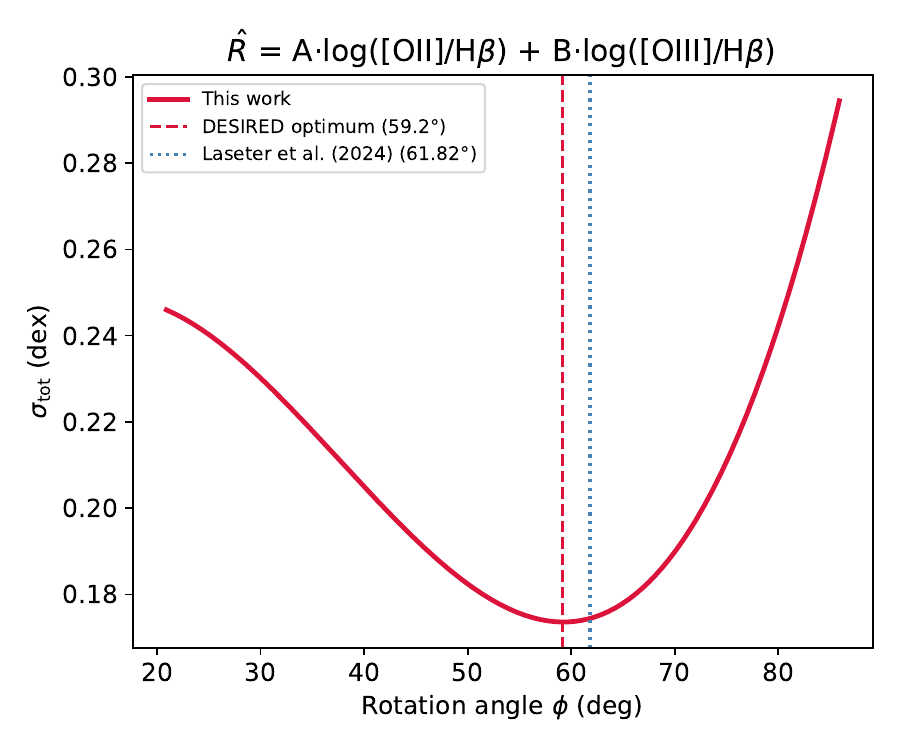}
    \includegraphics[width=0.9\linewidth]{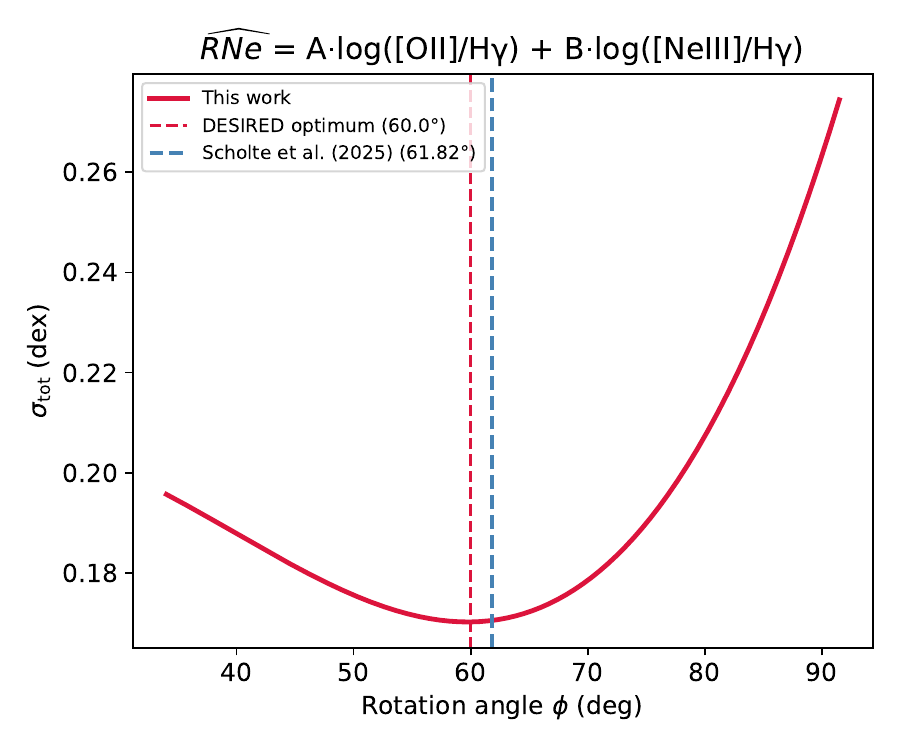}
    \caption{Total dispersion, $\sigma_{\rm tot}$, as a function of rotation angle $\phi$ for
    \Rhat\ (upper) and \RNe\ (lower panel), evaluated on the DESIRED calibration
    sample. For \Rhat\ the dashed blue line marks the \citet{Laseter:24} (and, for \RNe,
    \citealt{Scholte:25}) angle; the dotted red line mark the DESIRED-optimal angle. In both panels the minimum is a broad, shallow plateau rather than a sharp point.}
    \label{fig:angles}
\end{figure}

We find that the recovered optimum for both diagnostics is consistent with the values found by \citet{Laseter:24} and \citet{Scholte:25}: the minimum dispersion falls within a broad, shallow
plateau spanning approximately 57--63 degrees (Fig.~\ref{fig:angles}), comfortably including both the \citet{Laseter:24} angle (61.82 degrees, adopted by \citealt{Scholte:25} for \RNe\ as well) and the independent verification of 63.5 degrees ($\hat{R}$) and 63.9 degrees (\RNe) reported by \citet{Scholte:25} using their own sample. We take the mutual agreement of these three independent determinations as evidence that the adopted rotation reflects a genuine, stable feature of the R2--R3 (\oii/H$\gamma$--Ne3/H$\gamma$)--O/H geometry rather than an artefact of any single calibration sample. Concretely, the total dispersion achieved with the
DESIRED-specific rotation differs from that obtained with the adopted \citet{Laseter:24}/\citet{Scholte:25} coefficients by only $\sim0.01$~dex or less in either branch, for both diagnostics (Table~\ref{tab:angles}), a difference with no practical impact on the resulting metallicities. As with \citet{Scholte:25}, who verified but did not adopt an independently-derived projection for the same reason, we therefore retain the original
\citet{Laseter:24}/\citet{Scholte:25} coefficients in both calibrations, since adopting
a different rotation here would prevent the direct comparison with the literature calibrations (Sec. \ref{sec:comparison}, in exchange for a dispersion improvement of order $0.01$~dex.

Table~\ref{tab:angles} lists, for reference, the best-fit rotation and coefficients found for each diagnostic over the DESIRED sample, together with the corresponding dispersion in each branch, alongside the original \citet{Laseter:24}/\citet{Scholte:25} values for comparison.

\begin{table}
\centering
{\setlength{\tabcolsep}{4pt}  
\caption{Rotation angle and coefficients of $\hat{R}=A\cdot R2+B\cdot R3$ and
$\widehat{RNe}=A\cdot$\oii/H$\gamma$ + B$\cdot$Ne3/H$\gamma$, derived in this work and compared with those of \citet{Laseter:24} and \citet{Scholte:25}. The $\sigma_{\rm tot}$ column reports the total upper- and lower-branch dispersions evaluated over the DESIRED calibration sample.}
\begin{tabular}{llcccc}
\hline
Diagnostic & Source & $\phi$ (deg) & $A$ & $B$ & $\sigma_{\rm tot}$ (dex) \\
\hline\\[-5pt]
$\widehat{R}$        & \citet{Laseter:24}     & 61.82 & 0.470 & 0.880 & 0.175 / 0.178 \\
                 & This work              & 59.2  & 0.512 & 0.859 & 0.162 / 0.177 \\
\hline\\[-5pt]
$\widehat{RNe}$  & \citet{Scholte:25}     & 61.82 & 0.470 & 0.880 & 0.137 / 0.198 \\
                 & This work              & 60.0  & 0.500 & 0.866 & 0.130 / 0.190 \\
\hline
\end{tabular}}
\label{tab:angles}
\end{table}

\label{app:new_Rhat_RNe}

\section{Comparison with Pilyugin \& Thuan (2005)}

\begin{figure*}
	\includegraphics[width=0.8\textwidth]{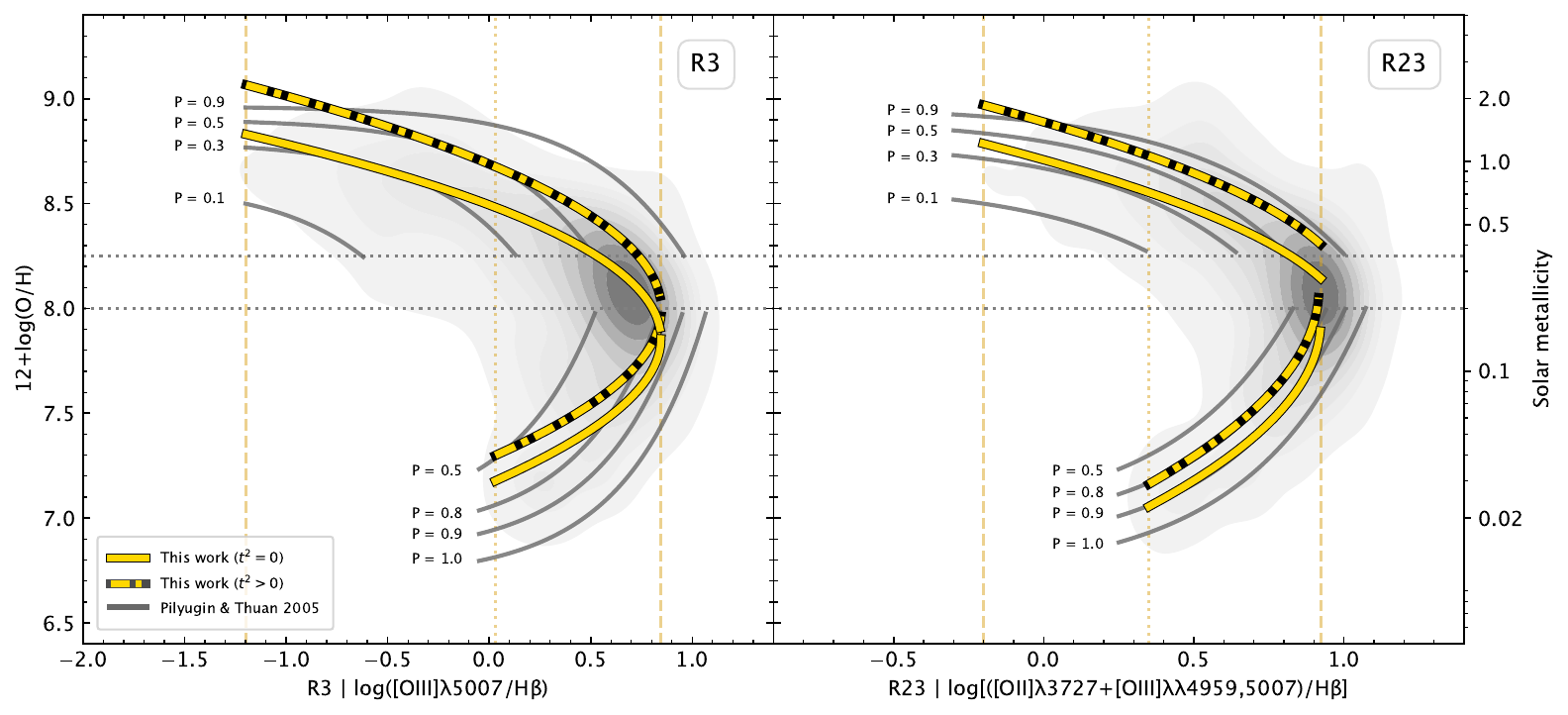}
    \caption{Comparison of the DESIRED metallicity calibrations for R3 and R23 shown in Fig.~\ref{fig:calib_1} with the family of relations $\text{O/H} = f(R3, P)$ (left) and O/H = $\text{O/H} = f(R23, P)$ (right) proposed by \citet{PilyuginThuan:05}, where the different curves correspond to distinct values of $P$, the excitation parameter. The dotted horizontal lines define the transition zone between the upper and lower metallicity branches, as defined by PT05. All definitions and plotting conventions are the same as in Fig.~\ref{fig:compar_1}.}
    \label{fig:compar_T05}
\end{figure*}

The calibration by \citet{PilyuginThuan:05} (hereafter PT05) is an updated version of the $P$ method \citep{Pilyugin:00,Pilyugin:01b}, based on the empirical relations between R3, R23 and $T_{\rm e}$-based metallicities, to account for the degeneracy inherent in the $R_{23}$ calibrations of \citet{Pagel:79}. It represents one of the most influential empirical strong-line metallicity calibrations of the past two decades, and has served as a benchmark for subsequent calibration efforts. 

This formulation introduces the excitation parameter $P$, defined as $P = R3/(R2+R3)$, which represents the fractional contribution of the high-ionisation emission (\oiii $\lambda\lambda$4959,5007) to the total oxygen line flux. Physically, $P$ traces the relative extent of the O$^{++}$ and O$^{+}$ zones within the nebula, and is therefore sensitive to the ionisation structure of the \hii\ region: high-excitation regions with hard ionising spectra or high ionisation parameters exhibit large values of $P$ ($P \gtrsim 0.7$), as most of the oxygen resides in the doubly ionised state; conversely, low-excitation, metal-rich \hii\ regions are characterised by $P \lesssim 0.3$, since the softer radiation field and more efficient cooling confine most of the oxygen to the singly ionised state. Thus, $P$ serves as an empirical proxy for the ionisation parameter, enabling a two-dimensional parametrisation of the form $\text{O/H} = f(R23, P)$ that partially breaks the well-known degeneracy between metallicity and ionisation conditions in strong-line diagnostics.

PT05 provide separate parametrisations for the upper and lower branches of both R3 and R23, valid for $T_{\rm e}$-based metallicities 12+log(O/H) > 8.25 and 12+log(O/H) < 8.0, respectively, expressed as polynomial functions of $P$ and the corresponding line ratio, with $P$ spanning values between 0 and 1. PT05 claim an agreement between calibration-based and $T_{\rm e}$-based abundances within
$\sim$0.1~dex across the explored range of $P$.

In Fig.~\ref{fig:compar_T05} we compare the DESIRED calibrations for R3 and R23 with the family of $\text{O/H} = f(R3, P)$ and $\text{O/H} = f(R23, P)$ curves from PT05, evaluated at different values of $P$. The grey contours represent the density distribution of the DESIRED calibration sample, which covers nearly the full parameter space originally defined by PT05 for both branches of R3 and R23 across the explored range of $P$. This broad coverage is a direct consequence of the substantially larger DESIRED compilation, which overcomes the statistical limitations explicitly acknowledged by PT05, particularly the scarcity of low-excitation ($P \lesssim 0.4$) objects in the upper branch and of regions with $P \lesssim 0.55$ in the lower branch.

For R3, in the upper-metallicity branch, the DESIRED calibration for the $t^2=0$ solutions is consistent within the intrinsic dispersion and falls in the PT05 parameter space corresponding to $P \sim 0.3-0.5$ across most of its extent. The $t^2>0$ solution, runs through the parameter space between $P \sim 0.5-0.9$, exceeding this value only at the highest-metallicity end. In the lower branch, both DESIRED calibrations trace the PT05 locus at $P \sim 0.5-0.9$, in both cases with a slight divergence of the functional forms of the two frameworks. 
For R23, in the upper-metallicity branch, the DESIRED $t^2=0$ calibration follows closely the PT05 relation for $P \sim 0.3$--$0.5$, while the $t^2>0$ solution spans a broader range of $P \sim 0.5-0.9$. In the lower branch, both DESIRED calibrations converge and remarkably follow the PT05 relations at $P \sim 0.8-0.9$, with a very similar functional form. In both panels, the turnover of the DESIRED calibrations occurs at a different index value than that of the PT05 relations, consistent with the different calibrating samples and metallicity scales underlying each framework.

The fact that the locus of the DESIRED calibration sample naturally falls within the PT05 curves for physically motivated values of $P$ constitutes a powerful \textit{a posteriori} validation of the $P$-method framework. Moreover, the fact that the DESIRED $T_{\rm e}$-based homogeneous abundances ––recomputed with a uniform methodology and updated atomic data–– populate the same region of parameter space reinforces the idea that the strong nebular oxygen lines carry sufficient information to constrain the oxygen abundance when the ionisation conditions are properly accounted for.

A particularly noteworthy result of this comparison is the systematic offset between the DESIRED $t^2=0$ and $t^2>0$ calibrations when projected onto the PT05 framework. The upward shift of the derived oxygen abundances considering temperature inhomogeneities in the $R3$-- and $R23$--O/H planes is geometrically equivalent to moving along the PT05 curves towards higher effective values of $P$. Since PT05 was constructed entirely under the assumption of a homogeneous temperature structure, its family of curves for different $P$ values effectively represents only one ``slice'' of the full parameter space that becomes accessible when temperature variations are taken into account. The DESIRED calibrations therefore extend the PT05 framework by demonstrating that the spread in $P$ required to encompass both the $t^2=0$ and $t^2>0$ solutions is physically consistent with the magnitude of the abundance corrections derived from the temperature inhomogeneities formalism \citep{MendezDelgado:23a}. This result also suggests that part of the scatter attributed by PT05 to variations in the ionisation parameter may, in fact, reflect the contribution of unaccounted temperature inhomogeneities to the derived $T_{\rm e}$-based abundances.

\label{app:pt05}

\section{Comparison of calibrations and $T_{\rm e}$-based metallicities}

\begin{figure*}
	\includegraphics[width=\textwidth]{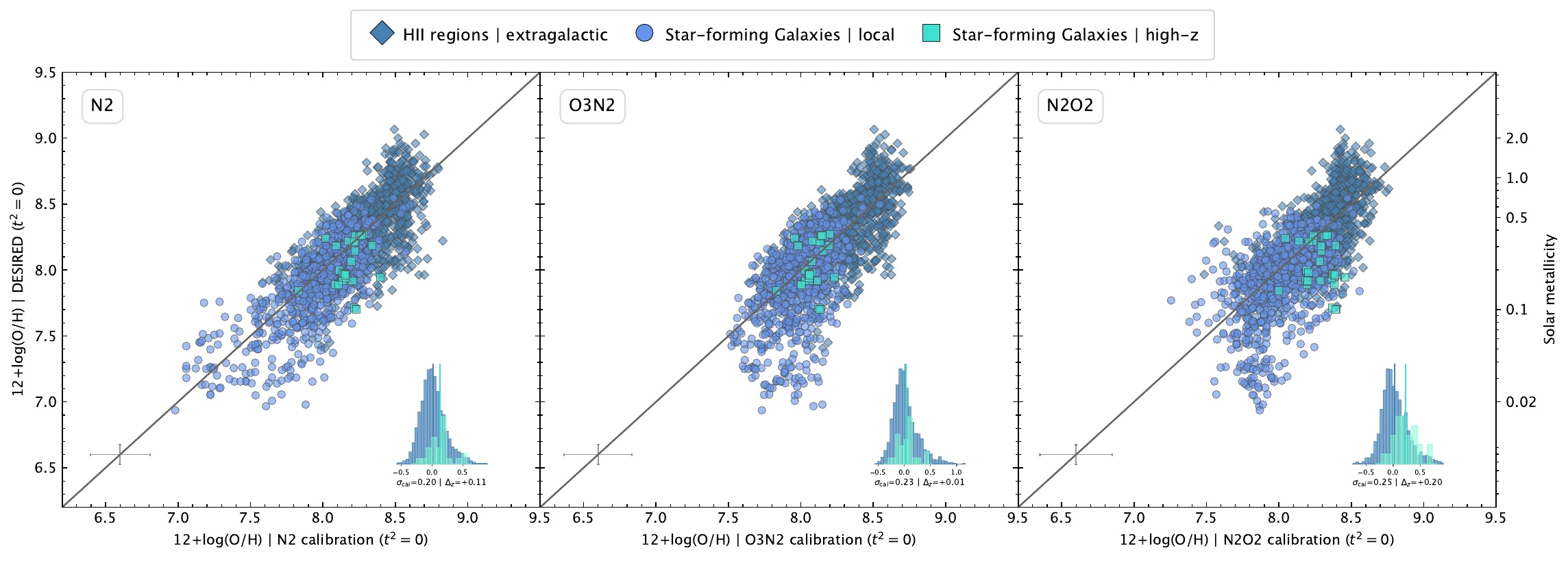}
	\includegraphics[width=\textwidth]{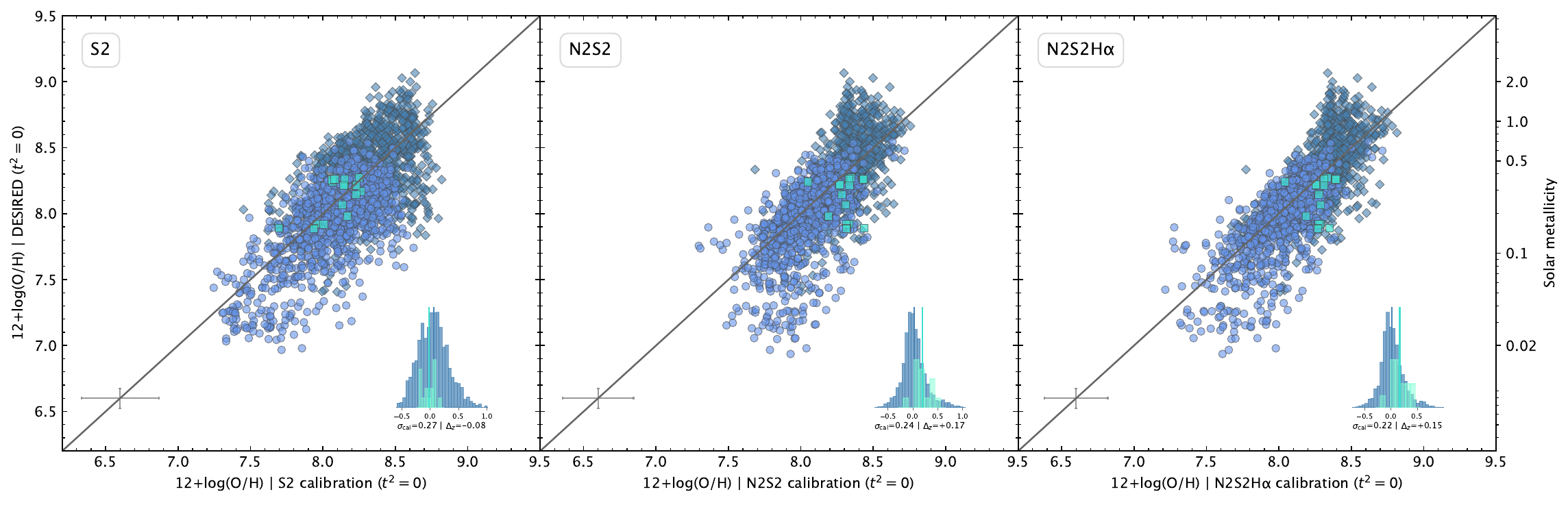}
	\includegraphics[width=\textwidth]{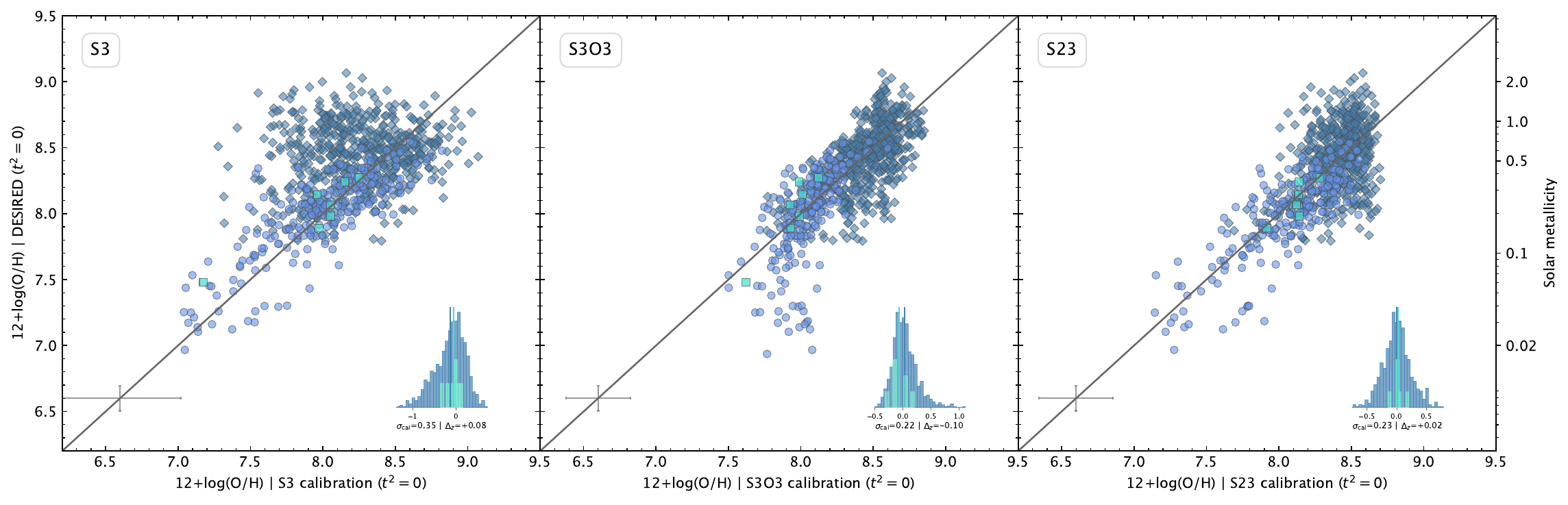}
    \caption{Comparison between the oxygen abundances inferred from the calibrations presented in Fig.~\ref{fig:calib_2} and the direct ($T_{\rm e}$-based, $t^2=0$) metallicities of the calibration sample. The panels correspond to the N2, O3N2, N2O2, S2, N2S2, N2S2H$\alpha$, S3, S3O3, and S23 diagnostics, arranged from top to bottom and left to right, as indicated.}
    \label{fig:sigma_2}
\end{figure*}

\begin{figure*}
	\includegraphics[width=\textwidth]{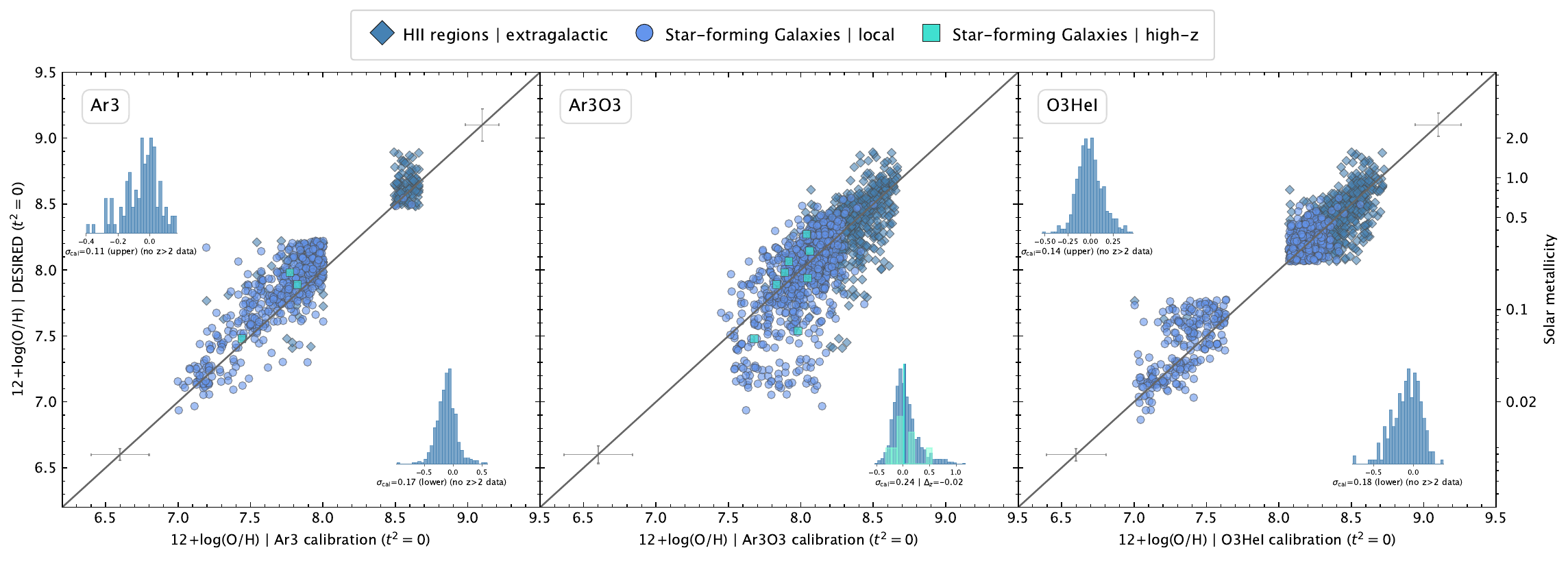}
	\includegraphics[width=\textwidth]{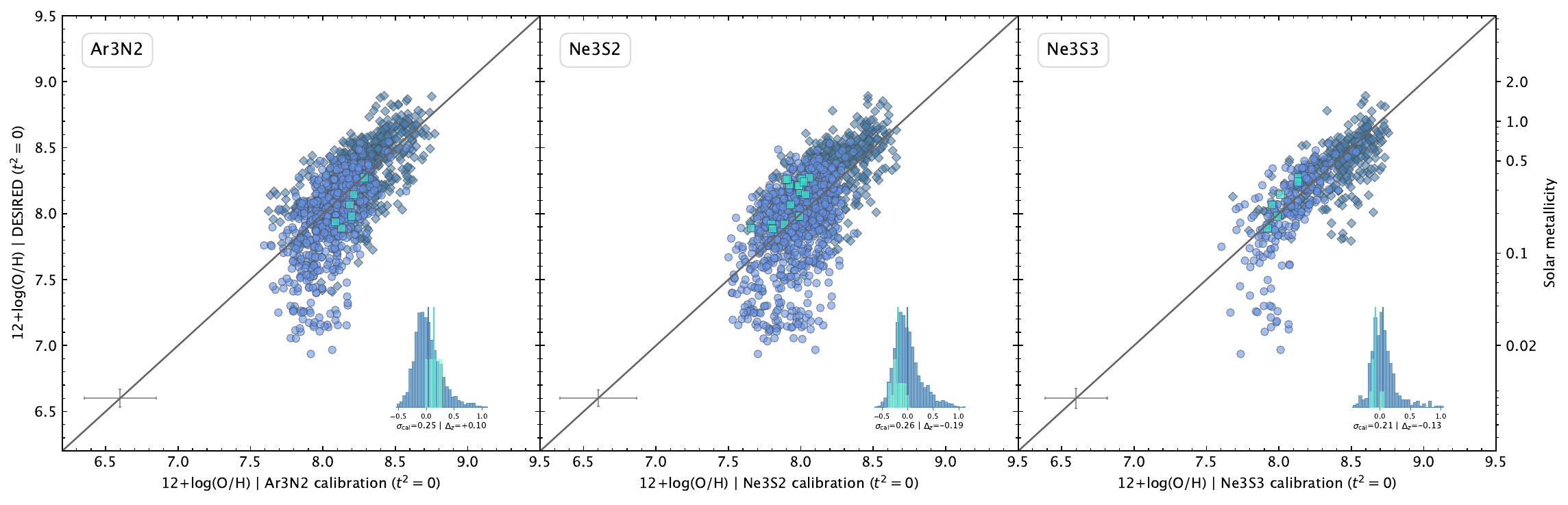}
	\includegraphics[width=\textwidth]{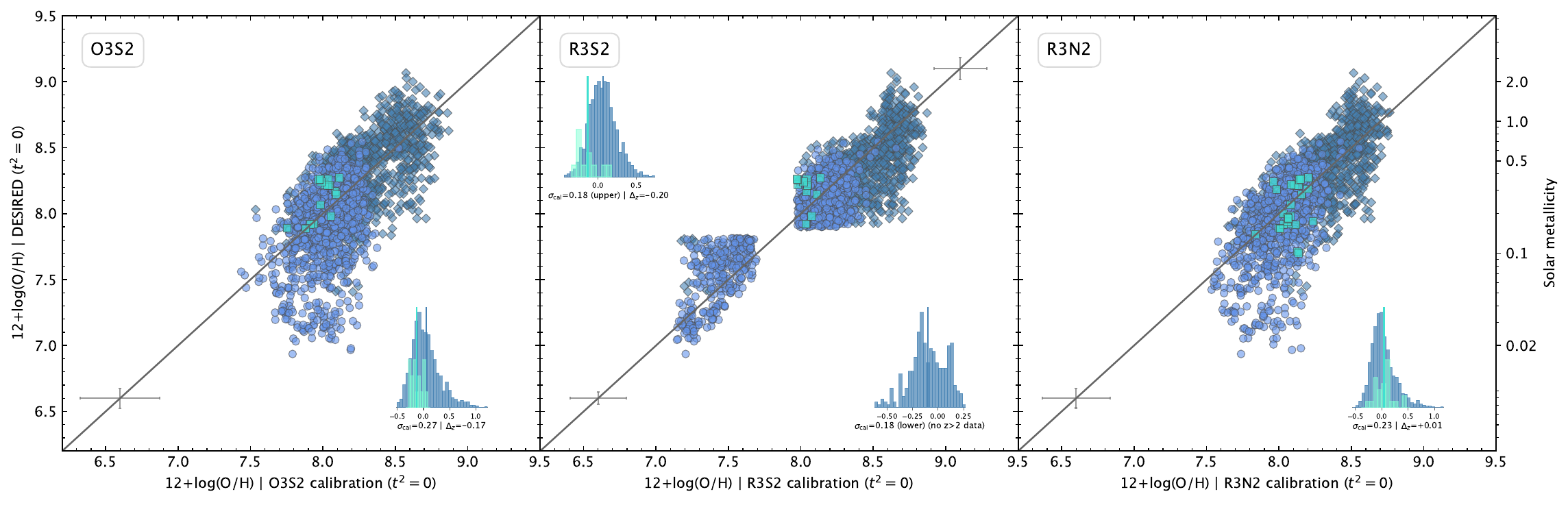}
    \caption{Comparison between the oxygen abundances inferred from the calibrations presented in Figs.~\ref{fig:calib_3} and \ref{fig:calib_1d}, and the direct ($T_{\rm e}$-based, $t^2=0$) metallicities of the calibration sample. The panels correspond to Ar3, Ar3O3, O3HeI, Ar3N2, Ne3S2, Ne3S3, O3S2 and R3S2 calibrations, arranged from top to bottom as indicated.}
    \label{fig:sigma_3}
\end{figure*}

\begin{figure*}
	\includegraphics[width=\textwidth]{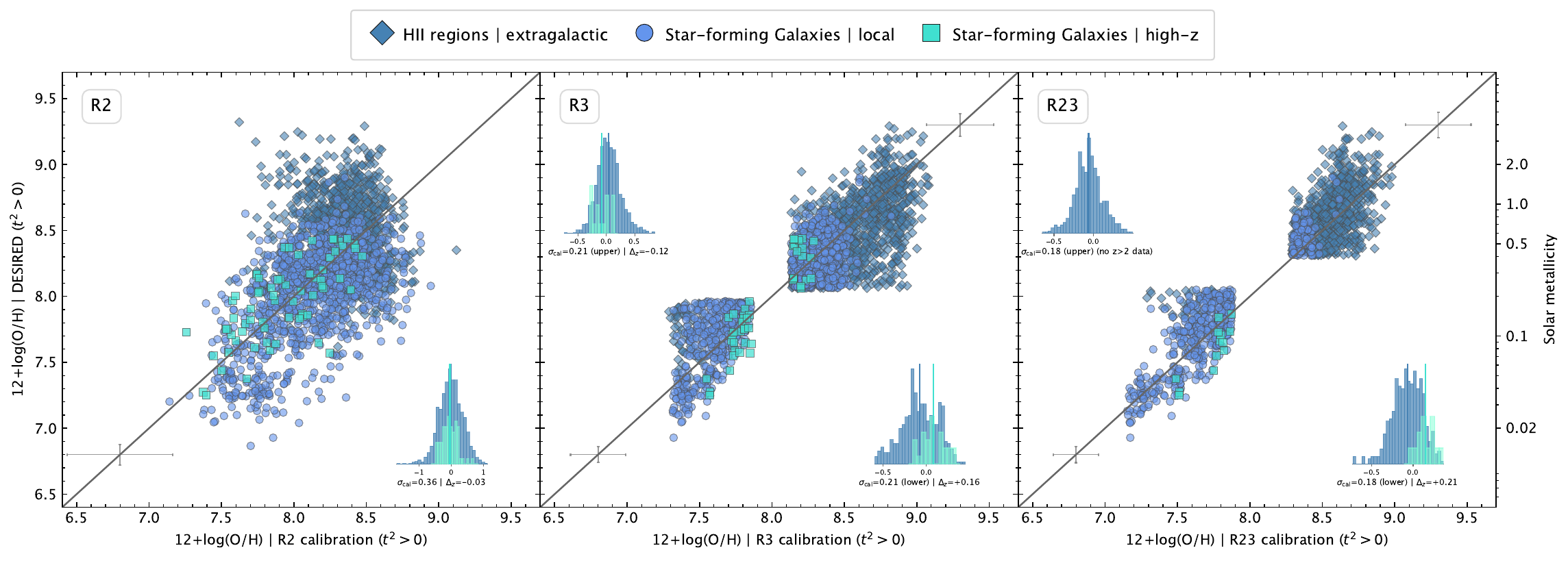}
	\includegraphics[width=\textwidth]{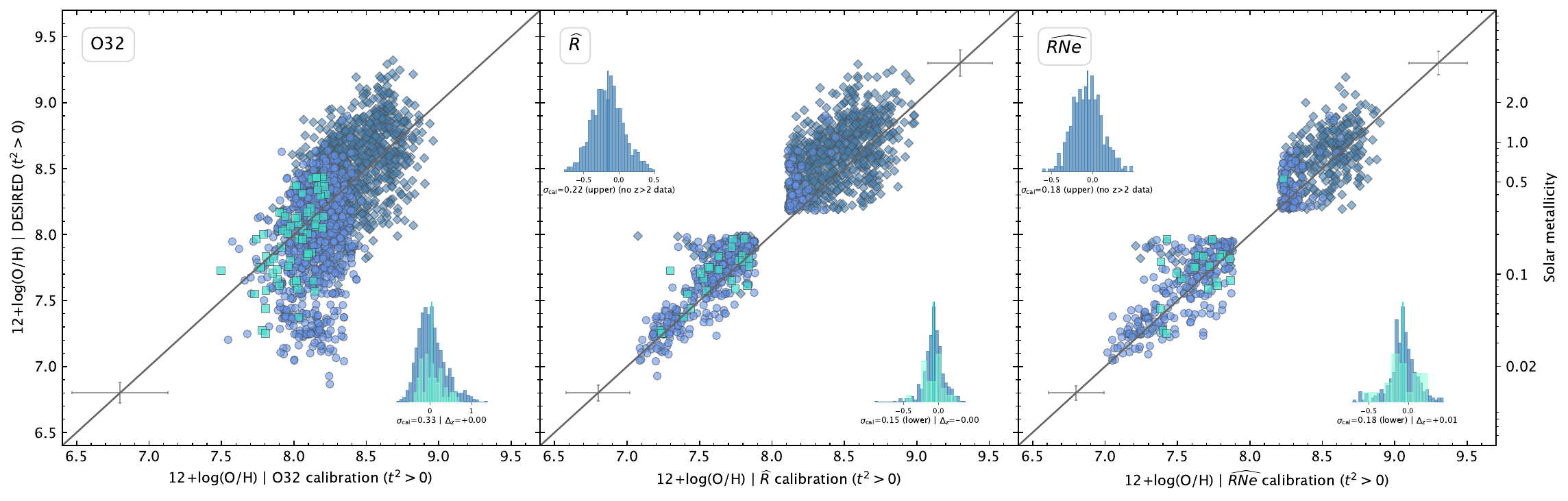}
	\includegraphics[width=\textwidth]{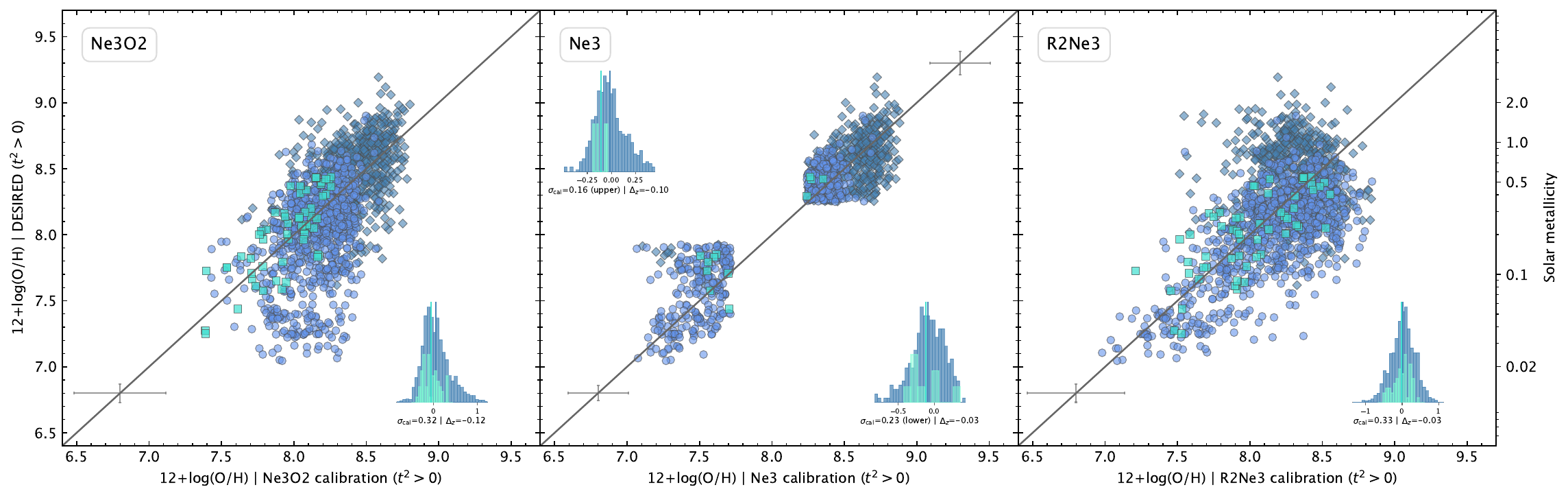}
    \caption{Same as Fig.~\ref{fig:sigma_1}, but for the calibrations and metallicities for the $t^2>0$ case. The values of the standard deviation of the metallicity residuals relative to the $T_{\rm e}$-based abundances, $\sigma_{\rm cal}$, shown in the inset histograms, are reported in Table \ref{tab:coeff_t2ge0} for the $t^2>0$ case.}
    \label{fig:sigma_t2_1}
\end{figure*}

\begin{figure*}
	\includegraphics[width=\textwidth]{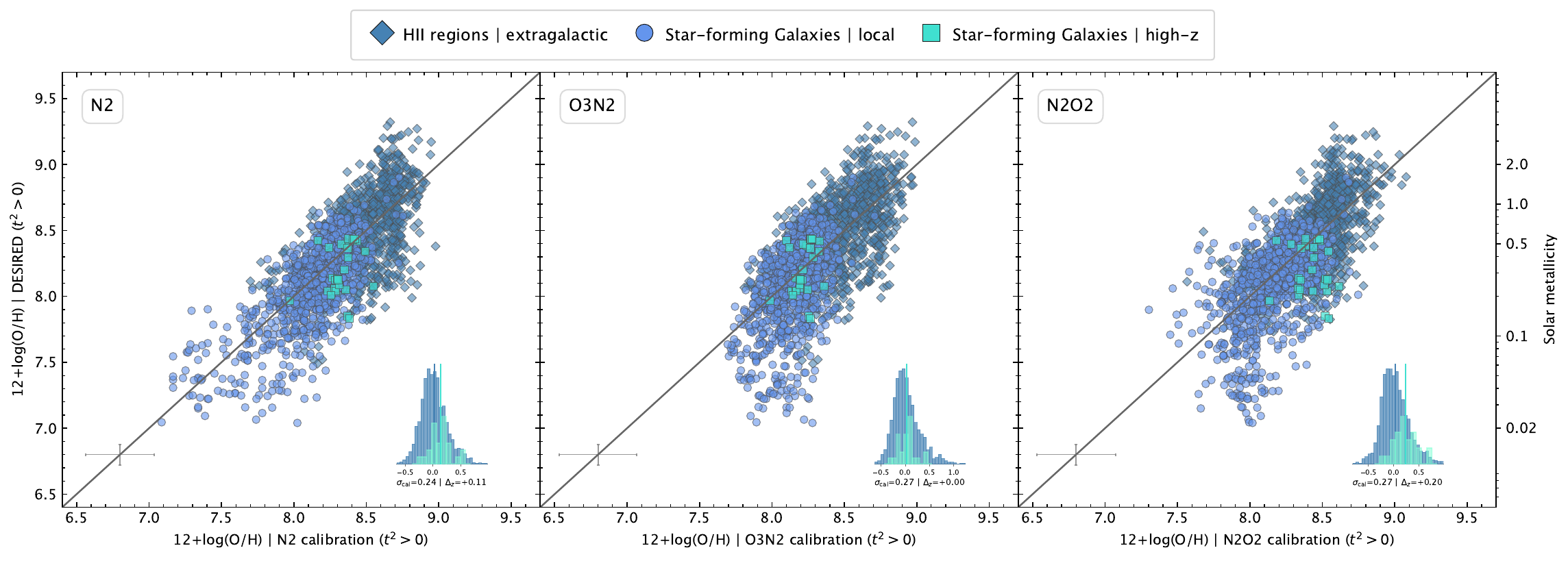}
	\includegraphics[width=\textwidth]{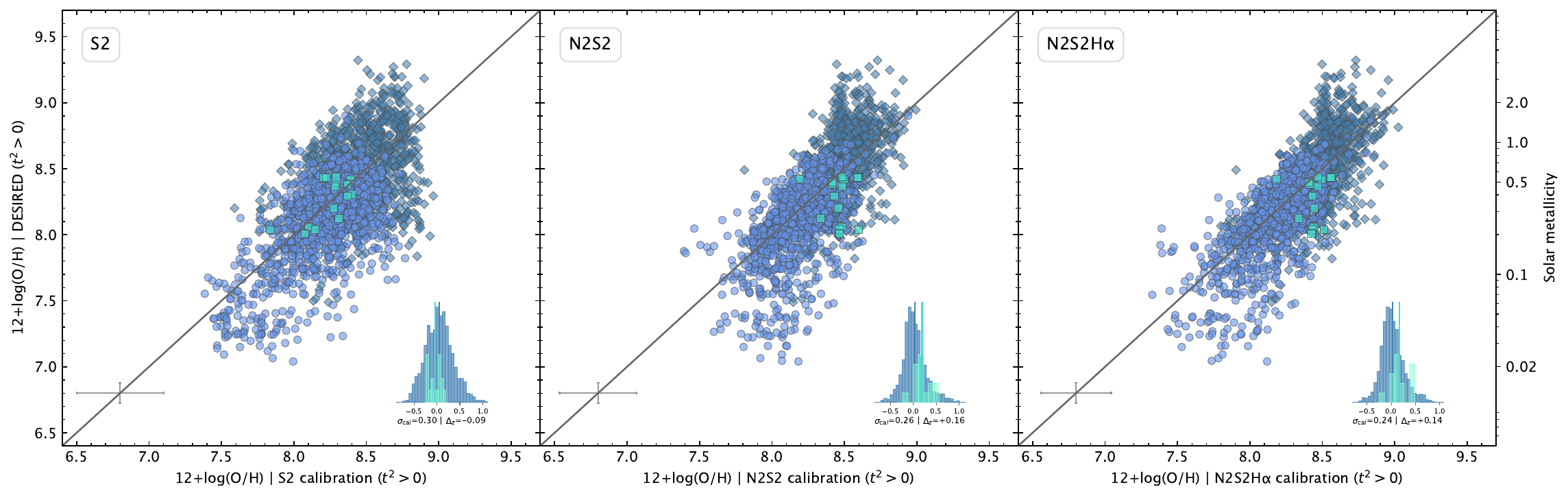}
	\includegraphics[width=\textwidth]{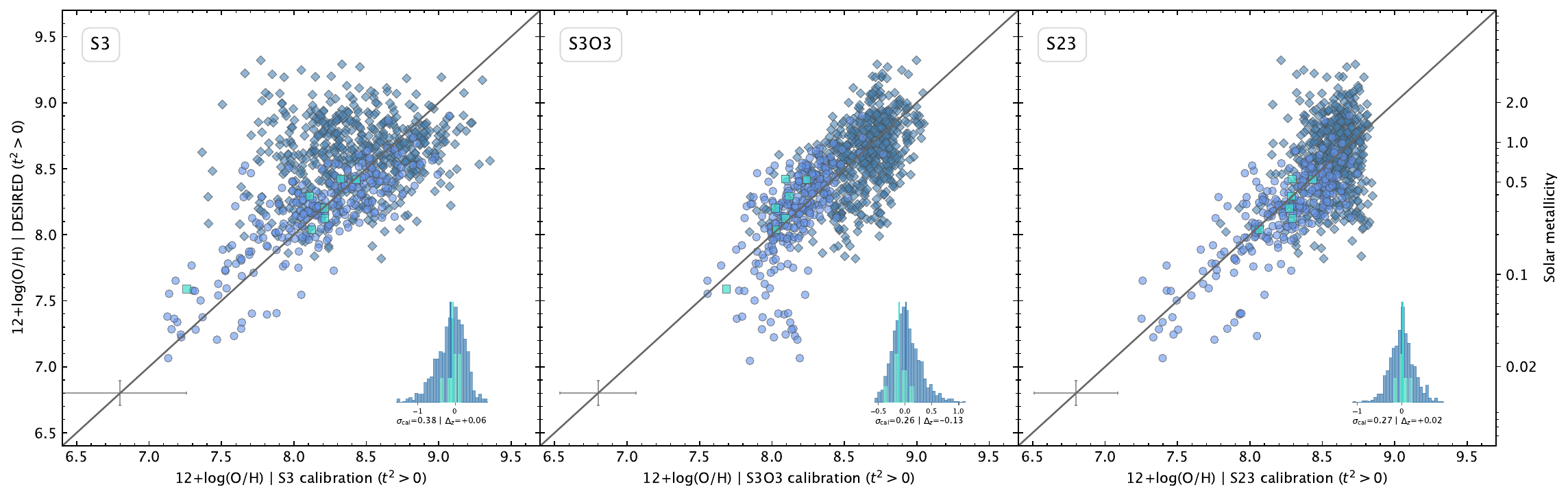}
    \caption{Same as Fig.~\ref{fig:sigma_2}, but for the calibrations and metallicities for the $t^2>0$ case.}
    \label{fig:sigma_t2_2}
\end{figure*}

\begin{figure*}
	\includegraphics[width=\textwidth]{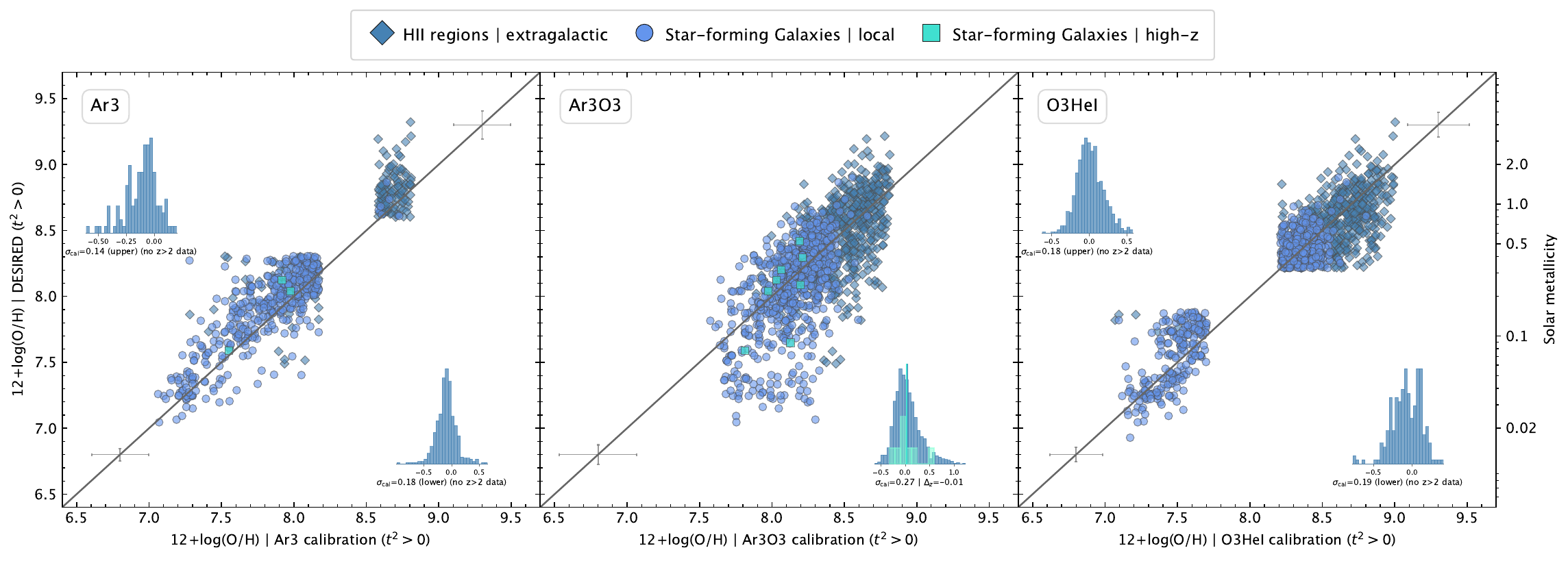}
	\includegraphics[width=\textwidth]{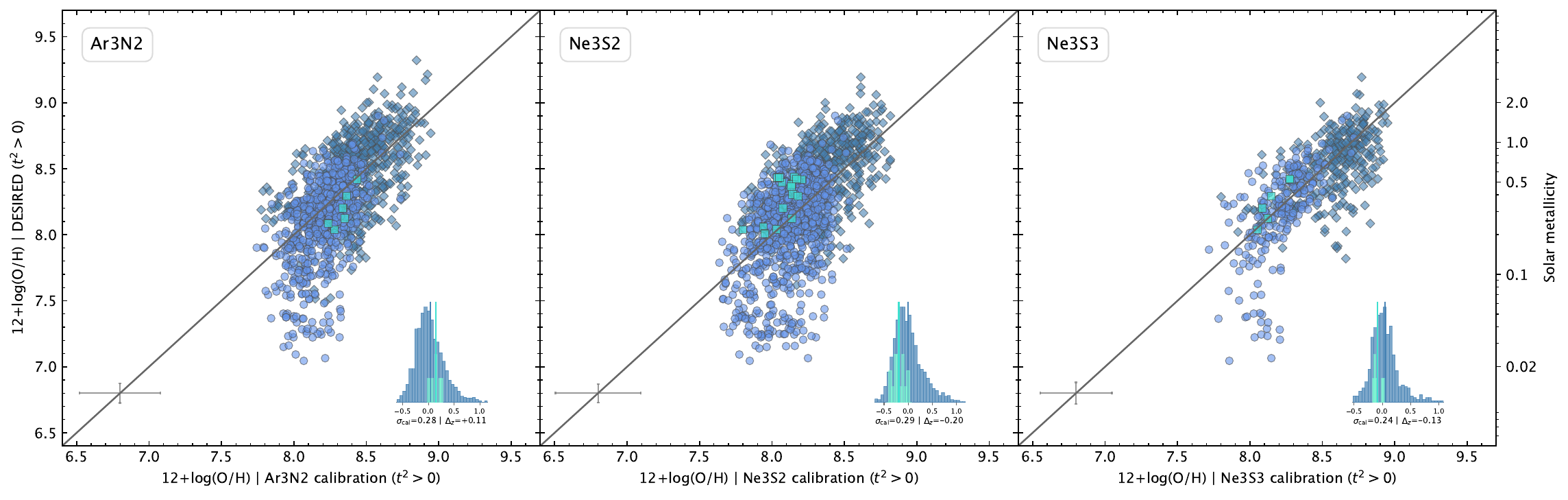}
	\includegraphics[width=\textwidth]{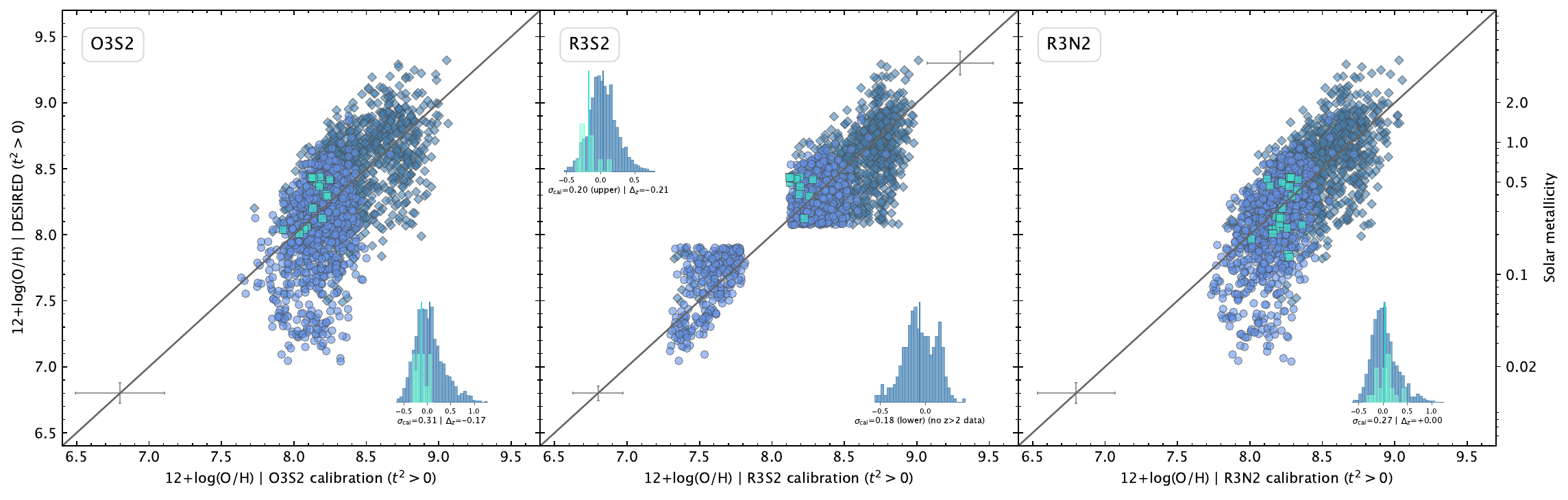}
    \caption{Same as Fig.~\ref{fig:sigma_3}, but for the calibrations and metallicities for the $t^2>0$ case.}
    \label{fig:sigma_t2_3}
\end{figure*}

Figs.~\ref{fig:sigma_2} to \ref{fig:sigma_t2_3} present the calibration–$T_{\rm e}$ comparisons for the remaining diagnostics discussed in Sec.~\ref{sec:sigma}, for both the $t^2=0$ and $t^2>0$ cases. The figures follow the same order adopted in Sec.~\ref{sec:calibrations}: Fig.~\ref{fig:sigma_2} corresponds to the nitrogen- and sulphur-based calibrations, and Fig.~\ref{fig:sigma_3} to the argon-based and newly introduced diagnostics, for the $t^2=0$ case. Figs.~\ref{fig:sigma_t2_1} to \ref{fig:sigma_t2_3} present the analogous comparisons for the $t^2>0$ case, in the same sequence.

Note that each calibration exhibits distinct behaviour across different metallicity regimes. As discussed in Sec.~\ref{sec:guide}, individual calibrations should be applied within the metallicity ranges where the comparison with the direct abundances shows tight clustering around the one-to-one relation, indicating negligible systematic offsets and limited intrinsic scatter.

The corresponding values of $\sigma_{\rm cal}$—defined as the standard deviation of the metallicity residuals relative to the $T_{\rm e}$-based abundances, and thus quantifying the predictive scatter of each calibration in 12 + log(O/H)—are reported in Tables~\ref{tab:coeff_t2eq0} and \ref{tab:coeff_t2ge0} for the $t^2=0$ and $t^2>0$ cases, respectively.


\label{app:sigma}

\section{The DESIRED calibration catalogue}

Table~\ref{tab:catalogue} presents a representative subset of 55 objects from the DESIRED calibration catalogue. For each entry, we list an identification number, the object classification (\hii\ region, local SFG, or high-$z$ SFG), the host galaxy and region name, the derived physical conditions, ionic and total oxygen abundances, and the corresponding bibliographic reference for the spectrum.

The complete DESIRED calibration catalogue, corresponding to Appendix~\ref{app:sample}, is made available online as Supplementary Material.


\begin{landscape}
\begin{table}
\centering
\caption{The DESIRED calibration catalogue (example). For each spectrum, the table lists: {\sc (i)} an identification number (ID); {\sc (ii)} the object classification (\hii\ region, local SFG, or high-$z$ SFG); {\sc (iii--iv)} the host galaxy and/or region name; {\sc (v)} the adopted electron density (in cm$^{-3}$); {\sc (vi--viii)} the derived electron temperatures $T_{3}$(\oiii), $T_{2}$(\nii), and/or $T_{\rm e}$(\siii), in units of 10$^4$ K; {\sc (ix--xi)} the ionic abundances O$^+$ and O$^{2+}$ in 12+log(X$^{i+}$/H$^+$) units (including O$^{2+}$ for the $t^2>0$ case); {\sc (xii-xiv)} the total oxygen abundances expressed as 12+log(O/H) for both $t^2=0$ and $t^2>0$, and those obtained via recombination lines (RL); {\sc (xv)} the corresponding bibliographic reference for the spectrum. The complete catalogue is provided online as Supplementary Material.}
\label{tab:catalogue}
\scriptsize

{\setlength{\tabcolsep}{5pt}  

\begin{tabular}{rcccccccccccccl} 
\toprule
ID & Type & Galaxy & Region & $n_e$ & $T_{3}$(\oiii) & $T_{2}$(\nii) & $T_{\rm e}$(\siii) & O$^+$ & O$^{2+}$ & O$^{2+}_{t^2>0}$ & 12+log(O/H)$_{t^2=0}$ & 12+log(O/H)$_{t^2>0}$ & 12+log(O/H)$_{\rm RL}$ & Reference \\
{\sc (i)} & {\sc (ii)} & {\sc (iii)} & {\sc (iv)} & {\sc (v)} & {\sc (vi)} & {\sc (vii)} & {\sc (viii)} & {\sc (ix)} & {\sc (x)} & {\sc (xi)} & {\sc (xii)} & {\sc (xiii)} & {\sc (xiv)} & {\sc (xv)} \\
\midrule

\#1 & HII & CG-1419 & 1440-53084-103 & 100 (100) & 1.44 (0.12) &  \NA  &  \NA  & 7.19 (0.28) & 7.38 (0.12) & 7.56 (0.12) & 7.60 (0.11) & 7.72 (0.10) &  \NA  &  \citet{Pilyugin:07} \\ 
\#2 & HII & CG-1419 & 1996-53436-330 & 100 (100) & 1.62 (0.12) &  \NA  &  \NA  & 7.07 (0.27) & 7.26 (0.09) & 7.41 (0.10) & 7.47 (0.11) & 7.58 (0.09) &  \NA  &  \citet{Pilyugin:07} \\ 
\#3 & HII & HS-1103+4346 & 1363-53053-138 & 100 (100) & 1.65 (0.12) &  \NA  &  \NA  & 7.01 (0.19) & 7.30 (0.07) & 7.45 (0.09) & 7.48 (0.07) & 7.58 (0.07) &  \NA  &  \citet{Pilyugin:07} \\ 
\#4 & HII & HS-1103+4346 & 1364-53061-295 & 100 (100) & 1.72 (0.08) &  \NA  &  \NA  & 6.96 (0.12) & 7.33 (0.04) & 7.47 (0.05) & 7.48 (0.04) & 7.58 (0.05) &  \NA  &  \citet{Pilyugin:07} \\ 
\#5 & HII & HS-1132+4416 & 1366-53063-083 & 100 (100) & 1.47 (0.07) &  \NA  &  \NA  & 7.14 (0.16) & 7.40 (0.06) & 7.57 (0.07) & 7.59 (0.06) & 7.71 (0.06) &  \NA  &  \citet{Pilyugin:07} \\ 
\#6 & HII & IC10 & HL90-120 & 135 (226) & 0.91 (0.04) &  \NA  &  \NA  & 7.36 (0.11) & 8.29 (0.09) & 8.57 (0.09) & 8.34 (0.07) & 8.60 (0.08) &  \NA  &  \citet{Magrini:09} \\ 
\#7 & HII & IC10 & HL90-29 & 158 (184) & 1.54 (0.24) &  \NA  &  \NA  & 7.38 (0.42) & 7.85 (0.28) & 8.01 (0.31) & 7.96 (0.18) & 8.09 (0.20) &  \NA  &  \citet{Magrini:09} \\ 
\#8 & HII & IC10 & HL90-50 & 228 (187) & 1.24 (0.04) &  \NA  &  \NA  & 7.32 (0.06) & 7.80 (0.04) & 8.00 (0.05) & 7.93 (0.04) & 8.09 (0.04) &  \NA  &  \citet{Magrini:09} \\ 
\#9 & HII & IC1613 & S3 & 100 (100) & 1.79 (0.07) &  \NA  &  \NA  & 6.79 (0.05) & 7.58 (0.04) & 7.71 (0.05) & 7.64 (0.03) & 7.76 (0.04) &  \NA  &  \citet{Bresolin:07b} \\ 
\#10 & HII & IC1613 & S7 & 100 (100) & 1.44 (0.07) &  \NA  &  \NA  & 7.54 (0.09) & 7.43 (0.06) & 7.60 (0.07) & 7.79 (0.05) & 7.87 (0.05) &  \NA  &  \citet{Bresolin:07b} \\ 
\#11 & HII & IC2458 & +002–006 & 100 (100) & 1.30 (0.05) &  \NA  &  \NA  & 7.58 (0.06) & 7.88 (0.04) & 8.08 (0.05) & 8.06 (0.03) & 8.20 (0.04) &  \NA  &  \citet{vanZee:98} \\ 
\#12 & HII & IC2458 & –033–007 & 225 (189) & 1.31 (0.10) &  \NA  &  \NA  & 7.28 (0.15) & 7.87 (0.12) & 8.05 (0.12) & 7.96 (0.09) & 8.13 (0.08) &  \NA  &  \citet{vanZee:98} \\ 
\#13 & HII & LMC & IC2111 & 272 (119) & 0.91 (0.02) & 0.98 (0.03) & 0.93 (0.03) & 8.06 (0.07) & 8.19 (0.03) & 8.39 (0.06) & 8.43 (0.03) & 8.56 (0.04) & 8.51 (0.07) &  \citet{DominguezGuzman:22} \\ 
\#14 & HII & LMC & N11B & 264 (90) & 0.91 (0.01) & 0.98 (0.01) & 1.05 (0.02) & 7.98 (0.03) & 8.19 (0.02) & 8.38 (0.03) & 8.40 (0.02) & 8.53 (0.02) & 8.52 (0.02) &  \citet{DominguezGuzman:22} \\ 
\#15 & HII & LMC & N44C & 149 (72) & 1.13 (0.02) & 1.05 (0.04) & 1.15 (0.03) & 7.31 (0.07) & 8.23 (0.02) & 8.58 (0.06) & 8.28 (0.02) & 8.61 (0.05) & 8.59 (0.04) &  \citet{DominguezGuzman:22} \\ 
\#16 & HII & LMC & NGC1714 & 405 (156) & 0.95 (0.02) & 1.01 (0.05) & 0.96 (0.04) & 7.68 (0.10) & 8.28 (0.03) & 8.47 (0.09) & 8.38 (0.03) & 8.53 (0.07) & 8.52 (0.06) &  \citet{DominguezGuzman:22} \\ 
\#17 & HII & M101 & 1323-52797-002 & 100 (100) & 1.18 (0.06) &  \NA  &  \NA  & 7.73 (0.14) & 7.56 (0.07) & 7.76 (0.08) & 7.95 (0.08) & 8.04 (0.08) &  \NA  &  \citet{Pilyugin:07} \\ 
\#18 & HII & M101 & 1323-52797-008 & 100 (100) & 1.13 (0.08) &  \NA  &  \NA  & 7.74 (0.21) & 7.51 (0.10) & 7.73 (0.12) & 7.94 (0.11) & 8.04 (0.09) &  \NA  &  \citet{Pilyugin:07} \\ 
\#19 & HII & M101 & 1324-53088-234 & 100 (100) & 1.94 (0.06) &  \NA  &  \NA  & 6.85 (0.08) & 7.10 (0.03) & 7.23 (0.03) & 7.30 (0.03) & 7.38 (0.03) &  \NA  &  \citet{Pilyugin:07} \\ 
\#20 & HII & M101 & 1324-53088-263 & 100 (100) & 1.68 (0.07) &  \NA  &  \NA  & 6.93 (0.15) & 7.28 (0.04) & 7.43 (0.05) & 7.45 (0.05) & 7.55 (0.05) &  \NA  &  \citet{Pilyugin:07} \\ 
\#21 & HII & M101 & 1324-53088-271 & 100 (100) & 1.28 (0.16) &  \NA  &  \NA  & 7.50 (0.50) & 7.21 (0.20) & 7.39 (0.27) & 7.68 (0.24) & 7.76 (0.19) &  \NA  &  \citet{Pilyugin:07} \\ 
\#22 & HII & M101 & 1324-53088-279 & 100 (100) & 1.55 (0.08) &  \NA  &  \NA  & 7.14 (0.14) & 7.23 (0.04) & 7.38 (0.05) & 7.48 (0.07) & 7.58 (0.06) &  \NA  &  \citet{Pilyugin:07} \\ 
\#23 & HII & M101 & 1325-52762-348 & 100 (100) & 1.31 (0.18) &  \NA  &  \NA  & 7.18 (0.47) & 7.30 (0.21) & 7.49 (0.24) & 7.54 (0.17) & 7.66 (0.15) &  \NA  &  \citet{Pilyugin:07} \\ 
\#24 & HII & M101 & 1325-52762-350 & 100 (100) & 1.48 (0.12) &  \NA  &  \NA  & 7.09 (0.24) & 7.37 (0.13) & 7.53 (0.13) & 7.55 (0.10) & 7.67 (0.10) &  \NA  &  \citet{Pilyugin:07} \\ 
\#25 & HII & M101 & 1325-52762-352 & 100 (100) & 1.22 (0.08) &  \NA  &  \NA  & 7.67 (0.19) & 7.55 (0.09) & 7.75 (0.10) & 7.91 (0.10) & 8.02 (0.09) &  \NA  &  \citet{Pilyugin:07} \\ 
\#26 & HII & M101 & H1013 & 100 (100) & 0.76 (0.02) & 0.81 (0.02) & 0.77 (0.06) & 8.33 (0.06) & 8.01 (0.05) & 8.44 (0.07) & 8.50 (0.04) & 8.69 (0.05) &  \NA  &  \citet{Bresolin:07a} \\ 
\#27 & HII & M101 & H1013 & 100 (100) & 0.74 (0.02) & 0.76 (0.02) & 0.69 (0.01) & 8.44 (0.07) & 8.05 (0.05) & 8.68 (0.07) & 8.59 (0.05) & 8.88 (0.04) &  \NA  &  \citet{Croxall:16} \\ 
\#28 & HII & M101 & H1013 & 100 (100) & 0.73 (0.06) & 0.77 (0.03) &  \NA  & 8.23 (0.11) & 8.07 (0.18) & 8.60 (0.09) & 8.47 (0.08) & 8.75 (0.06) &  \NA  &  \citet{Esteban:09} \\ 
\#29 & HII & M101 & H1013 & 100 (100) &  \NA  & 0.80 (0.06) & 0.67 (0.03) & 8.29 (0.23) & 8.13 (0.17) & 8.51 (0.19) & 8.52 (0.12) & 8.72 (0.12) &  \NA  &  \citet{Kennicutt:03} \\ 
\#30 & HII & M101 & H1018 & 100 (100) &  \NA  & 0.76 (0.04) & 0.77 (0.05) & 8.36 (0.16) & 7.80 (0.14) & 8.21 (0.15) & 8.46 (0.11) & 8.59 (0.10) &  \NA  &  \citet{Croxall:16} \\ 
\#31 & HII & M101 & H103 & 100 (100) & 0.90 (0.04) & 0.94 (0.10) & 1.12 (0.11) & 8.18 (0.26) & 7.89 (0.07) & 8.15 (0.30) & 8.36 (0.14) & 8.46 (0.16) &  \NA  &  \citet{Croxall:16} \\ 
\#32 & HII & M101 & H1040 & 100 (100) &  \NA  & 0.74 (0.06) & 0.75 (0.03) & 8.50 (0.18) & 8.08 (0.16) & 8.50 (0.14) & 8.64 (0.12) & 8.80 (0.10) &  \NA  &  \citet{Croxall:16} \\ 
\#33 & HII & M101 & H1045 & 100 (100) &  \NA  & 0.77 (0.04) & 0.73 (0.03) & 8.30 (0.12) & 8.00 (0.10) & 8.39 (0.10) & 8.47 (0.07) & 8.65 (0.07) &  \NA  &  \citet{Croxall:16} \\ 
\#34 & HII & M101 & H104 & 100 (100) & 1.08 (0.03) &  \NA  &  \NA  & 7.92 (0.05) & 7.56 (0.04) & 7.79 (0.04) & 8.08 (0.04) & 8.16 (0.03) &  \NA  &  \citet{Croxall:16} \\ 
\#35 & HII & M101 & H1052 & 210 (62) & 0.78 (0.01) & 0.84 (0.03) & 0.89 (0.02) & 8.13 (0.06) & 8.39 (0.02) & 8.74 (0.07) & 8.58 (0.03) & 8.84 (0.05) &  \NA  &  \citet{Croxall:16} \\ 
\#36 & HII & M101 & H1105 & 201 (97) & 0.89 (0.02) & 0.90 (0.08) & 0.79 (0.02) & 8.03 (0.27) & 8.25 (0.05) & 8.59 (0.23) & 8.45 (0.09) & 8.68 (0.16) &  \NA  &  \citet{Kennicutt:03} \\ 
\#37 & HII & M101 & H1118 & 100 (100) & 1.25 (0.06) &  \NA  &  \NA  & 7.62 (0.09) & 7.93 (0.07) & 8.13 (0.08) & 8.10 (0.05) & 8.24 (0.05) &  \NA  &  \citet{Esteban:20} \\ 
\#38 & HII & M101 & H1118 & 100 (100) & 1.17 (0.03) &  \NA  &  \NA  & 7.67 (0.05) & 8.09 (0.04) & 8.29 (0.04) & 8.23 (0.03) & 8.39 (0.04) &  \NA  &  \citet{Li:13} \\ 
\#39 & HII & M101 & H1122 & 100 (100) & 0.90 (0.02) &  \NA  & 0.90 (0.06) & 8.13 (0.04) & 7.92 (0.04) & 8.20 (0.04) & 8.34 (0.03) & 8.46 (0.03) &  \NA  &  \citet{Croxall:16} \\ 
\#40 & HII & M101 & H1125 & 100 (100) & 1.19 (0.02) &  \NA  &  \NA  & 7.75 (0.03) & 7.98 (0.02) & 8.19 (0.03) & 8.18 (0.02) & 8.32 (0.02) &  \NA  &  \citet{Croxall:16} \\ 
\#41 & HII & M101 & H1146 & 100 (100) & 1.07 (0.02) &  \NA  &  \NA  & 7.87 (0.03) & 8.10 (0.02) & 8.33 (0.03) & 8.30 (0.02) & 8.46 (0.02) &  \NA  &  \citet{Croxall:16} \\ 
\#42 & HII & M101 & H1146 & 100 (100) & 1.19 (0.10) &  \NA  &  \NA  & 7.56 (0.17) & 7.97 (0.14) & 8.19 (0.15) & 8.10 (0.09) & 8.28 (0.11) &  \NA  &  \citet{Esteban:20} \\ 
\#43 & HII & M101 & H1146 & 100 (100) & 1.01 (0.02) &  \NA  &  \NA  & 7.85 (0.04) & 8.21 (0.03) & 8.46 (0.04) & 8.37 (0.03) & 8.55 (0.03) &  \NA  &  \citet{Li:13} \\ 
\#44 & HII & M101 & H1148 & 100 (100) & 1.10 (0.03) &  \NA  &  \NA  & 7.92 (0.05) & 7.85 (0.04) & 8.07 (0.04) & 8.19 (0.03) & 8.30 (0.03) &  \NA  &  \citet{Croxall:16} \\ 
\#45 & HII & M101 & H1151 & 100 (100) & 1.08 (0.02) &  \NA  &  \NA  & 7.75 (0.03) & 8.01 (0.02) & 8.24 (0.03) & 8.20 (0.02) & 8.36 (0.02) &  \NA  &  \citet{Croxall:16} \\ 
\#46 & HII & M101 & H1159 & 100 (100) & 0.96 (0.08) &  \NA  & 0.92 (0.06) & 7.88 (0.17) & 8.12 (0.13) & 8.37 (0.15) & 8.32 (0.09) & 8.49 (0.10) &  \NA  &  \citet{Kennicutt:03} \\ 
\#47 & HII & M101 & H1170 & 100 (100) & 1.05 (0.06) &  \NA  & 0.78 (0.04) & 7.96 (0.09) & 7.79 (0.09) & 8.02 (0.10) & 8.18 (0.06) & 8.29 (0.06) &  \NA  &  \citet{Kennicutt:03} \\ 
\#48 & HII & M101 & H1176 & 100 (100) & 0.99 (0.04) &  \NA  & 0.86 (0.04) & 7.75 (0.09) & 8.14 (0.06) & 8.39 (0.07) & 8.29 (0.05) & 8.48 (0.06) &  \NA  &  \citet{Kennicutt:03} \\ 
\#49 & HII & M101 & H1216 & 100 (100) & 1.06 (0.01) & 1.02 (0.15) & 0.99 (0.03) & 7.83 (0.41) & 8.09 (0.02) & 8.42 (0.32) & 8.27 (0.11) & 8.52 (0.19) &  \NA  &  \citet{Croxall:16} \\ 
\#50 & HII & M101 & H1216 & 100 (100) & 1.06 (0.01) & 1.09 (0.03) &  \NA  & 7.60 (0.04) & 8.09 (0.02) & 8.26 (0.05) & 8.21 (0.02) & 8.35 (0.04) &  \NA  &  \citet{Esteban:20} \\ 
\#51 & HII & M101 & H1216 & 100 (100) & 1.15 (0.03) &  \NA  & 1.00 (0.04) & 7.53 (0.04) & 8.04 (0.04) & 8.25 (0.04) & 8.15 (0.03) & 8.32 (0.04) &  \NA  &  \citet{Kennicutt:03} \\ 
\#52 & HII & M101 & H1231 & 100 (100) & 1.40 (0.08) &  \NA  &  \NA  & 7.60 (0.10) & 7.88 (0.07) & 8.05 (0.09) & 8.06 (0.06) & 8.19 (0.06) &  \NA  &  \citet{Croxall:16} \\ 
\#53 & HII & M101 & H1248 & 100 (100) & 1.20 (0.02) &  \NA  & 1.33 (0.16) & 7.67 (0.04) & 7.87 (0.03) & 8.07 (0.03) & 8.08 (0.02) & 8.21 (0.02) &  \NA  &  \citet{Croxall:16} \\ 
\#54 & HII & M101 & H128 & 100 (100) & 0.88 (0.03) & 0.96 (0.12) & 0.84 (0.03) & 7.81 (0.43) & 8.35 (0.07) & 8.56 (0.36) & 8.46 (0.08) & 8.63 (0.23) &  \NA  &  \citet{Kennicutt:03} \\ 
\#55 & HII & M101 & H143 & 100 (100) & 0.97 (0.01) & 0.94 (0.04) & 1.12 (0.04) & 8.09 (0.09) & 8.02 (0.01) & 8.44 (0.09) & 8.36 (0.04) & 8.60 (0.07) &  \NA  &  \citet{Croxall:16} \\ 


\dots \\
\bottomrule
\end{tabular} }
\end{table}
\end{landscape}



\label{app:sample}


\bsp	
\label{lastpage}
\end{document}